\documentclass[journal]{IEEEtran}

\usepackage{times}
\usepackage{graphicx}
\usepackage{placeins}
\usepackage[export]{adjustbox}
\usepackage{ifxetex}
\usepackage{flafter}
\usepackage{amsmath}
\usepackage{amsthm}
\usepackage{amsfonts}
\usepackage{amssymb}
\usepackage[utf8]{inputenc}
\usepackage{xcolor}
\usepackage[space]{grffile} 
\usepackage{epstopdf}
\usepackage{subfloat}
\usepackage{subfig}
\usepackage{xspace}
\usepackage{mathtools}
\usepackage[labelfont=bf,font=small]{caption}
\usepackage{booktabs}
\usepackage{array,multirow}
\usepackage{bm}
\usepackage{bbm}
\usepackage{dsfont}
\usepackage{url}
\usepackage{hyperref}

\usepackage[square, sort, numbers]{natbib}
	
\usepackage{lineno}

\usepackage{expl3}
\ExplSyntaxOn
\cs_new_eq:NN \fppeval \fp_eval:n
\ExplSyntaxOff

\newtheorem{definition}{Definition}
\newtheorem{proposition}{Proposition}
\newtheorem*{proposition1}{Proposition 1}
\newtheorem*{proposition2}{Proposition 2}
\newtheorem*{proposition3}{Proposition 3}
\newtheorem*{proposition4}{Proposition 4}
\newtheorem*{proposition5}{Proposition 5}
\newtheorem*{proposition6}{Proposition 6}
\newtheorem{lemma}{Lemma}

\makeatletter
\DeclareRobustCommand*{\escapeus}[1]{%
    \begingroup\@activeus\scantokens{#1\endinput}\endgroup}
\begingroup\lccode`\~=`\_\relax
    \lowercase{\endgroup\def\@activeus{\catcode`\_=\active \let~\_}}
\makeatother

\makeatletter
\newcommand*\bigcdot{\mathpalette\bigcdot@{.5}}
\newcommand*\bigcdot@[2]{\mathbin{\vcenter{\hbox{\scalebox{#2}{$\m@th#1\bullet$}}}}}
\makeatother

\newcommand{\st} 			{\ \textup{s.t.}\ }

\newcommand{\overN}[1]	{_{#1= 1}^{N}}

\usepackage[normalem]{ulem}

\newcommand{\inlinetitle}[1]{\noindent\textbf{#1.}}

\newcounter{marginNoteCounter}

\newcommand{\Opinion}{X}
\newcommand{\minOpinion}{x}  
\newcommand{\maxOpinion}{X}  
\newcommand{\overbar}[1] {\mkern 1.5mu\overline{\mkern-3.5mu#1\mkern-1.0mu}\mkern 1.5mu}
\newcommand{\V}{V} 
     
\newcommand{\zero}       {\mathbf{0}}    
\newcommand{\ind}       {\mathds{1}}
\newcommand{\norm}[1]     {\lVert #1 \rVert}

\renewcommand*{\top}{{\mkern-1.5mu\mathsf{T}}}

\newcommand{\Appendix}       {Appendix} 

\newcommand{\Sec}[1]	{Sec.\,\ref{#1}}
\newcommand{\Eq}[1]	  {Eq.\,\ref{#1}}
\newcommand{\Fig}[1]	{Fig.\,\ref{#1}}
\newcommand{\Def}[1]	{Def.\,\ref{#1}}
\newcommand{\Tab}[1]	{Tab.\,\ref{#1}}

\def\modelname {GSM-DeGroot\xspace}

\usepackage{ulem}
\usepackage{contour}

\contourlength{0.8pt}

\title{\vspace{-3mm}\huge 
DeGroot-based opinion formation \\
under a global steering mechanism
}

\author{Ivan Conjeaud$^{*\diamond}$ \qquad Philipp Lorenz-Spreen$^+$ \qquad Argyris Kalogeratos$^{*\dagger}$\\
{\small $^*$Centre Borelli, ENS Paris-Saclay, Gif-sur-Yvette, France}\\
{\small $^\diamond$Paris School of Economics, Paris, France}\\
{\small $^+$Max Planck Institute for Human Development, Berlin, Germany}%
\thanks{\small $^\dagger$I.C. and A.K contributed equally to this work. \hfill ~\hfill Correspondence to: A.K.; email: argyris.kalogeratos@ens-paris-saclay.fr.
}
\thanks{Manuscript received December XX, XXXX; revised XXXX XX, XXXX.}%
}
\date{}%

\begin{document}

\maketitle

\begin{abstract}
This paper investigates how interacting agents arrive to a consensus or a polarized state. 
We study the opinion formation process under the effect of a global steering mechanism (GSM), which aggregates the opinion-driven stochastic agent states at the network level and feeds back to them a form of global information. We also propose a new two-layer agent-based opinion formation model, called \emph{\modelname}, that captures the coupled dynamics between agent-to-agent local interactions and the GSM's steering effect. This way, agents are subject to the effects of a DeGroot-like local opinion propagation, as well as to a wide variety of possible aggregated information that can affect their opinions, such as trending news feeds, press coverage, polls, elections, etc. Contrary to the standard DeGroot model, our model allows polarization to emerge by letting agents react to the global information in a stubborn differential way. Moreover, the introduced stochastic agent states produce event stream dynamics that can fit to real event data. We explore numerically the model dynamics to find regimes of qualitatively different behavior. We also challenge our model by fitting it to the dynamics of real topics that attracted the public attention and were recorded on Twitter. Our experiments show that the proposed model holds explanatory power, as it evidently captures real opinion formation dynamics via a relatively small set of interpretable parameters.

\smallskip
\end{abstract}

\section{Introduction}\label{sec:intro}

The explosive development of new electronic communication means is heavily impacting the self-organized social dynamics of opinion formation and political participation, in ways that are not fully-understood. Our limited view over the incurred changes is partly due to the fact that we lack expressive yet interpretable models that could account for the complex multilevel information pathways that become available through modern communication technology. To advance our understanding, what is mostly needed is rather simple models able to highlight a meaningful prototypical agent-based mechanism that drives opinion formation.

The landscape in which modern public debate takes place, includes national and international broadcasting media, and more recently online social networking platforms, which have altered the way and the speed with which people exchange information \cite{hilbert2011world, lorenz2019accelerating}. Especially for the exchanges on online platforms, these have substituted part of the physical interactions between individuals, and have led to a reshaping of the social network formed around each individual \cite{COP2020}, e.g.~by having a wider set of contacts including weak-ties and contacts that are geographically remote. The transition from one-to-many to many-to-many communication that these platforms allow has brought new attention to self-organized social behavior, like the new ways of political participation through digital media \cite{boulianne2020twenty}. Among the interesting related phenomena, one can find some that are emergent, such as price formation%
, panic buying, overnight formation of social movements, persistent rumors, and self-organized fake news circulation \cite{Bettencourt2006,Jin2013,del2016spreading, KALOGERATOS2018651}. A recent spur in modeling efforts for such phenomena from a complex system perspective, largely concerns opinion dynamics \cite{Review}. Several recent analyses and models have either focused on misinformation spreading or polarization dynamics \cite{Baumann, PhysRevX.11.011012}. Only few modeling efforts have explicitly studied the interplay between individual opinion dynamics, which are driven locally by social influence, and the correlation of different debate topics that co-evolve in a multidimensional space \cite{quattrociocchi2014opinion}.

The theoretical literature on modeling opinion dynamics comprises mainly two streams, one with models considering opinions as continuous variables, and another one considering them as binary or discrete variables. The first one contains models based on the DeGroot model \cite{degroot}, which is itself a generalization of French's seminal work \cite{French}. A great variety of generalizations and variations of this model have been proposed, mainly by relaxing the assumption that the influence between any two agents is fixed, and allowing instead to vary as functions of time or the opinion of the nodes \cite{FriedkinJohnsen,HegselmannKrause2002,HegselmannKrause2005}. Continuous modeling is not restricted to use DeGroot-Friedkin 
models \cite{degroot,friedkin,Friedkin-Formal-Theory} as a basis, but rather includes a variety of other models \cite{Review, Baumann, Gelation, stubborn2012}. The other stream of research, initiated by Granovetter \cite{Granovetter}, considers opinions as binary (or discrete) variables and frequently adopts a game theoretical approach, in which opinions are considered as strategies that give each time the best response to the state of the local environment \cite{Morris, GHADERI20143209}, or a physics-like approach in which opinions are states, with models adapted from physics to social sciences \cite{Sznajd,Grabowski}. Often, these models can be summed up to threshold models, where an opinion state is adopted when a sufficiently large proportion of a node's neighborhood has done so \cite{Watts-model2002}. Both these research streams have boosted the interest in understanding the opinion formation process, consensus formation \cite{degroot,golub}, maintenance of diversity despite increasing local resemblance \cite{axelrod1997dissemination}, with some attempts to model global interaction on top of the one at the local level \cite{Deffuant,Schulze}. Such models are limited as they define global interactions to be also peer-to-peer, whereas with other arbitrarily distant agents. 

\inlinetitle{DeGroot-based modeling}~%
At the core of many of the opinion formation studies is the DeGroot-based modeling \cite{degroot, friedkin}, which is also central in this work. The classic DeGroot model considers only local interactions between neighboring agents, and brings their opinions closer and closer. An agent can still be influenced by any other if there exists a path connecting them, but only through step-by-step bilateral interactions involving intermediaries. Essentially, this simulates the primordial idea that an agent's opinion is driven mainly by locally influential individuals \cite{katz} and her tendency to conform with her social environment. The DeGroot model is prototypical and insightful as a mechanism, but, it comes with a number of notable limitations, most of which have occupied the literature. %

First, the local smoothing of opinions, under weak assumptions, leads always to global consensus. Consequently, it is unable to generate opinion diversity or polarization (i.e.~multimodal consensus) on its own. To fill the gap, there have been conjectures and speculations about  mechanisms that could allow such phenomena to emerge. One idea is that polarization can come from \emph{stubborn agents} that are not eager to change their positions regardless the changes in their social surrounding, and therefore act as diverse attractors \cite{stubborn2012,Acemoglu-stubborn2013, TIAN2018213}. Another one, also at the local level, stipulates that \emph{signed networks}, which model local attraction-repulsion, can also lead to polarization \cite{nguyen2017opinion}. We discuss in technical terms that these approaches lead to limited polarization, specifically upper-bounded by the initial conditions (see \Sec{sec:polarization-sources}). One may point out that the attempts to explain opinion divergence introduce \emph{pre-inscribed features} to the system, either at the connectivity level, or at the agents' opinion update level. This implies that divergence is not really generated by the process itself%
, but is due to the pre-inscribed features that push the system to polarized states. The pre-inscribed features can be the result of deeper beliefs or psychological factors that do not change during a short-term debate, such as those taking place in social media. Few works have tried to include psychological factors that can cause an agent's behavior to change during the opinion formation, e.g.~the notion of \emph{tolerance} that makes an agent's opinion to saturate the more agreement there is in her neighborhood \cite{Topirceanu2016}.%

Second, by conceptualizing the opinion formation as taking place strictly through peer-to-peer interactions, it lacks any mechanism of broadcasting or aggregation of agents' opinions%
, or ways for agents to get feedback from the global state of the debate over the network; hence it leaves mass-media effects completely out of its scope. In reality, such mechanisms become more and more relevant due to the fact that it is natural for agents who operate under cognitive and time constraints to seek for summarized or filtered information sources. In the modern landscape there are new interacting entities and information pathways \cite{bernays,chomsky,Nadeau,McAllister}, as well as the increased coupling of local and global information flows (e.g.~mass media picking up on social media trends), which are usually in place simultaneously \cite{tsfati2020causes}. 

Third, a point of our criticism that is somewhat related to the previous one, the DeGroot-based modeling rarely considers political participation as an important aspect of the opinion formation. However, political participation has been found to be reliably associated with media usage, and especially social media \cite{Boulianne,boulianne2009does}. We accordingly argue that for an agent, public expression beyond her narrow social environment and political participation are intertwined with her opinion, which is a mostly overlooked feature in the literature. In this work, we regard agents as being in conversation with both their local environments and the global state-of-things represented by information aggregation. %
Furthermore, and related to the first point of criticism about polarization, we argue that the attraction or repulsion to information aggregation can be more important as a factor producing opinion diversity, compared to similar local level reactions, for several reasons. To mention a few: i) local reactions going against an agent's social surrounding is likely to be frictional and costly; ii) the effects of this kind of local disagreement can be negligible compared to the -usually more frequent- interactions with global information that is supposed to be more representative for the state of the debate at the whole network level; iii) for the same reason, information aggregation is likely to generate structured reactions, while local disagreement is not. 

Fourth, a point of general criticism to all the stream of classical opinion formation modeling is that it idealizes the process (e.g.~by assuming that opinions are visible and subject to direct exchange between agents, by considering simplistic opinion propagation and update rules, or by ignoring psychological aspects in agents' reactions) and does not offer in the end sufficient tools for addressing problems involving real data \cite{InfoDiff_KDD2012, MSD_WSDM2014, SLANT2016, peralta2022opinion,HP2017}.

\inlinetitle{Contribution}~%
In this paper, we present the \modelname model that aims at capturing the intertwined relationship between each agent's opinion (a continuous variable) and the publicly visible political expression or participation (e.g.~protest participation, posting on social networking platforms, etc.), which is represented by an opinion-dependent stochastic state. It is thereby a hybrid model that combines elements from different literature streams. %

The proposed model %
consists of three mechanisms, where the last two represent distinct but potentially contradicting forces: %
i) an \emph{event generation mechanism} (EGM) that introduces an opinion-based stochastic state (binary) for each agent corresponding to events of public manifestation or participation; \\%
ii) a typical \emph{local opinion propagation mechanism} (OPM) that is a \textit{converging} force %
making agents more and more alike; and iii) a \emph{global steering mechanism} (GSM) that is a \textit{polarizing} force acting at the global level, and can make agents moving apart from each other. More specifically, the GSM computes a summary of the agents' states and feeds it back to the agents, who are allowed to have stubborn differential reactions to it, hence contrasting opinion updates. %
Note that, the defined process can also be seen as a point-process over a graph, with the difference to existing processes (e.g.~like Hawkes process \cite{HP2017}) that here the agents do not interact directly through their states (i.e.~the occurring events) but through their opinions (latent variables) that drive the states. 

The originality of our approach is that it goes beyond the standard DeGroot-based modeling: on the top of a DeGroot-like idealization, \modelname accounts for information aggregation phenomena that can lead to structured agent reactions and polarization, while also builds a stochastic process that can fit to real event data and offer quantitative insight.

By both extensive numerical simulations and deriving mathematical properties, we show how the interaction of these mechanisms allow richer and more complex dynamics, such as disagreement, polarization, and radicalization. We show that %
there are areas of distinct behavior in different regions of the model's parameter space, and that the model offers interpretable descriptions of the associated dynamics. %
We also show that our model is capable of fitting to the approximate dynamics of several phenomena of recent collective movement or action recorded on Twitter. The model parameters allow the interpretation and comparison of different public events, or the same event across different linguistic areas, and this way to get insight about their characteristics. An improved fitting is achieved when combining our approach with (fully) stubborn agents. Contrary, %
when removing the proposed GSM, the remaining model equipped only with stubbornness cannot fit well to the event data. %

The organization of the rest is as follows: 
\Sec{sec:model} presents the proposed model. \Sec{sec:results} investigates some of its mathematical properties. In \Sec{sec:properties} we study empirically the model dynamics in synthetic scenarios. In \Sec{sec:data}, we fit our model to real event data and we highlight its interpretability. We give our conclusions in \Sec{sec:conclusions}. The Appendix provides technical proofs and additional material.

\begin{figure}[t]
\centering
\includegraphics[width=0.95\columnwidth, viewport=0pt 20pt 1180pt 1250pt, clip]{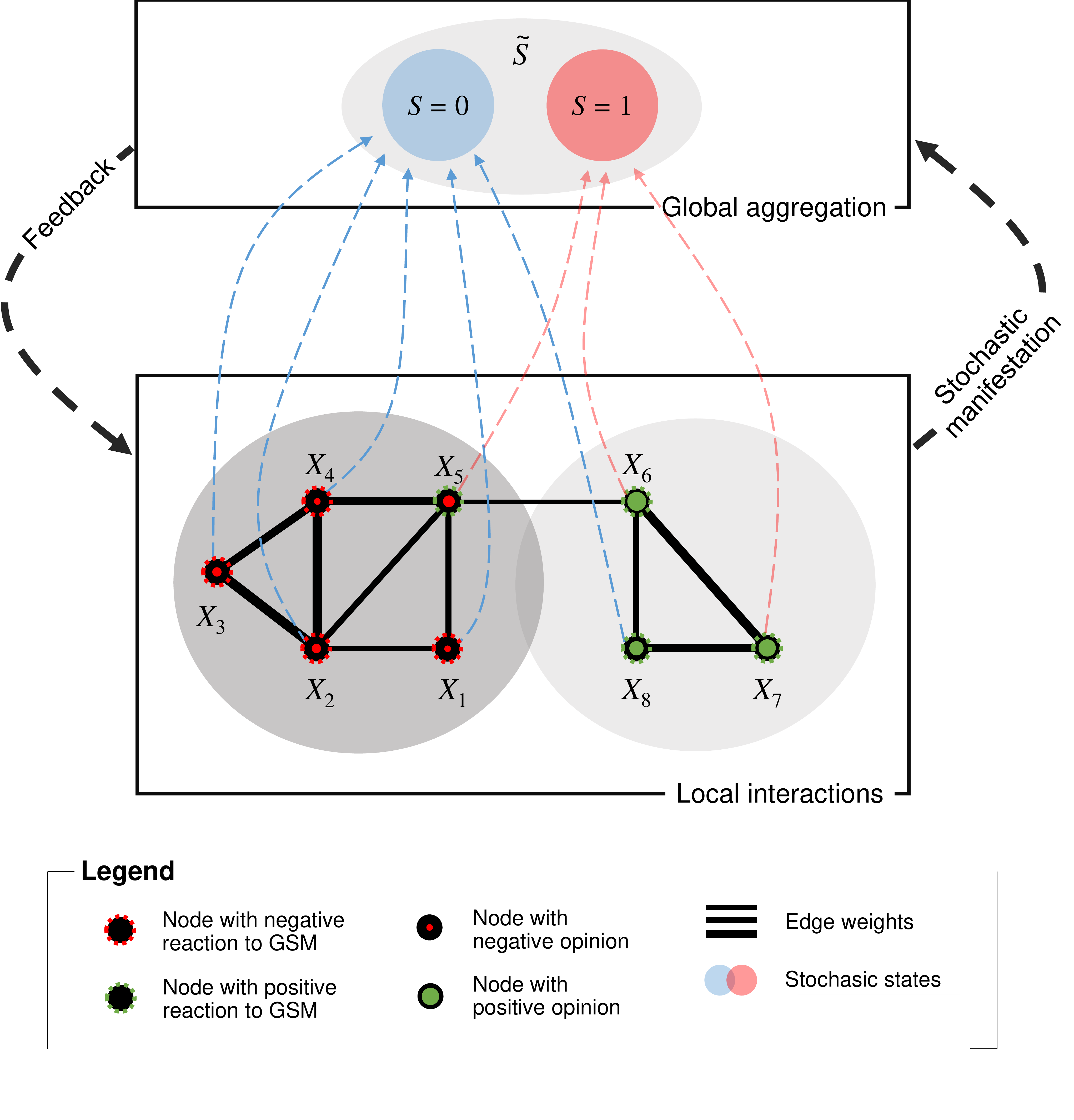}
\vspace{-1.2em}
\caption{\textbf{Scheme of the proposed two-layer \modelname model.} At the bottom there is the \emph{local interaction layer}, and at the top the \emph{global information aggregation layer}. We are at time $t$ (here omitted in the notations). The model assumes that the opinions $X_1,...,X_8$ (their value scale is shown as red or blue areas inside the nodes), are exchanged at the local level between connected agents through the opinion propagation mechanism (OPM). Then, according to the event generation mechanism (EGM), each agent $i$ enters stochastically a state $S_i = \{0,1\}$ depending on her opinion $X_i$. Next, the global steering mechanism (GSM) aggregates the states at a global level, and finally feeds back a view over this information to the agents. Each agent reacts to global information in a different but fixed way $\beta_i$, positive or negative (shown as dashed green or red node boundaries).}
\label{fig:GSM-DeGroot-model}
\end{figure}

\section{The enhanced \modelname model}\label{sec:model}

\subsection{Model statement}\label{sec:our_model}
$N$ agents are represented as nodes in a fixed, strongly connected, weighted digraph $G=(V,W)$, where $V = \{1,...,N\}$ is the set of node indexes. $W = \{w_{ji}\}_{i,j \in \V}$ is a matrix with normalized incoming edge weights, i.e.~$\forall i\in\V, \ \sum\overN{j} w_{ji} = 1$, where $w_{ji}$ indicates the influence level of agent $j$ to $i$. %

Each agent $i$ is characterized by: an opinion-dependent stochastic state $S_{i,t} \in \{0,1\}$, produced by the \emph{event generation mechanism} (EGM), indicating whether or not the agent generates an \emph{event} to manifest her views beyond her local environment; a time-dependent opinion $\Opinion_{i,t}\in \mathbb{R}$, which is exchanged locally with neighboring agents through an opinion propagation mechanism (OPM); and a fixed inherent (i.e.~stubborn) way $\beta_{i} \in \mathcal{B} \subseteq \mathbb{R}$, in a value range $\mathcal{B}$ around $0$, in which she responds to received global information. Moreover, we consider $g(S_{t})$ to be a function representing the global steering mechanism (GSM) that aggregates information from the network at a global level and feeds it back to the agents. %

Given agent $i$'s current opinion $\Opinion_{i,t}$, the discrete-time evolution of her state and opinion for time $t+1$ is given by:
\begin{equation}\label{eq:S_update}
\hspace{-1mm}\textnormal{\small\textbf{State update:}} \ \ \quad %
\underbrace{\phantom{\Big(} \!\!\!\!\!\!    S_{i,t} \sim \text{Bernoulli}}_{\substack{\text{event generation}}}\Big(\textstyle\frac{1}{1+\exp(-\lambda \Opinion_{i,t})}\Big)
\end{equation}
\newcommand{\dummysum}{\vphantom{\sum_1^N}}
\begin{equation}\label{eq:O_update}
\textnormal{\small\textbf{Opinion update:}} \ \quad %
\!\!\!\!\!\!\!\Opinion_{i,t+1} = \!
\underbrace{\beta_{i}\dummysum}_{\substack{\text{agent's}\\\text{reaction}}}\underbrace{g(S_{t})\dummysum}_{\substack{\text{global}\\\text{steering}}} \ + \ \underbrace{\sum\overN{j} w_{ji}\Opinion_{j,t}}_{\substack{\text{local opinion}\\\text{propagation}}}
\end{equation}

According to the EGM (\Eq{eq:S_update}), $\mathbb{P}(S_{i,t}=1)=\frac{1}{1+\exp(-\lambda \Opinion_{i,t})}$ and $\mathbb{P}(S_{i,t}=0) = 1 - \mathbb{P}(S_{i,t}=1)$, with %
$\lambda$ being a sensitivity parameter%
. In the rest,  %
we consider $g(S_{t}) = \gamma \tilde{S_{t}}$, where $\tilde{S_{t}} = \frac{1}{N}\sum\overN{i} \ind\{S_{i,t}=1\}$%
, and $\gamma \geq 0$ is a parameter expressing the GSM's scaling effect. We call the value of $g(S_{t})$ as the GSM's \emph{steering strength} at time $t$. %
Note that, by setting $\gamma=0$, the GSM is neutralized and leaves only the OPM in effect, thus this model becomes equivalent to the classical DeGroot model. \Fig{fig:GSM-DeGroot-model} shows schematically the elements of the proposed model.

\subsection{Model interpretation}%
\modelname introduces, for of each agent $i$, a stochastic state $S_{i,t}$ and a fixed predisposition $\beta_{i}$ over the received global information. These two additions to the classical DeGroot model \cite{degroot} are explained next. 

\inlinetitle{Opinion-dependent states.}~%
At time $t$, the EGM generates stochastically the state of agent $i$ as a function that is increasing with her current opinion value, and independently of her previous state. The \modelname model is a particular \emph{discrete-time stochastic process generating one-sided opinion-driven events} (i.e.~agents getting in state $1$) with variable probability intensity over time (i.e.~non-iid events). This is totally different to typical state-based models as there is no notion of agent's transition from one state to the other. The model could be seen as a discrete-time point-process over a graph, however the difference to existing processes (e.g.~Hawkes process \cite{HP2017}) is that the agents in our model do not interact directly through their states (i.e.~the events), but only through their opinions, which are latent variables driving the states; then, states affect the process only through the global aggregation of the GSM.%

State $1$ might be regarded as any kind of behavior or action induced by the agent's opinion. For instance, an opinion on a governmental policy can lead to protesting against it. Here, the one-sidedness of the process means that an agent remaining in state $0$ does not imply she protests in support of the policy. For cases in which protests are two-sided, an agent in state $1$ can be interpreted as protesting for one of  the side and her opinion as measuring how important the matter is to her. In this case, the number of agents in state $1$ should be interpreted as a measure of how controversial a topic is. A non-deterministic state means that the decision is taken considering additional factors that are external to the model, which are here assumed to be randomly distributed. E.g.~deciding an agent whether to participate in a protest can be a function of her view on the seriousness of a situation, psychological factors (e.g.~social pressure, fatigue), her whereabouts or time availability, which are not explicitly modeled. Instead, such factors are captured by the global parameter $\lambda$ that controls the opinion-driven actions (see \Sec{sec:model-generalization}). %

\inlinetitle{Steering mechanism and agents' reaction}~%
Beyond the assumption of most variations of the DeGroot model %
that an agent's opinion is only affected by her local social interactions, our model formalizes the idea that the global network state has also an important role in the opinion formation. The GSM represents any form of information aggregation that may modify agents' opinions over a topic of public debate. The underlying idea for $g(S_{t})$ summarizing the agents' states is that the steering mechanism relies on aggregated coarse information from the whole network, contrary to peer-to-peer interaction that is based on repeated social exchanges and allows for more nuance. The GSM is characterized by the proportion of positively-reacting agents in the population, $\beta = \frac{1}{N}\sum\overN{i} \ind\{\beta_i > 0\}$, and the function $g(\cdot)$ described by the parameter $\gamma$. Essentially, our model assumes that the agents have already formed their views, biases or predispositions, prior to the debate, and those determine the stubborn way they react to global information. This reaction can be due to either a sense of alignment, or as a reaction in opposition to what the agent perceives as the opponent `other' (e.g.~believing that the media have an agenda or corrupted and distort reality).%

\subsection{Model extensions}\label{sec:model-generalization}
\emph{Fully stubborn agents} with fixed opinions can be easily incorporated to the \modelname model by allowing them to 
skip the opinion update step at each iteration. We use this approach in our experiments. More generally, as in the Friedkin-Johnsen (FJ) model \cite{friedkin}, we could express by $(1-\xi_i)$, $\xi_i \in [0,1]$, the extent to which each agent is stubborn about her initial opinion: $\Opinion_{i,t+1} = \xi_i \left( \text{``opinion update \Eq{eq:O_update}''} \right) + (1-\xi_i) X_{i,0}$. %

Another direction to extend the model is to introduce time-dependency to some of its elements. For instance, the $\beta_i$'s and the graph structure may evolve with time, however, we suppose this takes place in a much longer time-scale compared to shorter-term opinion formation dynamics (e.g.~those observed in social media), and therefore can be ignored. Besides, an individual $\lambda_{i,t}$ for each agent $i$ could represent effects such as her engagement in the debate over time and saturation of interest. That would be an attractive feature, yet in this work we choose to keep the model simpler by assuming homogeneity across the population and no temporal variation, thus $\forall i,t,\, \lambda_{i,t} =\lambda$. Since events are proportional to the agent's opinion value (defined to be around $0)$, a notable implication of the chosen setting is that all opinions need to get very negative for no events to be generated (e.g.~at the beginning or the end of an information spread). In that sense, the opinion value $X_{i,t}$ should be perceived as a combination of agent's opinion and her interest to participate to the associated debate. Contrary, a time-dependent $\lambda_{i,t}$ would make fitting to real data more complex, but it would also allow agents to seize generating events while remaining with non-negative opinions. It would be interesting to also combine local features, such as psychological factors that vary individually an agent's behavior throughout the process (e.g.~the tolerance proposed in \cite{Topirceanu2016}), with the \modelname that emphasizes large-scale effects.

\section{Technical results}\label{sec:results}

\subsection{Distinct properties of the opinion propagation and the global steering mechanisms}\label{sec:separate_properties} \modelname's opinion update rule (see \Eq{eq:O_update}), incorporates formally two mechanisms. The second term  corresponds to the OPM's effect on agent $i$ through direct social influence. The first term corresponds to the GSM's effect%
, subject to the agent's $\beta_i$ reaction to it. Next, we discuss the distinct properties of these two mechanisms when considered separately, along with the default EGM. We show that each of them exhibits stereotypical behavior with a clear role: the OPM acts as a \emph{converging force}, whereas the GSM acts as a \emph{polarizing force}. All technical proofs are provided in \Appendix\,\ref{sec:appendix-proofs}. 

\inlinetitle{Opinion propagation mechanism (OPM)}~%
Let $I$ be the identity matrix and $\zero$ be the zero matrix, both of size $N\times N$. Taken separately (i.e.~$\gamma=0)$, the OPM is exactly the \emph{DeGroot opinion update rule} for agent $i$ at time $t$, $\Opinion_{i,t+1}=\sum_{j=1}^{N}w_{ji}\Opinion_{j,t}$, hence it makes the opinions more and more similar. Considering the vector $X_t$ of all the opinions at time $t$, and that $\dot{W} = W^\top$ is a fixed transition probability matrix (i.e.~rows with non-negative weights summing to $1$), we can express the DeGroot model as a Markov chain: 
\begin{equation}\label{eq:markovian-degroot}
X_{t+1} = \dot{W} X_t = \dot{W} (\dot{W} X_{t-1}) = \,\dots\, = \dot{W}^{t+1} X_0.
\end{equation}
The FJ model \cite{friedkin} (see \Sec{sec:model-generalization}) can be expressed in a matricial form by including a non-negative diagonal matrix $\Xi = \operatorname{diag}((\xi_1,...,\xi_N)^\top)$, representing agents' stubbornness: $X_{t+1} = \Xi \dot{W} X_t + (I - \Xi) X_0$. When $\Xi = I$, the FJ model reduces to the DeGroot model. A recursive form, as in \Eq{eq:markovian-degroot}, is obtained by the augmented system \cite{TIAN2018213}: $\hat{X}_{t+1} = \hat{W} \hat{X}_{t} = \hat{W}^{t+1} \hat{X}_{0}$, where $\hat{X}_{t+1} = [X^\top_0\ X^\top_t]^\top$, $\hat{W} = \left[\begin{smallmatrix}I & \zero\\ (I-\Xi) & \ \Xi \dot{W}\end{smallmatrix}\right]$.

Using a diagonal matrix $B = \operatorname{diag}((\beta_1,...,\beta_N)^\top)$, our \modelname model can be in turn written as: 
\begin{equation}\label{eq:GSM-DeGroot-matricial}
\!\!\!\!\!X_{t+1} = B\,g(S_t) + \dot{W} X_t = \Big(\!\sum_{\tau=0}^t \dot{W}^\tau \!B\,g(S_{t-\tau})\!\Big) + \dot{W}^{t+1}\!X_0.\!\!\!
\end{equation}
The above matricial formulations are insightful for how the structural properties of the network drive the opinion formation. The opinion update rules make the opinions more and more similar at each step by applying a smoothing operator, that is $\dot{W}$ (or $\hat {W}$). The power of $\dot{W}^{t+1}$ represents a random walk on the graph, and can be seen as an operator that smooths directly the initial opinions $X_0$ in \Eq{eq:markovian-degroot}. The \modelname's recursive form is more complicated to analyze due to the stochastic term $S_t$, yet it shows clearly how $\dot{W}^{t+1}$ smooths $X_0$, and also that each $\dot{W}^\tau$ acts on the reaction to the GSM and contributes to a cumulative term over time.

Under weak assumptions, the DeGroot model always converges and reaches \emph{consensus} \cite{degroot}, which essentially requires that the range of opinions narrow over time. %
\smallskip
\begin{definition}\label{def:convergence-consensus}%
{\textbf{Convergence and consensus. }}%
Convergence is reached when all opinions converge to finite values: $\forall i \in V$, $\exists C_{i}$\st ${\lim_{+\infty}\displaystyle}\mathbb{E}[\Opinion_{i,t}]=C_{i}$. %
Consensus is the global convergence of all opinions to the same finite value: $\exists C$\st $\forall i \in V$, $C_i=C$. Polarization is reached if $\exists C_i \neq C_j$.
\end{definition}
\smallskip
\begin{proposition}\label{prop:narrowing}
{\textbf{Narrowing behavior over time. }}
Under the weak assumption of normalized incoming edge weights for all nodes, it holds $\forall i \in V$: $\{\min_{i\in V} \Opinion_{i,t}\}_t$ is an increasing sequence, and $\{\max_{i \in V} \Opinion_{i,t}\}_t$ is a decreasing sequence.%
\end{proposition}

Thus, not only is it impossible for the DeGroot model to converge to a polarized state, but it also reduces monotonically the maximal diversity of the system, by bringing the two most ``extreme'' opinions closer to each other as time passes. This is a prototypical, but also simplistic, behavior that has been criticized in the literature (see \Sec{sec:polarization-sources}). %

The weak assumptions ensuring DeGroot's convergence 
concern the network structure, which needs to be \emph{strongly connected} (i.e.~there is a directed path connecting every ordered node pair $(i,j)$), and \emph{aperiodic} (i.e.~the greatest common denominator of the length of its cycles to be $1$), while the incoming edge weights of each node should be normalized (see \Sec{sec:our_model}). These conditions translate to requiring an irreducible and aperiodic random walk defined by $\dot{W}$, thus an ergodic Markov process. From the DeGroot update we can see that each agent's initial opinion $X_{j,0}$ affects the process proportionally to its out-degree. DeGroot convergence can be analyzed also in situations that the above assumptions are relaxed \cite{golub}. Same for when stubbornness is present \cite{TIAN2018213}, although things get more complicated due to the positioning of the stubborn agents over the network structure. In this work, we discuss the properties of the \modelname model by restricting ourselves to the standard set of assumptions.

\inlinetitle{Global steering mechanism (GSM)}~%
Let us consider the special case where only the GSM is in effect (also the default EGM), i.e.~$\forall i,j \in V$ with $j \ne i$, $w_{ji}=0$, $w_{ii}=1$. At each time $t+1$, the opinion update rule for agent $i$ reduces to: 
\begin{equation}\label{eq:pure-gsm}
\Opinion_{i,t+1}= \beta_{i} \, g(S_{t}) + \Opinion_{i,t}.
\end{equation}%
\vspace{-4mm}
\smallskip
\begin{proposition}\label{prop:two-groups-divergence}
{\textbf{Two diverging groups under pure GSM. }}
Let $\beta^+ = \{i\in\V : \beta_{i}=1\} \ne \emptyset$ %
 and $\beta^-= \{i\in V \backslash \beta^+ : \beta_{i}=-1\}$%
, the two sets with opposite reaction to global information. Then:
\begin{equation}
\lim_{+\infty}\mathbb{E}[\Opinion_{i,t}]=
\left\{
\begin{array}{ll}
+\infty & \ i \in \beta^+; \\
-\infty & \ i \in \beta^-.
\end{array}
\right. 
\end{equation}
\end{proposition}
Therefore, in the absence of local interactions, the two groups of agents will get farther and farther away from each other over time, as they react oppositely to global information. %

\subsection{Interplay between the OPM and the GSM}\label{sec:interplay}
In \Sec{sec:separate_properties}, we considered the GSM and the OPM separately. Here, we investigate their interplay and its implication for the model dynamics. At any time $t$, let the average opinion $\overbar{\Opinion}\!_{t}$, and the \emph{maximum opinion diversity} between pairs of agents that characterizes the maximal polarization at that time:
\begin{equation} 
D_{\max,t} = \max_{i\in V} X_{i,t} - \min_{j\in V} X_{j,t}.
\end{equation}
We consider two polarization indices: the \emph{final polarization} %
$D_{\max,\infty} = \max_{i\in V} C_{i} - \min_{j\in V} C_{j}$, where $C_i$ is $i$'s converging opinion (see \Def{def:convergence-consensus}), and the \emph{maximal polarization} $D_{\max} = \max_t D_{\max,t}$ recorded throughout the process. %

From \Eq{eq:O_update}, we can derive the following interesting relation between the average opinions at two subsequent time steps:
\begin{equation}
\overbar{\Opinion}\!_{t+1} =  \big({\textstyle\sum\overN{i} \beta_i}\big)g(S_{t}) + \overbar{\Opinion}\!_{t}.
\label{eq: avg op}
\end{equation}
For simplicity, we consider $g(S_{t}) = \gamma \tilde{S_{t}}$ (see \Sec{sec:our_model}), and the agent reactions to GSM to be $\beta_i \in \{-1,1\}, \forall i \in \V$. Recall the proportion of positively reacting agents 
$\beta = \frac{1}{N}\sum\overN{i} \ind\{\beta_i > 0\}$. Thus, \Eq{eq: avg op} boils down to:
\begin{equation}
\overbar{\Opinion}\!_{t+1} = (2\beta-1)\gamma\tilde{S}_{t} + \overbar{\Opinion}\!_{t}.
\end{equation}
\newpage
Qualitatively, this identifies two different regimes:
\begin{itemize}
\item \textit{\textbf{self-cooling}} for $\beta<\frac{1}{2}$: the average opinion decreases; any additional agent entering state $1$ yields a negative effect for the majority of agents, thus reducing further the probability for the majority to enter in state $1$. %
\item \textit{\textbf{self-exciting}} for $\beta>\frac{1}{2}$: the average opinion increases; any additional agent entering state $1$ yields a positive effect for a majority of agents, thus increasing further the probability for the majority to enter in state $1$.%
\end{itemize}

\inlinetitle{Effects on maximal polarization}~%
While the OPM works against allowing the GSM to drive the two groups in two opposite directions, the latter \emph{always} prevents the former from making agents alike to the point that they reach consensus. The technical results are in the two following propositions. %
\smallskip
\begin{proposition}\label{prop:no-consensus}
{\textbf{No consensus under GSM. }}%
If the GSM is in effect (if $\gamma>0$), consensus is impossible to be reached.
\end{proposition}
\smallskip
\begin{definition}\label{def:e-consensus}%
{\textbf{$\varepsilon$-consensus. }}%
Let $C_i$, $\forall i\in V$, the finite opinion values to which the agents converge (see \Def{def:convergence-consensus}). We call an $\varepsilon$-consensus iff: $D_{\max,\infty} = \max_{i\in V} C_{i} - \min_{j\in V} C_{j}  \ \le \ \varepsilon$. 
\end{definition}
\smallskip
\begin{proposition}\label{prop:boundary-on-e-consensus}
{\textbf{Boundary on $\varepsilon$-consensus. }}%
For a strictly increasing $g(S_t)$ function, reaching a $(\lim_{+\infty}\mathbb{E}\big[g(S_t)\big])$-consensus is not possible. For the special GSM form of $g(S_t) = \gamma \tilde{S}_t$, this corresponds to a $(\gamma\lim_{+\infty}\mathbb{E}[\tilde{S_t}])$-consensus.%
\end{proposition}
The first result is trivial: the GSM always prevents consensus, whereas the OPM alone would yield consensus (under weak assumptions). The second one implies that extreme opinions cannot get closer than what the GSM strength allows. Simply put, the stronger the steering is, the larger will be the difference between the extremes at the end. Although this is true at $t\rightarrow \infty$ and in the case of individual convergence to fixed values, yet it provides insight to the GSM's role.

\subsection{Comparing GSM with other possible polarization sources} %
\label{sec:polarization-sources}
One of the main questions studied in the work is which mechanisms allow opinion divergence to emerge so that the final opinion state of a network is polarized. As opinions are the outcome of a complex process, there should be a more intrinsic underlying mechanism that generates divergence among agents. So far, we have been referring to networks with non-negative weights and we showed how, under the \modelname model, polarization can emerge as a result of the effects of the two main mechanisms of the model alone, and in particular the stubborn differential way agents react to global information. Here, we put this conjecture in perspective with the two other possible sources of polarization mentioned in \Sec{sec:intro}%
: agents' stubbornness \cite{stubborn2012,Acemoglu-stubborn2013}, and signed networks \cite{nguyen2017opinion}. We discuss their conceptual basis, and we explain that they offer a more limited view compared to the GSM.

Stubborn agents with diverse opinions act as opinion sources or (attractors) that impose polarization to the network. In a strongly connected network, a single fully stubborn agent makes the network converge to her opinion. If there are two fully stubborn agents with fixed opinions $X_{i,\cdot}$ and $X_{j,\cdot}$, their opinion distance predetermines also the maximum polarization level in the limit, $D_{\max,\infty} = |X_{i,\cdot} - X_{j,\cdot}|$. %
\smallskip
\begin{proposition}
{\textbf{Bounded polarization under full stubbornness. }}
Let a strongly connected digraph $G=\{V,W\}$ with $N$ nodes, with normalized non-negative weights w.r.t the incoming edges of each node, and $\forall k, w_{kk} > 0$. Suppose a DeGroot model with a set $\mathcal{S}\subset V$ of fully stubborn agents (i.e.~$\gamma = 0$ in our model definition to neutralize GSM). Then, the polarization of the system over time will be eventually bounded by $D_{\max,\infty} = \max_{k\in \mathcal{S}}X_{k,\cdot}  - \min_{k\in \mathcal{S}}X_{k,\cdot}$, where $X_{k,\cdot}$ is a fixed stubborn opinion such that $\forall t, \,X_{k,t} = X_{k,\cdot}$.
\end{proposition}
The proportion of agents whose opinion will converge to $X_{i,\cdot}$ (and not $X_{j,\cdot}$) depends on a number of factors, such as the network structure, the initial opinion distribution across the network, and the positions of the two sources in the network. For agents with variable level of stubbornness, the maximum distance between stubborn opinions can get larger in the limit, however, it is easy to see that the minimum and maximum stubborn opinions will still bound the polarization (yet, in a non-predetermined way). Model extensions introducing psychological factors that modulate each agent's participation in the process, yet without producing opinion radicalization, present similar behavior (e.g.~the tolerance defined in \cite{nguyen2017opinion} can be seen as a varying level of stubbornness). Despite this limitation, which is not present in the \modelname, there is no doubt that stubbornness can generate polarized final network states. Nevertheless, the main concern is whether this angle can also generate dynamics that are close to real-world processes. This is empirically studied in \Sec{sec:data}. %

It has been shown that polarization can also emerge over signed networks \cite{nguyen2017opinion}, however, as we show below, the radicalization is also limited by the initial opinion state, and there is one side that cannot get more radical than it initially was. 
\smallskip
\begin{proposition}\label{prop:limited-polarization-sign-nets}
{\textbf{Limited polarization in signed networks. }}
Consider the DeGroot update rule over a signed and possibly time-dependent network:
$\Opinion_{i,t+1}=\sum\overN{j} w_{ji,t}\Opinion_{j,t}$,  %
where $w_{ji,t}\in [-1,1]$ and $\sum\overN{j} |w_{ji,t}|=1$. %
Then, $\forall t$:
\begin{enumerate}
\item $\max_{(i,j\in V)} (\Opinion_{i,t}-\Opinion_{j,t})\le 2\max_{i\in V}|\Opinion_{i,0}|$; 
\item and either $\max_{i\in V} \Opinion_{i,t} \le \max_{i \in V} \Opinion_{i,0}$, \\\phantom{iiiiiiiiiii} or\, $\min_{i\in V} \Opinion_{i,t} \ge \min_{i\in V} \Opinion_{i,0}$.
\end{enumerate}
\end{proposition}
All these polarization mechanisms, including our view for divergence at a large-scale through GSM, are not mutually exclusive, and in reality they could be simultaneously in effect. However, our point here is that these mechanisms have effects at different scales, and therefore one could argue that global steering can lead to broader and more structured polarization compared to what local mechanisms seem to achieve.

\section{Empirical investigation of model properties} %
\label{sec:properties}
Synthetic scenarios and numerical experiments are used to demonstrate the aforementioned model properties, which are related to the interplay between the OPM and the GSM. In all cases, the initial opinions were drawn by $X_{i,0}\!\!\!~\sim$Normal($\mu$, $\sigma=1)$, $\forall i \in V$ ($\sigma$ had little or no effect), where $\mu$ is the mean initial opinion that gives the magnitude of the initial shock.

\inlinetitle{Random graph models}~We conduct our analysis on synthetic networks using the Barab\'asi-Albert (BA) scale-free model \cite{newman2002} and the Stochastic Block Model (SBM) \cite{SBM-JMLR-2018}. Given a generated graph structure, we produce a randomly weighted network with normalized %
weights as follows: for each node $i$, we give equal initial weights $\frac{1}{\text{indegree}(i)}$ to all its incoming edges. Then, we pick two of those edges uniformly at random, and we transfer half of the first edge weight to the second edge. This is repeated $10$ times per node.

For SBMs, we generate networks with forming two clusters of fixed size ratios $c^{(1)}=0.7$, $c^{(2)}=0.3$, and fixed proportions of agents with positive reaction to global information $\beta^{(1)} = 0.3$, $\beta^{(2)}=0.7$. Since $c^{(1)}\beta^{(1)}+c^{(2)}\beta^{(2)}=0.42<\frac{1}{2}$, the scenario lies in the self-cooling regime where dynamics do not explode (see \Sec{sec:interplay}). We are interested in the SBM's $r$ parameter that controls the connection probability for two agents from different clusters, i.e.~a smaller value indicates a more well-clustered network. While the choice of the specific values of the other parameters can be regarded as somehow arbitrary, the main idea here is to use an SBM with two clusters of opposite majority reaction to information as a \emph{population surrogate} to model agents forming opinions in two distinct communities.

We have validated empirically (see Appendix\,\ref{sec: supplement}.II) that \modelname's behavior does not change much with the network scale, or by the variation of the %
parameters generating the network. This allows us to run frugal simulations on small graphs of $N=100$ nodes that still give relevant insight about processes that may %
take place over larger networks. %

\begin{figure}[t]
    \centering
		\vspace{-1em}
    \subfloat[$D_{\max}$\,in the self-cooling regime]{ \includegraphics[width=0.48\columnwidth,viewport=5 0 390 255,clip]{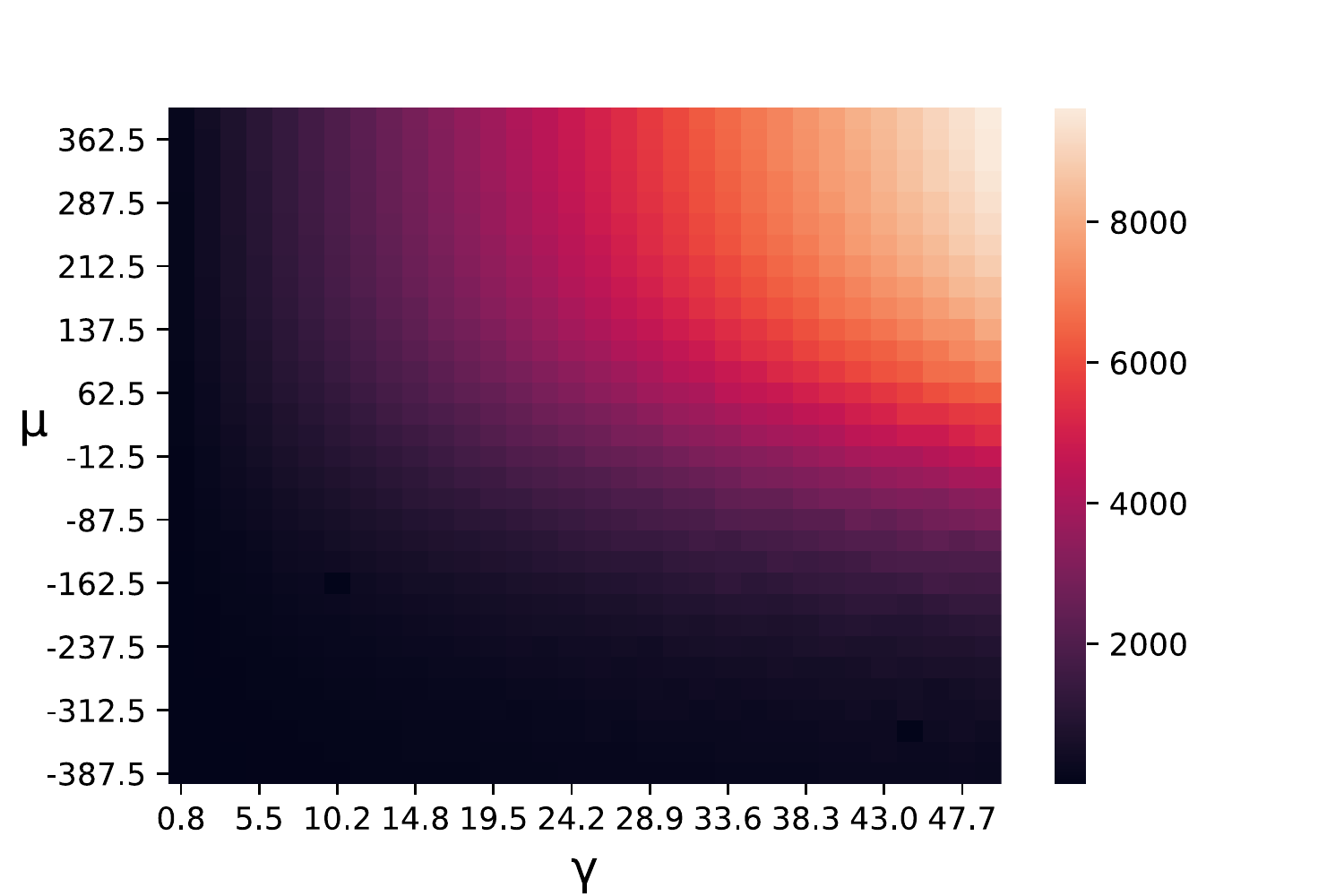} \label{fig:2a}}%
    \subfloat[$D_{\max}$\,in the self-exciting regime]{ \includegraphics[width=0.48\columnwidth,viewport=5 0 390 255,clip]{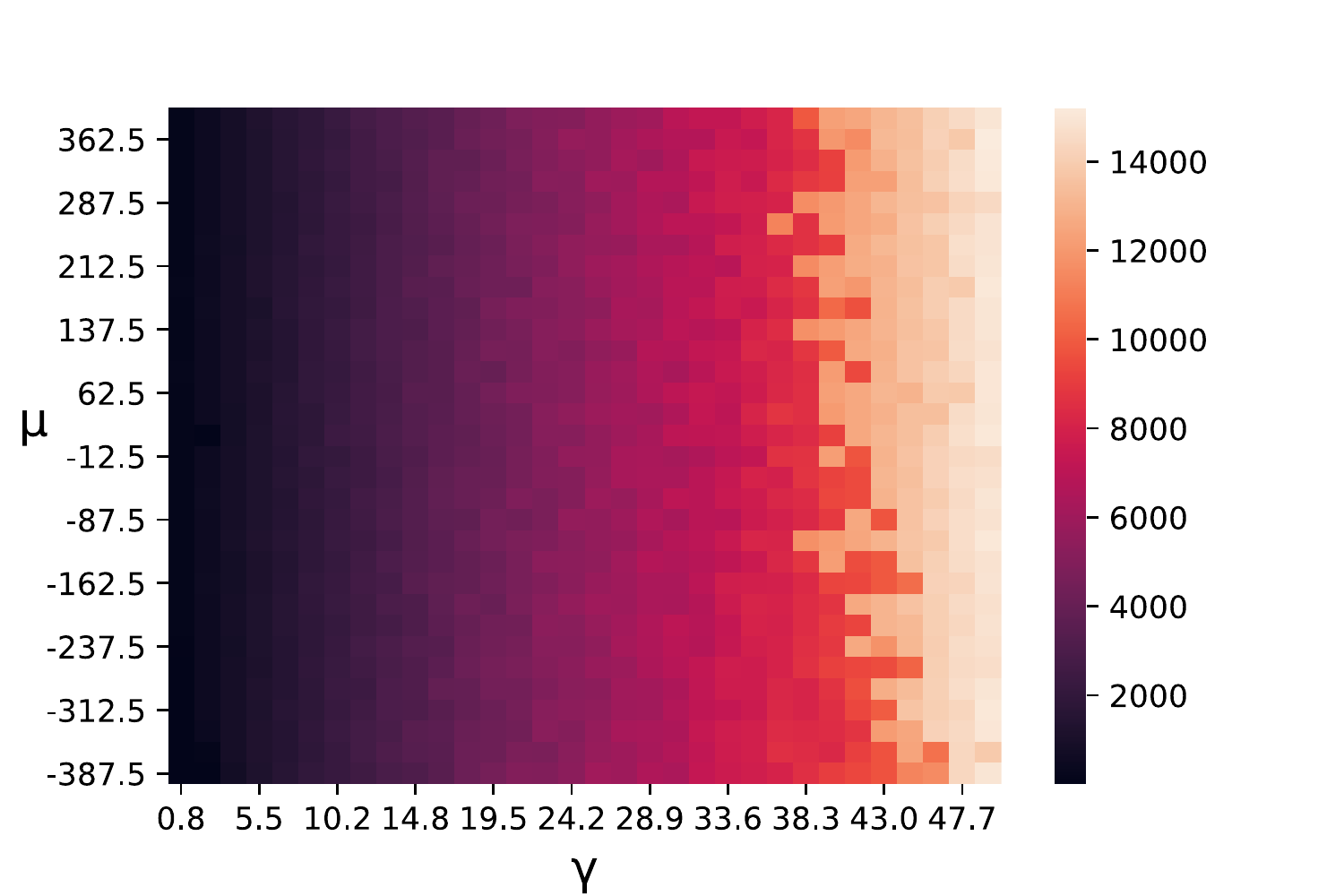} \label{fig:2b}}\\
    \subfloat[$D_{\max,\infty}$\,in the self-cooling regime]{ \includegraphics[width=0.48\columnwidth,viewport=5 0 390 255,clip]{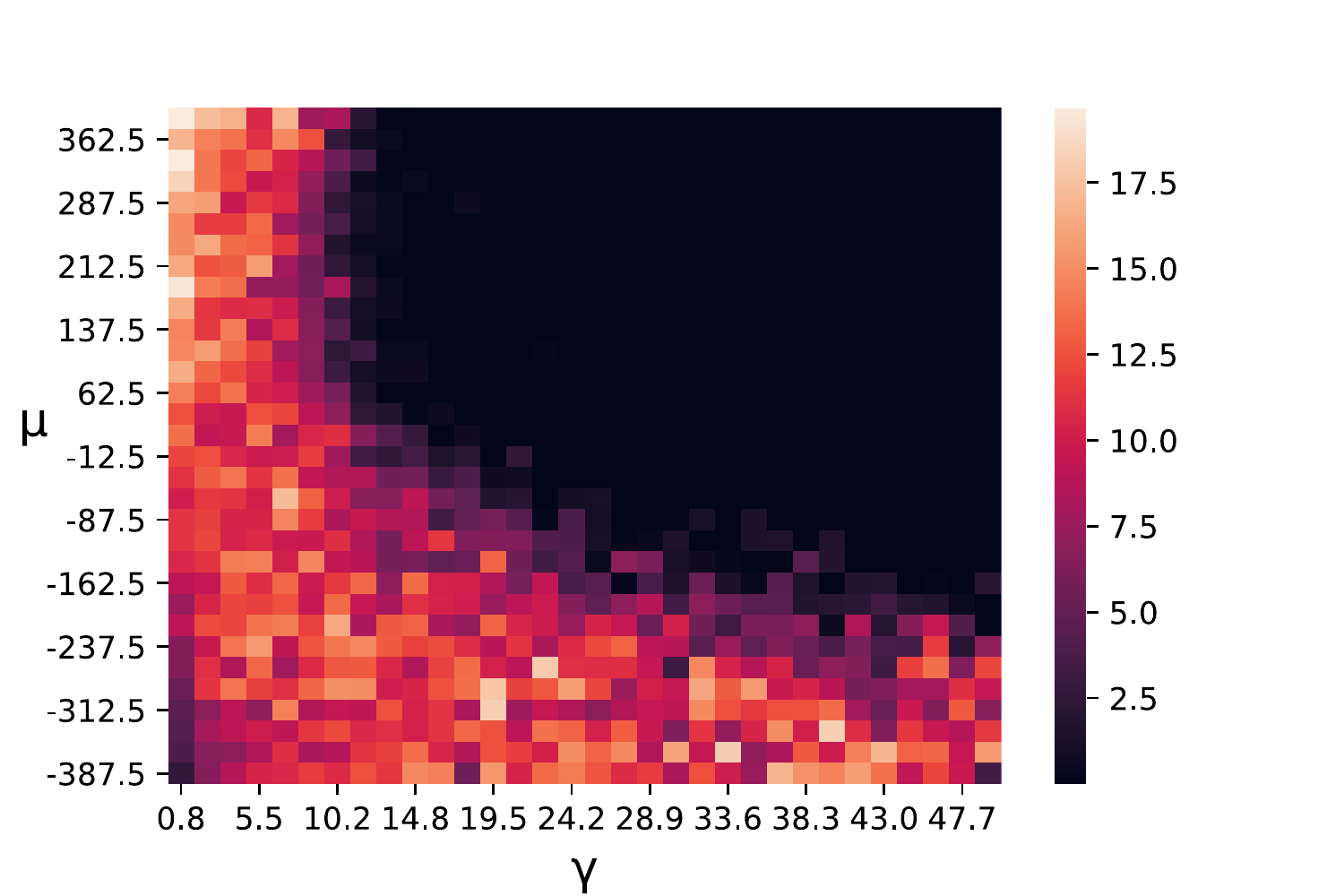} \label{fig:2c}}%
    \subfloat[$D_{\max,\infty}$\,in the self-exciting regime]{ \includegraphics[width=0.48\columnwidth,viewport=5 0 390 255,clip]{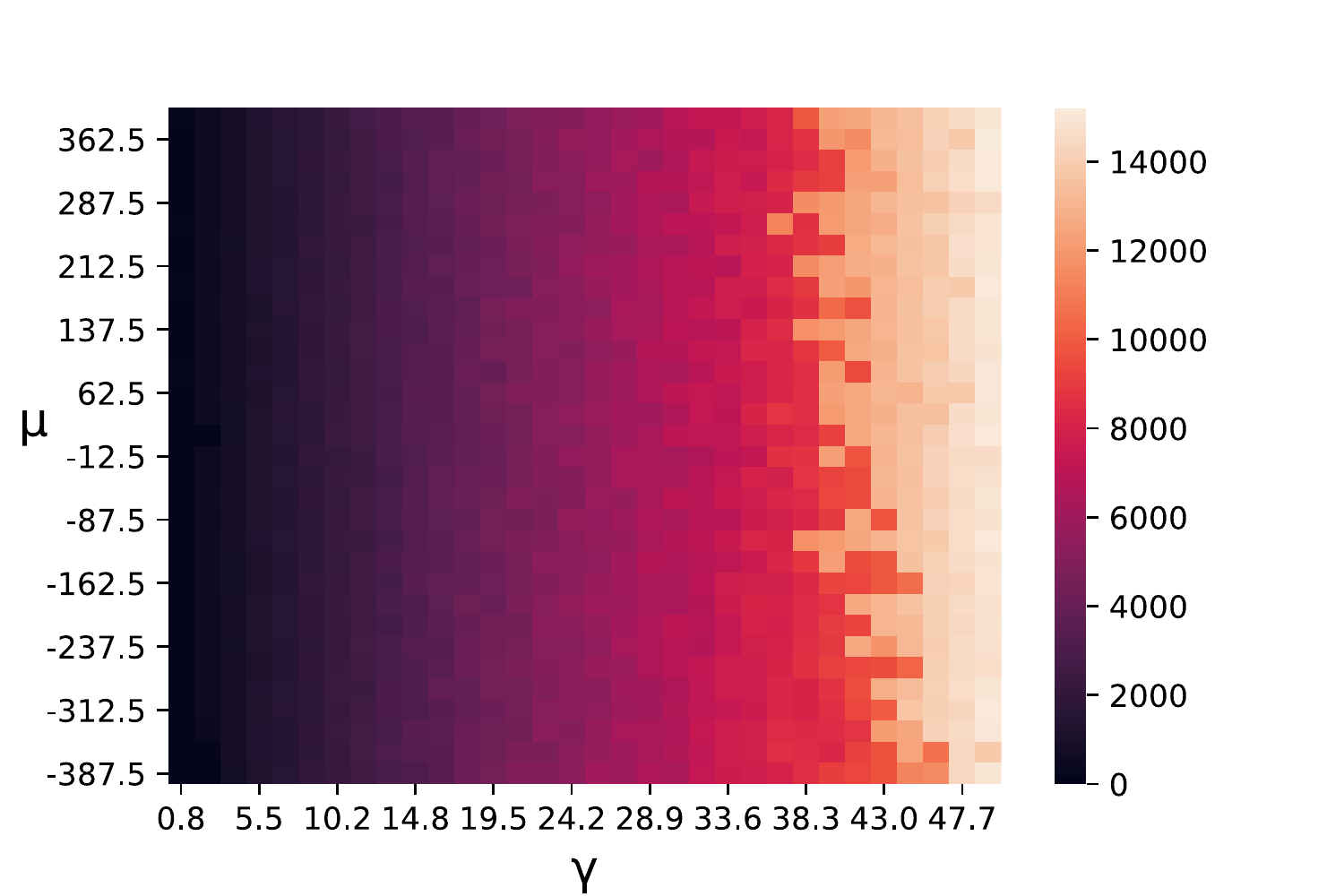} \label{fig:2d}}
    \caption{Effect of $\mu$ (controls indirectly the initial number of agents in state $1$) and $\gamma$ (GSM's scaling parameter) on two polarization indexes: the maximal polarization $D_{\max}$ and the final polarization $D_{\max,\infty}$. We show one instance in the self-cooling ($\beta=0.05$) regime, and one instance in the self-exciting ($\beta=0.95$) regime.}
 \label{fig:polarization}
\end{figure}

\vspace{4mm}
\inlinetitle{Effects of self-cooling and self-excitation on the final polarization}~%
We simulate our model on BA networks and create the heatmaps of \Fig{fig:polarization}, which illustrate the maximal polarization $D_{\max}$ and the final polarization $D_{\max,\infty}$ as a function of $\gamma$ and $\mu$ (it controls indirectly the initial GSM steering strength). On top of interacting with the OPM, the GSM interacts \textit{with itself} in the sense that in $g(S_t)=\gamma \tilde{S}_t$, clearly, $\tilde{S}_t$ depends on previous values $\{\gamma\tilde{S}_{\tau}\}_{\tau<t }$. In the self-cooling regime, higher previous steering strength tends to decrease $\tilde{S}_t$, and the opposite holds in the self-exciting regime. %
This essentially explains the difference observed between the figures in the left column and those in the right column of \Fig{fig:polarization}. Both polarization indexes get way lower in the self-cooling regime (the steering strength vanishes over time) compared to the self-exciting case, where a higher steering strength induces larger polarization over time ($D_{\max} = D_{\max,\infty}$, hence \Fig{fig:2b}, \Fig{fig:2d} are identical). In the self-cooling regime, it is the opposite: increasing $\mu$ and $\gamma$ leads to a decrease in $D_{\max,\infty}$, as explained earlier, by yielding higher steering strength in the early steps, which leads to further reduction in polarization in the next steps. This interpretation is supported by the behavior of $D_{\max}$: in the self-cooling regime, $D_{\max}$ is higher for higher strength of the steering mechanism, causing a reduction in polarization later and at the end of the simulation.

\begin{figure}[t]
\vspace{-2em}
    \centering
		\subfloat[Effect of $\mu$]{%
		\!\!\!\!\!\includegraphics[width=0.364\columnwidth, viewport=0 0 266 260, clip]{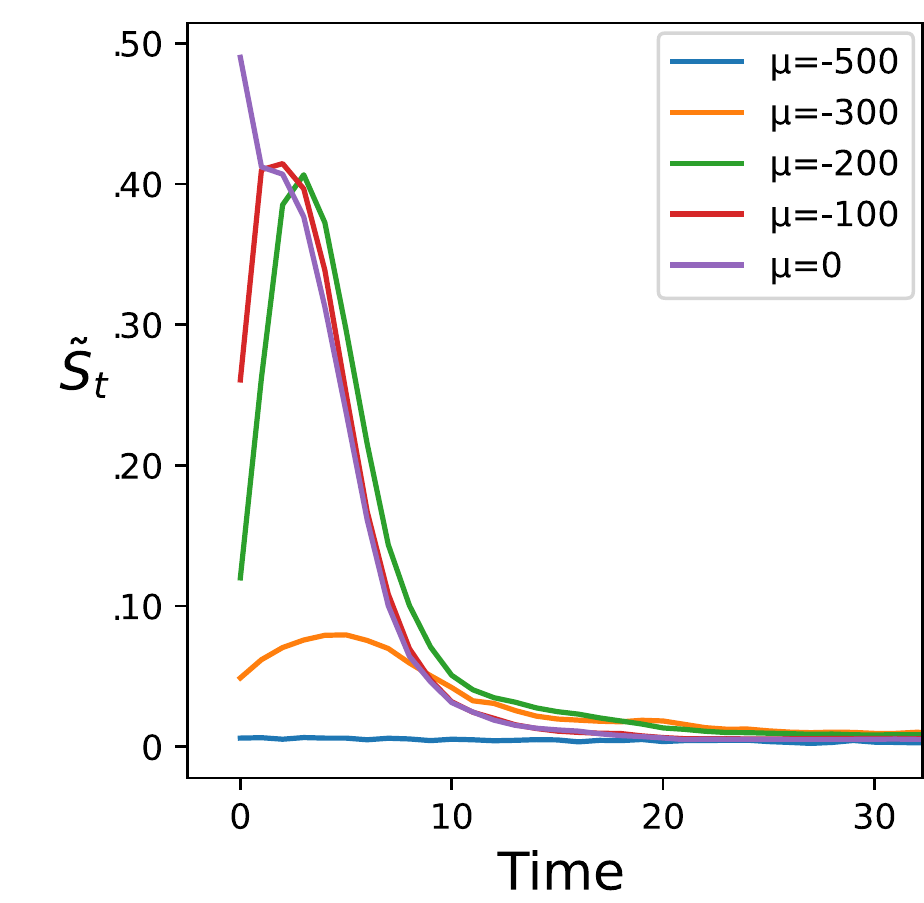}}
		\ 
    \subfloat[Effect of $\gamma$]{\includegraphics[width=0.32\columnwidth, viewport=32 0 266 260, clip]{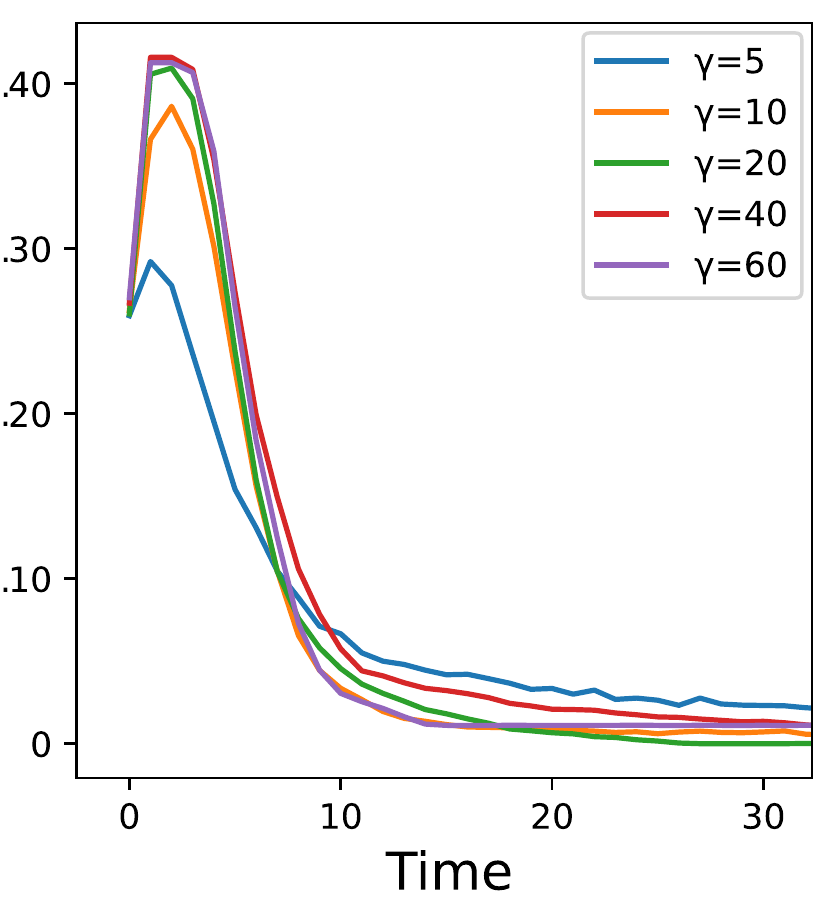}}
		\ 
    \subfloat[Effect of $r$]{\includegraphics[width=0.32\columnwidth, viewport=32 0 266 260, clip]{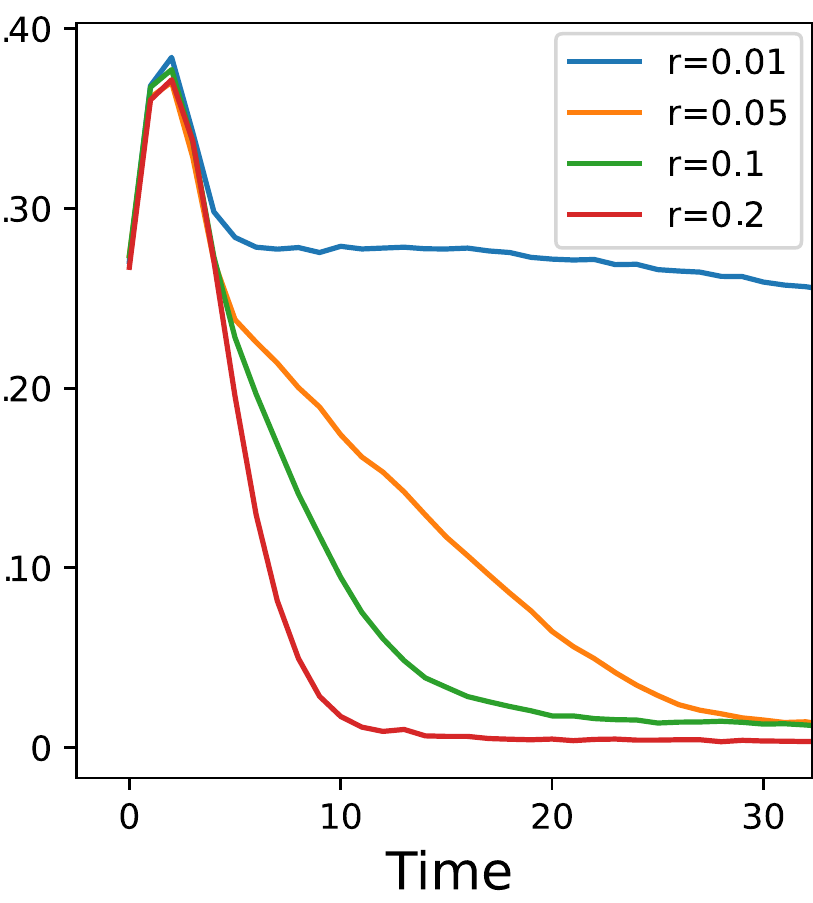}}
    \caption{Effect of $\mu$ (controls indirectly the initial number of agents in state $1$), $\gamma$ (GSM's scaling parameter), and $r$ (controls the inter-community connectivity) on the proportion of agents in state $1$ ($\tilde{S}_{t}$).}
     \label{fig:selected_parameters}
\end{figure}

\inlinetitle{%
Effects on the number of events}~%
We simulate our model on SBM networks (see details in \Appendix\,\ref{sec: supplement}), and look at the effects of three parameters with interpretative power, $\mu$, $\gamma$, $r$, on the proportion of agents getting in state $1$ ($\tilde{S}_t$). Recall, smaller $r$ values indicate more well-clustered SBM networks. \Fig{fig:selected_parameters} presents the simulation results. The effect of $\mu$ is two-fold: it controls the probability for agents to initially be in state $1$, and thus gives the departure point of the curves, while also ignites the GSM due to the opposite reaction of the agents with $\beta_i < 0$. As a consequence, the peak of the curve and the time at which it is reached depend on $\mu$: the lower $\mu$ is, the longer it will take for the GSM to amplify the initial shock, and also the lower the peak will be. Concerning $\gamma$, one can see that it controls also  the height of the peak (the higher $\gamma$, the stronger the steering, and hence, the higher the peak) as well as for the time at which the curve peaks (the higher $\gamma$, the longer it takes for the peak to be reached). It also seems that $\gamma$ has an effect on the decrease after the peak, since high $\gamma$ values associate to quicker decrease. This can be interpreted as the decrease being sharper when the GSM is stronger, in the self-cooling regime. Finally, $r$ affects the speed in which the model ``cools down". In this specific scenario, $r$ reduces the effect of the GSM as it mixes the minor community (positively reacting) with the major one (negatively reacting). More generally, in self-cooling, for any graph model, increasing the network connectivity is expected to decrease the height of the curve peak and/or accelerate the convergence. Additional simulations about the impact of extreme opinions to the dynamics, are reported in \Appendix\,\ref{sec: supplement}.

\section{Confronting real-world data}
\label{sec:data}

In this section, we assess the \modelname's capacity to model real-world data dynamics. Our goal is not necessarily to achieve the best possible fit, but rather to offer a quantitative interpretation of the phenomena appearing in real situations. The implementation of our model, the fitting process, as well as the datasets used in this study are publicly available\footnote{Online material: \scriptsize\texttt{https://kalogeratos.com/psite/gsm-degroot/}}%

\inlinetitle{Dealing with online social network data}~%
Acquiring access to data concerning the topology of a physical or online social network, and the actual opinions of individuals over a subject, is very hard. However, the \modelname model allows us to consider opinions as latent variables, and then use the agents' states (the $S_i$'s) as observed behaviors generated by those underlying opinions. We use Twitter data from StoryWrangler \cite{storyWrangler}\footnote{Online tool: {\scriptsize\texttt{https://storywrangling.org/}}}, and specifically data concerning the frequency of use of a specific characteristic \emph{term} in each use-case, namely a hashtag or symbol appearing in the tweets written in a certain language. We thereby interpret the use of such a term by agent $i$ as a publicly visible behavior manifesting her opinion-dependent state $S_i$. %

We focus on three use-cases: the \emph{Black Lives Matter} (BLM) and the \emph{MeToo} movements, as well as the geopolitical conflict and the subsequent military invasion of Ukraine by Russia in February 2022. We pick as representative terms the \#BlackLivesMatter (denoted in the rest by \#BLM), \#MeToo, and the emoji of the Ukrainian flag (EUF). The choice of the topics is motivated by two facts: i) they had world-wide attention%
, which enables the comparison between countries; ii) they fit with the timeline considered in our model: an event\footnote{\emph{BLM:} the event was the murder of George Floyd by police officer Derek Chauvin. The BLM movement was actually initiated years before, after a similar incident that took place in Ferguson. However, here we focus on the second wave of the movement, which was triggered in Minneapolis and went viral worldwide.~{--}~\emph{MeToo:} the Jeffrey Epstein case, who took advantage of her position as a Hollywood producer and sexually  abused several women.~{--}~\emph{Invasion of Ukraine:} the initiation of the ``special military operation'' that was the invasion of Ukraine by Russia on the 24th of February 2022.} triggers both a collective process of debate and protests, and wide media coverage. Our objective is to fit our model with data concerning several languages, and then compare the outcomes.

\newcommand{\clabel}[1] {\multicolumn{1}{c}{#1}} %
\renewcommand{\toprule}{\noalign{\hrule height 1.2pt}}
\renewcommand{\bottomrule}{\noalign{\hrule height 1.2pt}}
\renewcommand{\midrule}{\noalign{\hrule height 0.9pt}}

\newcolumntype{L}[1]{>{\raggedright\let\newline\\\arraybackslash\hspace{0pt}}m{#1}}
\newcolumntype{C}[1]{>{\centering\let\newline\\\arraybackslash\hspace{0pt}}m{#1}}
\newcolumntype{R}[1]{>{\raggedleft\let\newline\\\arraybackslash\hspace{0pt}}m{#1}}

\renewcommand*{\arraystretch}{0.95}
\setlength\tabcolsep{4pt}

\begin{table*}[ht]
\scriptsize
\centering
\vspace{-1mm}
\begin{tabular}{L{2.4cm} L{2.2cm} C{1.3cm} R{1.15cm} R{1cm} R{0.85cm} R{1.15cm} R{1cm} R{0.85cm}}  
\toprule
\multirow{2}{*}{\textbf{Category}} 
& \multirow{2}{*}{\textbf{Language}} & \multirow{2}{*}{\textbf{Error}} & \multicolumn{3}{c}{\textbf{Best parameters}} 
&\multicolumn{3}{c}{\textbf{Category average}} 
\\
 & & & \clabel{\ ${\bm \mu}^*$} & \clabel{\ ${\bm\gamma}^*$} & \clabel{\ ${\bm r}^*$} & \clabel{$\overline{\bm\mu}$} & \clabel{$\overline{\bm\gamma}$} & \clabel{\ $\overline{{\bm r}}$}
\\
\midrule
\multirow{9}{*}{European}
&Swedish&0.251&-150.899&0.296&0.071&\multirow{9}{*}{-98.423}&\multirow{9}{*}{0.194}&\multirow{9}{*}{0.161}\\
&French&0.331&-97.398&0.303&0.122\\
&German&0.361&-87.777&0.176&0.379\\
&Portuguese&0.429&-120.259&0.093&0.238\\
&Dutch &0.435&-175.000&0.275&0.025\\
&English & 0.454&-120.219&0.067&0.249\\
&Spanish&0.472&-148.704&0.124&0.167\\
&Greek&0.472&-110.551&0.089&0.022\\
&Catalan&0.588&125.000&0.325& 0.175\\
\hline
\multirow{7}{*}{Non-European}&Arabic &0.344&-147.889&0.151&0.256&\multirow{7}{*}{-112.824}&\multirow{7}{*}{0.145}&\multirow{7}{*}{0.234}\\
&Tamil&0.372&-100.820&0.204& 0.061\\
&Kannada&0.534&-200.527&0.101&0.032\\
&Korea&0.538&-100.593&0.198&0.198\\
&Turkish&0.552& -132.869&0.175&0.314\\
&Hindi&0.583&-68.538&0.107&0.427\\
&Persian&0.651& -38.534&0.079&0.350\\
\bottomrule
\end{tabular}
\vspace{-1mm}
\caption{\textbf{\#MeToo social movement~--}~Results obtained by our optimization process using Twitter data. The best model parameters ($\mu^*, \gamma^*, r^*$) estimated for each language and the corresponding fitting error are reported. The languages are grouped into geographic categories, within which their order is from lower to higher fitting error.}
\vspace{-2mm}
\label{table: MeToo}
\end{table*}
\begin{table*}[h!]\footnotesize\centering
\scriptsize
\centering
\vspace{2mm}
\begin{tabular}{L{2.4cm} L{2.2cm} C{1.3cm} R{1.15cm} R{1cm} R{0.85cm} R{1.15cm} R{1cm} R{0.85cm}}  
\toprule
\multirow{2}{*}{\textbf{Category}} 
& \multirow{2}{*}{\textbf{Language}} & \multirow{2}{*}{\textbf{Error}} & \multicolumn{3}{c}{\textbf{Best parameters}} 
&\multicolumn{3}{c}{\textbf{Category average}} 
\\
 & & & \clabel{\ ${\bm \mu}^*$} & \clabel{\ ${\bm\gamma}^*$} & \clabel{\ ${\bm r}^*$} & \clabel{$\overline{\bm\mu}$} & \clabel{$\overline{\bm\gamma}$} & \clabel{\ $\overline{{\bm r}}$}
\\
\midrule
\multirow{9}{*}{Western Europe}&English&0.274&-199.503&0.285&0.057&\multirow{9}{*}{-182.275}&\multirow{9}{*}{0.243}&\multirow{9}{*}{0.120}\\
&French&0.363&-168.266&0.172&0.253\\
&Italian&0.391&-162.113&0.122&0.071\\
&German&0.415&-66.471&0.134&0.086%
\\
& Dutch&0.437&-275.000&0.175&0.075\\
&Portuguese&0.527&-275.000&0.225&0.325\\
&Esperanto&0.538&-186.561&0.438&0.054\\
& Spanish&0.585&-275.000&0.475&0.075\\
&Catalan&0.785&-32.559&0.162&0.092\\
\hline
\multirow{6}{*}{Eastern Europe}&Ukrainian&0.381&-160.021&0.129&0.225&\multirow{6}{*}{-114.62}&\multirow{6}{*}{0.252}&\multirow{6}{*}{0.249%
}\\
&Greek&0.390&-155.085&0.171&0.122\\
& Russian&0.417&-80.443&0.443&0.485\\
&Hungarian&0.428&-275.000&0.175&0.325\\
&Czech&0.602&45.326&0.397&0.267\\
&Serbo-Croatian&0.709&-62.503&0.196&0.068\\
\hline
\multirow{4}{*}{Northern Europe}&Swedish&0.411&-158.172&0.195&0.493&\multirow{4}{*}{-179.522}&\multirow{4}{*}{0.201}&\multirow{4}{*}{0.209}\\
&Finnish&0.446&-166.834&0.193&0.099\\
&Norwegian&0.522&-194.055&0.218&0.073\\
&Danish&0.668&-199.027&0.199&0.171\\
\hline
\multirow{10}{*}{Non Europe}&Tagalog&0.318&-147.633&0.494 &0.137&\multirow{10}{*}{-179.522}&\multirow{10}{*}{0.224}&\multirow{10}{*}{0.191%
}\\
&Arabic&0.348&-275.000&0.175& 0.125\\
&Persian&0.352&-161.536&0.144& 0.058\\
&Hindi&0.441&-173.580&0.298& 0.071\\
&Urdu&0.492&-87.499&0.205& 0.441\\
&Turkish&0.524&-275.000&0.225& 0.075\\
&Cebuano&0.578&-139.896&0.157& 0.103\\
&Sinhala&0.689&-98.418&0.068& 0.446\\
&Swahili&0.757&-21.647&0.098&0.275\\
&Indonesian&0.802&-225.000&0.375& 0.175\\
\bottomrule
\end{tabular}
\vspace{-1mm}
\caption{\textbf{\#BlackLivesMatter social movement~--}~Results obtained by our optimization process using Twitter data. The best model parameters ($\mu^*, \gamma^*, r^*$) estimated for each language and the corresponding fitting error are reported. The languages are grouped into geographic categories, within which their order is from lower to higher fitting error.}
\label{table: BLM}
\end{table*}

\inlinetitle{Fitting \modelname to event data}~%
We have at our disposal the aggregate frequency of use over time of a specific term in the tweets written in a given language. To overcome the limitation of not knowing the actual interaction networks, we generate a \emph{synthetic surrogate network} in each case. Although small and rather prototypical, it can still be used for studying which parameterization allows the model to best reproduce real observed behaviors. We use %
a typical SBM network with two clusters, as described earlier in \Sec{sec:properties}. With this structure, we intend to model the interaction of (roughly) two adversarial communities, as it can often be the case in social media\footnote{We are aware that this is a simplification. For instance, in 2018 Gaumont, Panahi, and Chavalarias reported five communities in the French Twitter political landscape \cite{gaumont2018reconstruction}. However, we argued earlier that this is sufficient for a demonstration that does not aim to be perfectly accurate, but rather to highlight the potential of the \modelname model.}. Same as earlier, $r$ is the degree of cross-community edges in the synthetic network. Generating better surrogate networks could increase the precision of future empirical studies.

The \modelname fitting process for a given use-case and a given language is as follows: we first generate an SBM network, and then simulate the opinion formation using different $\{\mu, \gamma, r\}$ parameters from a grid. We use a scale-invariant comparison between the real time-series and the time-series produced by the simulation of our model (the proportion of agents in state $1$, $\tilde{S}_t$). The outcome of the optimization process is the triplet $(\mu^*,\gamma^*,r^*)$ associated to the best model fitting.

A subtle, but important, issue to note concerns the fact that after the outbreak of a subject in social media, the activity peaks and then usually fades dramatically to almost zero. Fitting the \modelname model to such shapes would require all opinions to become negative after a point in time, so agents tend to remain in state $0$ without generating events, and the GSM has also less effect. In such a case, a proper way to interpret the model is by perceiving each $X_{i,t}$ as a combination of agent's opinion and her interest to participate in the associated debate, where the latter becomes the dominant factor as time passes (e.g.~by shifting all the opinions to lower values).

\begin{figure}[t!]
\centering
\vspace{-0.5em}
\includegraphics[width=\columnwidth]{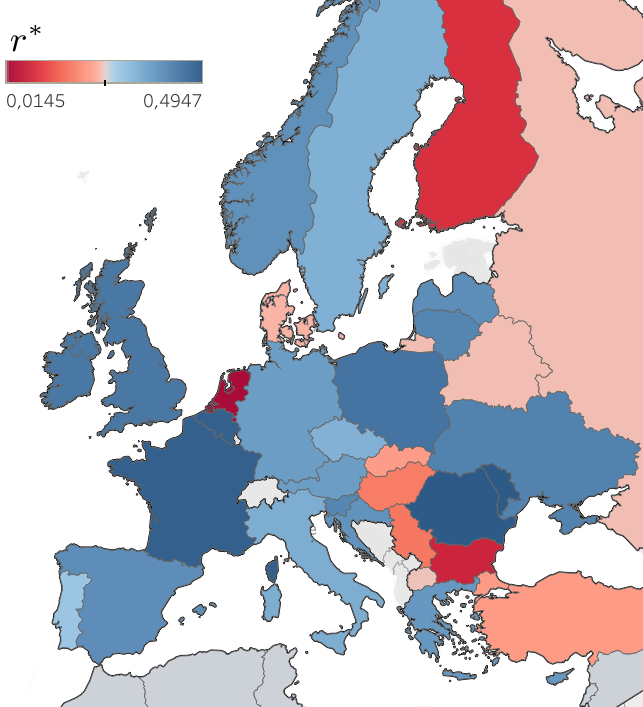}
\caption{Visualization of the country-wise estimation of the $r^*$ over a map of Europe for the debate over the military invasion of Ukraine by Russia in February 2022.}
\label{fig: map ukraine}
\end{figure}

\inlinetitle{Results}~%
The triplet $(\mu^*,\gamma^*,r^*)$ for each considered language is reported in a tabular format. For \#MeToo (see \Tab{table: MeToo}), we consider two language groups, ``European'' (E) and ``Non-European'' (NE)\footnote{``Non-European'' is not a meaningful category \textit{per se}; our aim here is simply to highlight a Europe and America's specificity concerning the topic, not to pin down a ``non-European'' specificity.}. For \#BLM (see \Tab{table: BLM}), on the other hand, we look at language categories such as the ``West-European'' and ``Eastern-Central European''\footnote{Admittedly, this uses a rule of thumb and may not be the best grouping.}. The detailed tabular results for the UEF are left for the Appendix (see \Tab{table: Ukraine}), and here we plot the results on a map of Europe (see \Fig{fig: map ukraine}) to identify salient differences between countries and geographical regions.

First, we observe that in the vast majority of cases, $\mu^*$ is negative, because the model tries to fit to the lag between the starting low activity of the series and the moment of the peak activity (see \Sec{sec:properties}). Then, $\gamma^*$ is also non-zero, which means that the GSM plays an active role in the fitting. Looking at the mean values of the estimated parameters for each language category enables the comparison of the phenomena taking place in those linguistic-geographical areas. For \#MeToo, the centroid\footnote{The unweighted average of the findings for each language in a category.} of the category ``European'' language is $\{\mu_{\text{E}}^*=-98.423, \gamma_{\text{E}}^*=0.194, r_{\text{E}}^*=0.161\}$, whereas the one of ``Non-European'' languages is $\{\mu_{\text{NE}}^*=-112.824, \gamma_{\text{NE}}^*=0.145, r_{\text{NE}}^*=0.234\}$. So, $\mu_{\text{E}}^*>\mu_{\text{NE}}^{*}$, which can be interpreted as the informational shock being larger for European languages. This would correspond to the Weinstein case having initially more impact in Europe and America than elsewhere. Also, $\gamma_{\text{E}}^*>\gamma_{\text{NE}}^*$, which corresponds to a more influential GSM, and thus can be interpreted as the fact that higher media attention was given to the \#MeToo movement in Europe and America compared to other places. However, it is important to note that because of the use of a scale-invariant distance, it is possible -but not obvious- that those two parameters cannot be interpreted as easily, in particular for \#BLM. The parameter in which we are mostly interested, since it is probably not affected by the scaling-invariance, is $r$ (see \Sec{sec:properties}). In simple terms, $r$ represents how clustered the synthetic network is, and in particular low values of $r$ indicate two distinct groups alimenting a controversy, and thus an important polemic subject that can lead to polarization. Therefore, note that $r^*_{\text{NE}}>r_{\text{E}}^*$.

Concerning the values for $\gamma^*$ and $\mu^*$ for \#BLM%
, %
the differences between language groups are not easy to interpret, and are probably uninformative, which may make one doubt about the interpretation given previously for \#MeToo. However, the difference here for $r^*$ is pretty meaningful: this value for Eastern Europe is more than double than the one for Western Europe, which makes sense with the interpretation given to low values of parameter $r$. This would indicate that in Eastern Europe, the BLM topic has not been really divisive, and thus has not led to much debate and activities or actions. 

The map in \Fig{fig: map ukraine} shows the outcome of the fitting process for the Ukrainian use-case, and associates colors to values of $r^*$ for countries in Europe in which the corresponding language is spoken (see detailed results in \Tab{table: Ukraine}). Shades of red indicate the subject being rather polemic in the associated country, whereas shades of blue indicate the opposite. First thing we note is that the topic has been way more consensual in Ukraine than in Russia (and Russian-speaking Belarus). This can be explained by the anti-war movement inside Russia, but also the fact that most probably the debate inside Ukraine is underestimated, since its the Russian-speaking population rather counts for the result associated to Russia. Second, we note the results to be particularly relevant for Finland (due to the historical ties with Russia, its proximity, and its recent application to become a NATO member), Bulgaria (which undergone a heated debate concerning whether Ukraine should be supported), Serbia (has historical and religious ties with Russia), Hungary, and Turkey (due to strong ties with Russia and their ambivalent position). On the other hand, the result concerning the Netherlands is surprising and difficult to interpret. This can also be due to data quality problems or/and bad model fitting. Finally, the topic is more consensual in the UK, France, and Spain compared to Germany, Italy, and Czech Republic, which have higher dependence on Russian gas.

\begin{table}[t]
\centering\scriptsize
\begin{tabular}{ p{1.77cm}  p{0.55cm} p{0.5cm} p{0.5cm} p{0.5cm}    p{0.5cm} p{0.5cm} p{0.5cm} p{0.5cm} p{0.4cm} }
\toprule
\multirow{2}{*}{\textbf{Hashtag}} & \multicolumn{5}{c}{\textbf{\modelname\!\,+\,stubbornness}} & \multicolumn{4}{c}{\textbf{DeGroot\,+\,stubbornness}} \\
  & \textbf{Error} & {$\bm \mu^*$} & {$\bm \gamma^*$} & {$\bm r^*$} & {$\bm p^*$} & \textbf{Error} & {$\bm \mu^*$} & {$\bm r^*$} & {$\bm p^*$}\\
 \midrule
\#BLM English    &0.39 & -400    &50.00&   0.12&0.06 &0.79& 0    &0.34&   0\\
 \#BLM Russian &0.41 &   -311  & 50.00   &0.89&0.08&0.83&   0  & 0.01   &0\\
 \#BLM Portuguese\!&0.47 &-355 & 41.11&  0.23&0&0.88&0 & 0.12&  0\\
 \#MeToo Arabic    &0.29 &-222 &32.22 & 0.23&0&0.79&0 &0.56 & 0.17\\
 \#MeToo French   &0.33 &   -222  & 41.11&0.12&0.02&0.80&   0  & 1&0\\
 \#MeToo Swedish\!\!  &0.22 & -133  & 27.77   &0.12&0.06&0.61& 0  & 1   &0.02\\
 UEF Portuguese   &0.18 & 0  & 27.77&0.23&0.04& 0.85&-133  & 0.01&0\\
 UEF Bulgarian & 0.35 & -311 & 36.66 & 0.11 & 0.06 & 0.60& -177 & 0.01 & 0   \\
 UEF French & 0.24 & 0 & 14.44 & 1 & 0 & 0.78 & 0 & 1 & 0 \\
\bottomrule
\end{tabular}
\vspace{-1mm}
\caption{Fitting results in selected cases (hashtag and language), for \modelname and simple DeGroot, where in both cases we include a proportion of $p^*$ fully stubborn agents.}
\label{tab:stubbornness}
\end{table}

\smallskip

\inlinetitle{Incorporating stubbornness at the opinion level} 
We now investigate the effect of full opinion stubbornness when incorporated in the \modelname model (see \Sec{sec:model-generalization}). We also test how competitive is a DeGroot model with full stubborn agents in fitting to event data; this is the FJ model simulated by our model with $\gamma=0$ that neutralizes the GSM, while keeping the proposed stochastic EGM that is necessary for data fitting. Specifically, we incorporate a proportion $p$ of \emph{fully stubborn} agents, uniformly distributed over the network, where $p$ is to be estimated in each case by the optimization process along with the rest of the model parameters. To deal with the incurred increase of the fitting complexity, we use a coarser grid search for each parameter, which means that the fitting error is not directly comparable with what has been reported previously in Tabs.\,\ref{table: MeToo},\,\ref{table: BLM}, and \ref{table: Ukraine} in the Appendix. 

\Tab{tab:stubbornness} reports the results for a selection of three languages per use-case (hashtag). A first finding is that DeGroot+stubbornness fits poorly to the curves' shapes. The higher the fitting error, the less meaningful it is to interpret the estimated parameters. Moreover, in all but two cases, the best fit is attained for no stubborn agents ($p^*=0$). On the other hand, \modelname\!\!+stubbornness has about $50\%$ to $80\%$ lower fitting error than that of stubborn agents only, and is achieved with reasonable proportion $p^*\leq 0.08$. An important finding is that the ranking of the fitted $r^*$ values for the used cases is stable when compared with the values reported in the other associated tables. This attests that our previous interpretation and the comparison between languages using \modelname remain valid even after introducing stubborn agents. Values typically change though, especially when $p^*$ is found to be higher. It is likely that $p$ interacts in a non-trivial manner with $r$ and other model parameters. This could have an interesting interpretation, particularly if one sees stubborn agents as agents or media organizations with a fixed agenda. Further research could provide additional insight. %

Note that, due to the additional model flexibility it offers, stubbornness improves also the fitting of the plain \modelname, when optimized with the same grid search (omitted results). In conclusion, in our experimental setting, full stubbornness does not seem to be a sufficient stand-alone feature able to explain well real debates, but it can be rather considered as a valid add-on feature to \modelname.%

\section{Conclusions}\label{sec:conclusions}

In this work we presented the two-layer opinion formation model called \modelname. In our approach, each agent is characterized by a continuous opinion variable as well as a binary state accounting for actions or behaviors induced stochastically by the agent's opinion. Our model features three distinct mechanisms: the event generation mechanism (EGM) that generates the state of each agent; the opinion propagation mechanism (OPM) that formally corresponds to the classic DeGroot model and acts as a converging force; and the global steering mechanism (GSM) that brings network-level information  as feedback to agents, and acts as a diverging force as agents are allowed to have differential reaction to it. By investigating their interplay theoretically as well as using numerical simulations, we show that: i) the GSM prevents agents from reaching a consensus and imposes a bound on reachable $\varepsilon$-consensus; ii) the GSM has a differentiated effect on limiting the maximal diversity depending on which of the two identified regimes (self-cooling or self-exciting) does the process lie on; These findings offer a new way to explain how various polarization phenomena can emerge through the differential interactions of agents with global information.%

One of the important characteristics of the proposed \modelname model is that its stochastic EGM produces a stream of discrete events, and hence enables addressing real event data. To the best of our knowledge, this is exceptional for a DeGroot-based model. In our experimental study, we fit our model to Twitter event data concerning the frequency of use of specific symbols and hashtags in tweets (implying agents being in state $1$ of our model) written in different languages. We investigated the role of interpretable parameters to show that our model is able to reproduce meaningful differences between geographical regions or countries. We also show that agents' stubbornness can be used in combination with \modelname for improving the data fitting, although stubbornness alone (i.e.~our model without the GSM) has lower expressiveness and it does not fit well to real data. %

As part of future investigations, there can be interesting refinements of the model (e.g.~see \Sec{sec:model-generalization}), namely to combine it with complementary features, such as different agent roles (e.g.~influencers or stubborn agents) and signed networks, or other features such as introducing psychological factors, influence saturation or thresholding. Furthermore, the analysis of real data can be greatly benefited by further improvements to the model fitting procedure.

\section*{Authors' contribution and Acknowledgments}
Ivan Conjeaud and Argyris Kalogeratos contributed equally in this work regarding the conception of the idea, the design and analysis of the model and the experiments, co-authoring the paper. I.C. implemented most of the codes of the experimental study, and handled the review process. Philipp Lorenz-Spreen contributed in the second phase of development of the work by reviewing some of its parts, by designing the experiments and also co-authoring the paper. A.K. was the principal scientific responsible and coordinator. %

P.L.-S. acknowledges financial support from the Volkswagen Foundation (grant ``Reclaiming individual autonomy and democratic discourse online: How to rebalance human and algorithmic decision making''). A.K. acknowledges support from the Industrial Data Analytics and Machine Learning Chair hosted at ENS Paris-Saclay, University Paris-Saclay. 

 \appendices

\section{Proofs of technical results}\label{sec:appendix-proofs}

In the rest, we refer to the following quantities: let the minimum and maximum opinion at time $t$ be denoted by $\minOpinion_{t} = \min_{i\in V} \Opinion_{i,t}$, and $\maxOpinion_{t} = \max_{i\in V} \Opinion_{i,t}$ (note that, in this case, $\minOpinion_{t}$ and $\maxOpinion_{t}$ are indexed only by the time).

\inlinetitle{I.~Importance of the normalized weights assumption}~%
We suppose the graph is fixed over time, and we consider that $\forall i \in \V$, $\sum\overN{j} w_{ji}=\alpha$, with $\alpha \geq 0$. Then, the magnitude of each opinion in the limit can be:\
\begin{equation}
\lim_{+\infty} |\Opinion_{i,t}| = 
\left\{
\begin{array}{ll}
+\infty & \ \text{when } \alpha > 1; \\
0 & \ \text{when } 0\le\alpha < 1.
\end{array}
\right. 
\end{equation}
Therefore, same as for the DeGroot model, in order for the \modelname model to be meaningful and interpretable, it is required to have normalized edge weights induced by $\alpha=1$.
\vspace{-0.5em}
\begin{proof}
At each time $t$, for each agent $i$, we have:
\begin{equation}\label{eq:narrowing}
\hspace{-0.1em}
{\small\left\{\hspace{-0.4em}
\begin{array}{lccl}
\text{the update rule gives} & \ \alpha   \minOpinion_{t} \le &  \Opinion_{i,t+1} & \le \ \alpha  \maxOpinion_{t}, \\
\text{which implies that} & \ \alpha \minOpinion_{t} \le & \minOpinion_{t+1} \leq \maxOpinion_{t+1} & \leq \ \alpha \maxOpinion_{t}, \\
\text{by induction yields} &
\alpha^{t} \minOpinion_{0} \leq & \minOpinion_{t+1} \leq 
\maxOpinion_{t+1} & \le \alpha^t \maxOpinion_{0},\\
\text{and by def. yields } &  \alpha^{t}\minOpinion_{0} \le & \Opinion_{i,t+1} & \le \alpha^t \maxOpinion_{0}.
\end{array}%
\right.\!\!\!\!\!\!\!\!\!}
\end{equation}
And finally, using the last expression (for $\Opinion_{i,t}$), the result follows naturally from the following inequality:
\begin{equation}
\alpha^{t} \min( |\minOpinion_{0}|, |\maxOpinion_{0}|) \le |\Opinion_{i,t}| \le \alpha^{t} \max( |\minOpinion_{0}|, |\maxOpinion_{0}|).
\end{equation}
\end{proof}
\vspace{-1em}
As a consequence, the assumption of normalized edge weights (i.e.~$\alpha=1$) is needed to ensure that the models are informative and interpretable. %

\inlinetitle{II.~Proofs for Propositions}~In what follows, we provide proofs for the Propositions stated earlier in the main text. Recall that $g(S_t) = \gamma \tilde{S}_t = \frac{\gamma}{N}\sum\overN{i} S_{i,t} = \frac{\gamma}{N} \norm{S_{t}}_1$, since $S_t$ is a binary vector. We define for convenience $\dot{g}(k) = \{g(S)=\gamma\frac{k}{N} : S\in\{0,1\}^{N\times 1}\!,\,\norm{S}_1 = k\}$, which maps to the same value all binary vectors of dimension $N$ with $k$ non-zero entries.\!\!
\smallskip
\begin{proposition1}
{\textbf{Narrowing behavior over time. }}%
Under the weak assumption of normalized incoming edge weights for all nodes, it holds $\forall i \in V$: $\{\min_{i\in V} \Opinion_{i,t}\}_t$ is an increasing sequence, and $\{\max_{i \in V} \Opinion_{i,t}\}_t$ is a decreasing sequence.%
\end{proposition1}
\vspace{-0.5em}
\begin{proof}
The proof relies on the same arguments we used previously. Using normalized edge weights ($\alpha=1$), the result comes from \Eq{eq:narrowing}: $\alpha \minOpinion_{t} \le \minOpinion_{t+1} \leq \maxOpinion_{t+1}  \leq \ \alpha \maxOpinion_{t}$.
\end{proof}

\begin{proposition2}
{\textbf{Two diverging groups under pure GSM. }}
Let $\beta^+ = \{i\in\V : \beta_{i}=1\} \ne \emptyset$ %
 and $\beta^-= \{i\in V \backslash \beta^+ : \beta_{i}=-1\}$%
, the two sets with opposite reaction to global information. Then:
\begin{equation}
\lim_{+\infty}\mathbb{E}[\Opinion_{i,t}]=
\left\{
\begin{array}{ll}
+\infty & \ i \in \beta^+; \\
-\infty & \ i \in \beta^-.
\end{array}
\right. 
\end{equation}
\end{proposition2}
\begin{proof}
Let ${i,j} \in \V : \beta_i=1, \beta_j=-1$. The update rule for agent $i$ is $\Opinion_{i,t+1} = g(S_{t}) + \Opinion_{i,t}$, yielding: \\
$\Opinion_{i,t+1}-\Opinion_{i,t}=g(S_{t})\ge 0 \,\Rightarrow\, 
\mathbb{P}(S_{i,t+1}=1)\ge \mathbb{P}(S_{i,t}=1)$. %
Using the definition of $\mathbb{E}[x]=\sum_x\mathbb{P}(x)x$, we can write:
\begin{align}
\!\!\!\!\!\!\mathbb{E}[g(S_{t})]>\mathbb{P}(\norm{S_t}_1=1)\dot{g}(1) &\ge \mathbb{P}(S_{i,t}=1)\dot{g}(1)\\&\ge \mathbb{P}(S_{i,0}=1)\dot{g}(1)>0,
\end{align}
so that $\exists \eta > 0 \text{ s.t. } \forall t,\ \mathbb{E}[g(S_t)]>\eta$. Taking the expectation of the update rule for $i$ and $j$, we get:
\begin{equation}
\mathbb{E}[\Opinion_{i,t}]=\mathbb{E}[X_{i,0}]+\sum_{\tau=0}^{t-1}\mathbb{E}[g(S_t)]\ge \mathbb{E}[X_{i,0}]+\eta t,
\end{equation}
\begin{equation}
\mathbb{E}[\Opinion_{j,t}]=\mathbb{E}[\Opinion_{j,0}]+\sum_{\tau=0}^{t-1}\mathbb{E}[g(S_t)]\le \mathbb{E}[\Opinion_{j,0}]-\eta t .
\end{equation}
The result is obtained by taking the limit on the right side of each of the above inequalities.
\end{proof}
\begin{lemma}
Let $g : \{0,1\}^{N} \rightarrow \mathbb{R}^{+}$ be an increasing function in the sum of $S_{i,t}$'s\st $g(S_t)=0 \,\Leftrightarrow\, S_t=(0,...,0)^\top$, and $\gamma>0$. Then:
\begin{equation}
\lim_{+\infty}\mathbb{E}\big{[}g(S_{t})\big{]}=0 \ \Rightarrow \ \forall i \in \V,\ \lim_{+\infty}\mathbb{E}[\Opinion_{i,t}]=-\infty.
\end{equation}\label{lemma: div. inf.}
\end{lemma}%
\vspace{-1.8em}
\begin{proof}
We start from the assumption on the left side: \\
$0 = \lim_{+\infty}\mathbb{E}[g(S_{t})] %
= \frac{\gamma}{N}\lim_{+\infty}\mathbb{E}[\norm{S_{t}}_1]$\\
$\phantom{0 }= \frac{\gamma}{N}\lim_{+\infty} \sum_{k=0}^{N}\mathbb{P}(\norm{S_{t}}_1 = k) k$\\
$\phantom{0 }= \frac{\gamma}{N}\lim_{+\infty} \sum_{k=0}^{N}\mathbb{P}(\sum\overN{i} S_{i,t} = k) k = 0$\\
$\Rightarrow\,\forall i\in V,\ \lim_{+\infty}\mathbb{E}[S_{i,t}]=0$, and therefore:
\begin{equation}
\lim_{+\infty} \mathbb{E}\Big{[}
{\textstyle\frac{1}{1+\exp(-\lambda \Opinion_{i,t})}}
\Big{]}=0.
\end{equation}
Assume $\exists M,\, \eta$\st $\forall t_{0}, \exists t>t_{0}$ for which  $\mathbb{P}(\Opinion_{i,t}>M)>\eta$, then, by denoting the density function of $\Opinion_{i,t}$ by $f_{i,t}$, we write: 
\begin{align}
\!\!\!\!\!\!\!\!\!\int_{-\infty}^{+\infty}{\!\!\!\!\!\textstyle\frac{1}{1+\exp(-\lambda x)}}f_{i,t}(x)dx =
& \int_{-\infty}^{M}\!\!\!{\textstyle\frac{1}{1+\exp(-\lambda \Opinion_{i,t})}}f_{i,t}(x)dx  \nonumber\\
&+ \int_{M}^{+\infty}\!\!\!\!\!\!{\textstyle\frac{1}{1+\exp(-\lambda x)}}f_{i,t}(x)dx,\!\!
\end{align}
so that %
$\mathbb{E}[\tilde{S}_{t}]\ge {\textstyle\frac{1}{N}} \mathbb{E}[S_{i,t}]>{\textstyle\frac{1}{1+\exp(-\lambda M)}}\cdot{\textstyle\frac{\eta}{N}}$, %
which contradicts $\lim_{+\infty}\mathbb{E}[\tilde{S}_{t}]=0$. Thus, we conclude that $\forall M, \,\lim_{+\infty}\mathbb{P}( \Opinion_{i,t}>M)=0$. Now we need to show:
\begin{equation}
\forall M, \, \lim_{+\infty}\mathbb{P}( \Opinion_{i,t}>M)=0 \,\implies\, \lim_{+\infty}\mathbb{E}[\Opinion_{i,t}]=-\infty.
\end{equation}
Assume $\forall M, \,\lim_{+\infty}\mathbb{P}( \Opinion_{i,t}>M)=0$. Assume also that $\exists [a,b] \subset [0,+\infty[$  s.t.  $\forall t_{0}, \exists t>t_{0}, f_{i,t}(x)\ge f_{i,0}(x)$ almost everywhere. Then clearly:
\begin{equation*}
\mathbb{P}(\Opinion_{i,t}\ge a) \ge \mathbb{P}(b \ge \Opinion_{i,t} \ge a)=\int_{a}^{b}\!\!f_{i,t}(x)dx \ge \int_{a}^{b}\!\!f_{i,0}(x)dx,
\end{equation*}
which yields a contradiction. Thus, $\forall [a,b]\subset [0, +\infty[, \,\exists t_{0}$\st $f_{i,t}(x)<f_{i,0}(x)$ almost everywhere. Furthermore, $\forall M,\,\forall \varepsilon>0, \exists t_0 \text{ s.t. } \forall t>t_{0}, \,\mathbb{P}(\Opinion_{i,t}\le M)\ge 1-\varepsilon$.
Let $M<0$, one has $\mathbb{E}[\Opinion_{i,t}]=\int_{-\infty}^{M}xf_{i,t}(x)dx+\int_{M}^{0}xf_{i,t}(x)dx+\int_{0}^{+\infty}xf_{i,t}(x)dx$, and thus, for $t$ large enough:
\begin{equation}
\mathbb{E}[\Opinion_{i,t}]\le M(1-\varepsilon) + \int_{0}^{+\infty}xf_{i,0}(x)dx.
\end{equation}
Thus $\forall M<0, \,\exists t_{0} \text{ s.t. } \forall t>t_{0},\, \mathbb{E}[\Opinion_{i,t}]\le M$, so finally, $\lim_{+\infty}\mathbb{E}[\Opinion_{i,t}]=-\infty$. 
\end{proof}
\begin{lemma}
If there exists $i$\st $\forall j\in\mathcal{N}_{i}\cap\{i\}$, $\exists C_{j}$ 
for which $\lim_{+\infty}\mathbb{E}[\Opinion_{j,t}]=C_{j}$, then $\lim_{+\infty}\mathbb{E}\big{[}g(S_{t})\big{]}$ exists and is finite. The proof of this proposition is obtained by manipulating trivially the update rule.
\label{lemma: limit}
\end{lemma}
\smallskip
\begin{lemma}
For any $i$, if $\forall i, \,\exists C_{i}$ 
for which $\lim_{+\infty}\mathbb{E}[\Opinion_{i,t}]=C_{i}$, then $\nexists C_{i}\in\mathbb{R}$ s.t. $\forall j\in\mathcal{N}_{i}\cup \{i\}$, $\lim_{+\infty}{\mathbb{E}[\Opinion_{j,t}]=C_{i}}$.
\label{lemma: local no consensus}
\end{lemma}
\vspace{-0.5em}
\begin{proof}
Suppose there is one $i$ such that $\forall j$ in $i$'s neighborhood $\lim_{+\infty}\mathbb{E}[\Opinion_{j,t}]=C_{i}$ and $\lim_{+\infty}\mathbb{E}[\Opinion_{i,t}]=C_{i}$. Then, the expectation of the update rule yields:
\begin{equation}\label{eq:one_more}
\mathbb{E}[\Opinion_{i,t+1}]-\sum\overN{j} w_{ji}\mathbb{E}[\Opinion_{j,t}]=\beta_{i}\mathbb{E}\big{[}g(S_{t})\big{]}.
\end{equation}
Lemma\,\ref{lemma: limit} allows us to take the limit in \Eq{eq:one_more}, and get:
\begin{equation}
C_{i}-C_{i}\sum\overN{j} w_{ji}=\beta_{i}\lim_{+\infty}\mathbb{E}\big{[}g(S_{t})\big{]},
\end{equation}
which, by the normalized-weights assumption, yields:
\begin{equation}
\lim_{+\infty}\mathbb{E}\big{[}g(S_{t})\big{]}=0,
\end{equation}
and together with Lemma \ref{lemma: div. inf.} yields a contradiction.
\end{proof}
\begin{proposition3}
{\textbf{No consensus under GSM. }}%
If the GSM is in effect (if $\gamma>0$), consensus is impossible to be reached.
\end{proposition3}
\vspace{-0.5em}
\begin{proof}
No consensus is a natural corollary of Proposition 3, by taking $\forall i \in \V, \,C_{i}=C$.
\end{proof}
\begin{lemma}
Suppose $\forall i \in \V, \,\exists C_{i}$ 
for which $\lim_{+\infty}\mathbb{E}[\Opinion_{i,t}]=C_{i}$. Then:
\vspace{-0.5em}
\begin{equation}
\lim_{+\infty}\mathbb{E}[\Opinion_{i,t}]=\min_{j\in\V} C_{j} \Rightarrow \beta_{i}=-1,
\end{equation}
\begin{equation}
\!\!\!\lim_{+\infty}\mathbb{E}[\Opinion_{i,t}]=\max_{j\in\V} C_{j} \,\Rightarrow\, \beta_{i}=1.
\end{equation}\label{lemma: min-max opinions}
\end{lemma}
\vspace{-1.8em}
\begin{proof}
Let $i\in \V$ such that $C_{i}=\min_{j\in\V} C_{j}$. Then, taking the limit of the update rule for agent $i$ gives:
\begin{equation}
\!\!\!\!\!\!\!C_{i}=\beta_{i}\lim_{+\infty}\mathbb{E}\big{[}g(S_t)\big{]} + \sum_{j\in\mathcal{N}_{i}}\!w_{ji}C_{j} 
\ge 
\beta_{i}\lim_{+\infty}\mathbb{E}\big{[}g(S_t)\big{]} + C_{i}\!.\!\!\!\!\!\!\!
\end{equation}
We know from Lemma \ref{lemma: div. inf.} that $\lim_{+\infty}\mathbb{E}\big{[}g(S_t)\big{]}\ne 0$, so we need $\beta_{i}<0$ for the above inequality to hold, thus $\beta_{i}=-1$. The proof for the second part of the proposition is similar.
\end{proof}

\begin{proposition4}
{\textbf{Boundary on $\varepsilon$-consensus. }}%
For a strictly increasing $g(S_t)$ function, reaching a $(\lim_{+\infty}\mathbb{E}\big[g(S_t)\big])$-consensus is not possible. For the special GSM form of $g(S_t) = \gamma \tilde{S}_t$, this corresponds to a $(\gamma\lim_{+\infty}\mathbb{E}[\tilde{S_t}])$-consensus.%
\end{proposition4}
\vspace{-0.5em}
\begin{proof}
Suppose we have an $\epsilon$-consensus, then clearly:
\begin{equation}
\forall i \in \V, \, \,C_{i} \le \min_{j\in\V} C_{j} + \epsilon.
\end{equation}
Let $i^{*}\in\V$ be s.t. $C_{i^{*}}=\min_{j\in\V} C_{j}$. 
From Proposition\,\ref{prop:two-groups-divergence}, we know that $\beta_{i}=-1$, so that, using the update rule we can write:
\begin{align}
C_{i^{*}} &\le C_{i^{*}}+\epsilon - \lim_{+\infty}\mathbb{E}\big{[}g(S_t)\big{]},\\
\Rightarrow \ \lim_{+\infty}\mathbb{E}\big{[}g(S_t)\big{]} &\le \epsilon.
\end{align}
This gives Proposition\,\ref{prop:boundary-on-e-consensus} by contraposition.
\end{proof}

\begin{proposition5}
{\textbf{Bounded polarization under full stubbornness. }}
Let a strongly connected digraph $G=\{V,W\}$ with $N$ nodes, with normalized non-negative weights w.r.t the incoming edges of each node, and $\forall k, w_{kk} > 0$. Suppose a DeGroot model with a set $\mathcal{S}\subset V$ of fully stubborn agents (i.e.~$\gamma = 0$ in our model definition to neutralize GSM). Then, the polarization of the system over time will be eventually bounded by $D_{\max,\infty} = \max_{k\in \mathcal{S}}X_{k,\cdot}  - \min_{k\in \mathcal{S}}X_{k,\cdot}$, where $X_{k,\cdot}$ is a fixed stubborn opinion such that $\forall t, \,X_{k,t} = X_{k,\cdot}$.
\end{proposition5}
\newcommand{\mytau}{(\tau)}%
\vspace{-1em}
\begin{proof}
It is sufficient to show that the process always reaches one of the two absorbent states: \\
\phantom{iiii}i)~$\exists t \text{ s.t. } \forall i \in V, \, \min_{k\in \mathcal{S}}X_{k,\cdot} \le X_{i,t}\le \max_{k\in \mathcal{S}}X_{k,\cdot}$; \\
\phantom{iii}ii)~$\lim_{+\infty}\max_{i\notin \mathcal{S}}X_{i,t}= \max_{k\in \mathcal{S}}X_{k,\cdot} \text{ and } $ \\
\phantom{iiiii)~}$\lim_{+\infty}\min_{i\notin\mathcal{S}}X_{i,t}= \min_{k\in \mathcal{S}}X_{k,\cdot}$.\\
The above states are absorbent as all opinions have come in a closed value range, and there is nothing (e.g. the GSM) that could lead them to challenge these bounds. 

    By contradiction to what the right side of case (i) states, assume %
        $\forall t, \exists i \notin \mathcal{S} \text{ s.t. } X_{i,t}>\max_{k\in \mathcal{S}}X_{k,\cdot}$. %
    Then, $\max_{i \notin S}X_{i,t}=\max_{j\in V}X_{j,t}$ is decreasing (Proposition\,\ref{prop:narrowing}) and bounded, by our hypothesis, thus %
converges to some $x=\lim_{+\infty}\max_{i \notin S}X_{i,t}$. Now, assume $x > \max_{k\in \mathcal{S}}X_{k,\cdot}$. Since the network is strongly connected, there exists $\tau\ge 0$ so that, for $t$ large enough where $X_{j,t} = X_{j,t-\tau}$, the update rule of any $i \notin \mathcal{S}$ will be:
    \begin{equation}
        X_{i,t+1}=\sum\overN{j}w_{ji}^{\mytau}X_{j,t}=\sum\overN{j}w_{ji}^{\mytau}X_{j,t-\tau},
    \end{equation}
    where, by design, $\sum\overN{j}w_{ji}^{\mytau}=1$ and $\forall i\notin \mathcal{S}, \forall j \in V, \,w_{ji}>0$. Then for all $i \notin \mathcal{S}$:
    \begin{equation}
    \begin{split}
     \!\!\!   \exists t \text{ s.t. } X_{i,t+1}&=\sum_{k \in \mathcal{S}}w_{ki}^{\mytau}X_{k,t-\tau} + \sum_{j \notin \mathcal{S}}w_{ji}^{\mytau}X_{j,t-\tau} \\
        &\le \sum_{k \in \mathcal{S}}w_{ki}^{\mytau}X_{k,\cdot}+ \Big{(}\sum_{j \notin \mathcal{S}}w_{ji}^{\mytau}\Big{)}\max_{j\notin \mathcal{S}}X_{j,t-\tau}.
    \end{split}
    \end{equation}
Taking the limit, w.r.t $t$, of the right-most term of the above inequality, yields:
\begin{equation}
    \exists t \text{ s.t. } X_{i,t+1}<\sum_{k\in \mathcal{S}}w_{ki}^{\mytau}X_{k,\cdot} + \Big{(}\sum_{j\notin \mathcal{S}}w_{ji}^{\mytau}\Big{)} x. 
\end{equation}
Since this is true for any $i\notin \mathcal{S}$, we get:
\begin{equation}
    \exists t \text{ s.t. } \max_{i\notin \mathcal{S}}X_{i,t+1}<\sum_{k\in \mathcal{S}}w_{ki}^{\mytau}X_{k,\cdot}+\Big{(}\sum_{j\notin \mathcal{S}}w_{ji}^{\mytau}\Big{)} x \leq x,
\end{equation}
where $\max_{i\notin \mathcal{S}}X_{i,t+1} \leq x$ gives a contradiction. With similar reasoning, we can get the result also for $\min_{k\in \mathcal{S}}X_{k,\cdot}$ (the left side of case (i)). 
\end{proof}

\begin{proposition6}
{\textbf{Limited polarization in signed networks. }}
Consider the DeGroot update rule over a signed and possibly time-dependent network,
$\Opinion_{i,t+1}=\sum\overN{j} w_{ji,t}\Opinion_{j,t}$,  %
where $w_{ji,t}\in [-1,1]$ and $\sum\overN{j} |w_{ji,t}|=1$. %
Then, $\forall t$:%
\begin{enumerate}
\item $\max_{(i,j\in V)} (\Opinion_{i,t}-\Opinion_{j,t})\le 2\max_{i\in V}|\Opinion_{i,0}|$; 
\item and either $\max_{i\in V} \Opinion_{i,t} \le \max_{i \in V} \Opinion_{i,0}$, \\\phantom{iiiiii }\, \,\ \ or\, $\min_{i\in V} \Opinion_{i,t} \ge \min_{i\in V} \Opinion_{i,0}$.
\end{enumerate}
\end{proposition6}
\begin{proof}
Taking the absolute value of the update rule, gives:
\begin{align}
|\Opinion_{i,t+1}|=|\sum\overN{j} w_{ji,t}\Opinion_{i,t}| &\le \sum\overN{j} |w_{ji,t}|\,|\Opinion_{i,t}| \nonumber\\
&\le \max_{i\in V}|\Opinion_{i,t}\underbrace{|\sum\overN{j} |w_{ji,t}|}_{=1}.
\end{align}
Thus, $\forall t, \max_{i\in V}|\Opinion_{i,t+1}|\le \max_{i\in V}|\Opinion_{i,t}|$. The result follows naturally from this observation.
\end{proof}

\section{Supplementary material}
\label{sec: supplement}

In this section, we discuss several interesting properties of the model, and we also present additional experimental results. Before proceeding, we provide in \Tab{table: Ukraine} the detailed fitted model parameters for the case of Russian invasion to Ukraine, which was earlier illustrated in the map of \Fig{fig: map ukraine}.

\begin{table}[t]
\scriptsize
\centering
\begin{tabular}{L{2.2cm} C{1.3cm} R{1.15cm} R{1cm} R{0.85cm}} 
\toprule
 \multirow{2}{*}{\textbf{Language}} & \multirow{2}{*}{\textbf{Error}} & \multicolumn{3}{c}{\textbf{Best parameters}}\\
&& ${\bm\mu}^*\ \ $ & ${\bm \gamma}^*\ \ $ & ${\bm r}^*\ \ $\\
\midrule
Portuguese &0.079&-15.032&0.436&0.293\\
 Danish&0.103&37.500&0.063&0.237\\
 French &0.116&-143.229&0.047&0.483\\
 Arabic &0.171&-187.500&0.063&0.262\\
 English&0.171&-237.500&0.088&0.437\\
 Polish&0.173&-129.998&0.078&0.444\\
 Ukrainian&0.179&-105.263&0.020&0.419\\
 Catalan&0.207& -148.104&0.140&0.408\\
 Turkish&0.216&-96.103&0.290&0.211\\
 Russian&0.216&-176.576&0.009&0.242\\
 Italian&0.228&-112.500&0.038&0.337\\
 Azerbaijani&0.238&79.507&0.042&0.442\\
 German&0.242&-74.876&0.098&0.366\\
 Mongolian&0.299&-112.500&0.088&0.337\\
 Croatian&0.309&114.992&0.067&0.385\\
 Indonesian&0.312& -135.514&0.088&0.393\\
 Spanish&0.315& -89.986&0.081&0.396\\
 Swedish&0.323& -181.669&0.108&0.333\\
 Greek&0.325& 76.365&0.038&0.376\\
 Serbian&0.326& -180.659&0.103&0.165\\
 Finnish&0.344&-115.637&0.127&0.080\\
 Bulgarian&0.344& -167.765&0.371&0.062\\
 Korean&0.354& -77.711&0.050&0.468\\
 Dutch&0.358& 40.817&0.466&0.014\\
 Norwegian&0.365& -187.746&0.117&0.388\\
 Slovenian&0.369& -155.508&0.138&0.411\\
 Serbo-Croatian&0.377& -174.138&0.052&0.156\\
 Hindi&0.383& -122.453&0.482&0.034\\
 Latvian&0.390& -147.599&0.054&0.395\\
 Romanian&0.404& -166.494&0.082&0.495\\
 Cebuano&0.421& -87.263&0.111&0.270\\
 Czech&0.426& -128.022&0.129&0.331\\
 Hungarian&0.510& -101.473&0.130&0.172\\
 Hebrew&0.524& 111.431&0.998&0.409\\
 Slovak&0.538& -287.500&0.113&0.212\\
 Lithuanian&0.539& 47.500&0.088&0.412\\
 Macedonian&0.572& -133.952&0.199&0.245\\
 Persian&0.716& -187.500&0.213&0.312\\
\bottomrule
\end{tabular}
\vspace{-1mm}
\caption{\textbf{Ukrainian flag emoji~--}~Results obtained by our optimization process using Twitter data related to the first period of the war in Ukraine. The best model parameters ($\mu^*, \gamma^*, r^*$) estimated for each language and the corresponding fitting error are reported. The languages are ordered from the lowest to the highest fitting error.}
\label{table: Ukraine}
\end{table}

\begin{figure}[t]
    \centering
	\subfloat[$X_{\max}$ for $\beta=0.05$]{\includegraphics[width=0.48\columnwidth,viewport=5 0 390 255,clip]{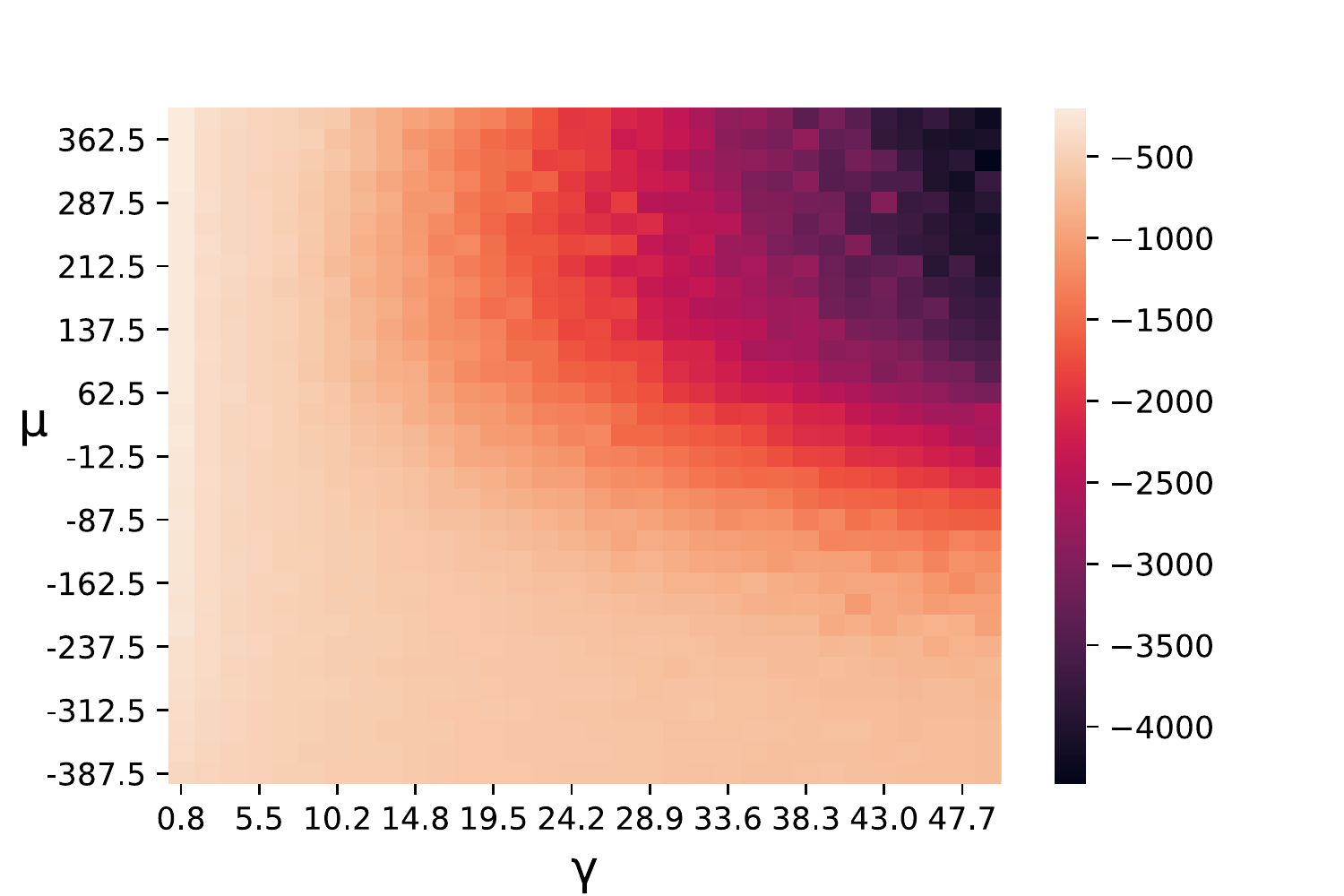} }
		\hfill 
    \subfloat[$X_{\max}$ for $\beta=0.95$]{\includegraphics[width=0.48\columnwidth,viewport=5 0 390 255,clip]{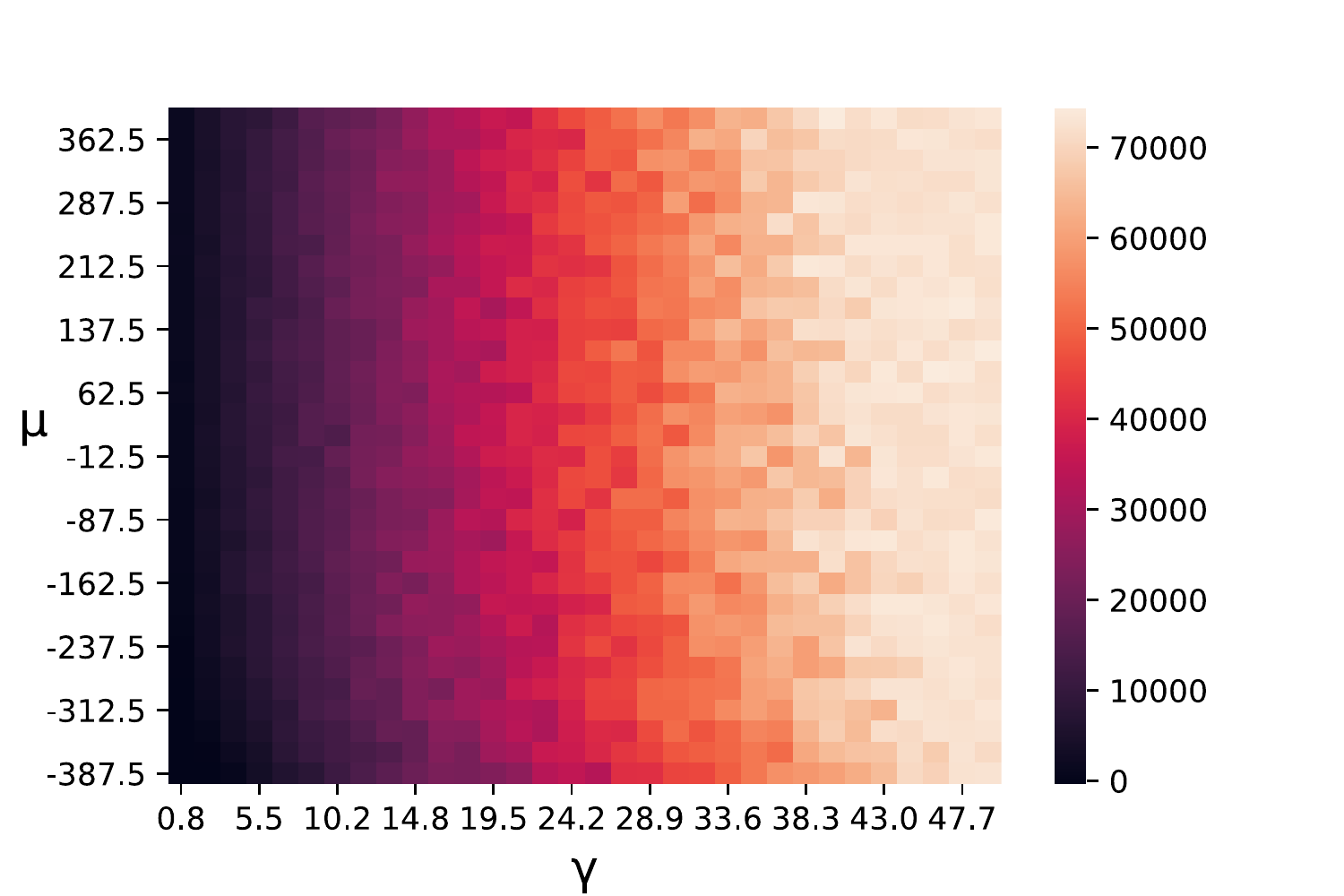} }\\
    \subfloat[$X_{\min}$ for $\beta=0.05$]{\includegraphics[width=0.48\columnwidth,viewport=5 0 390 255,clip]{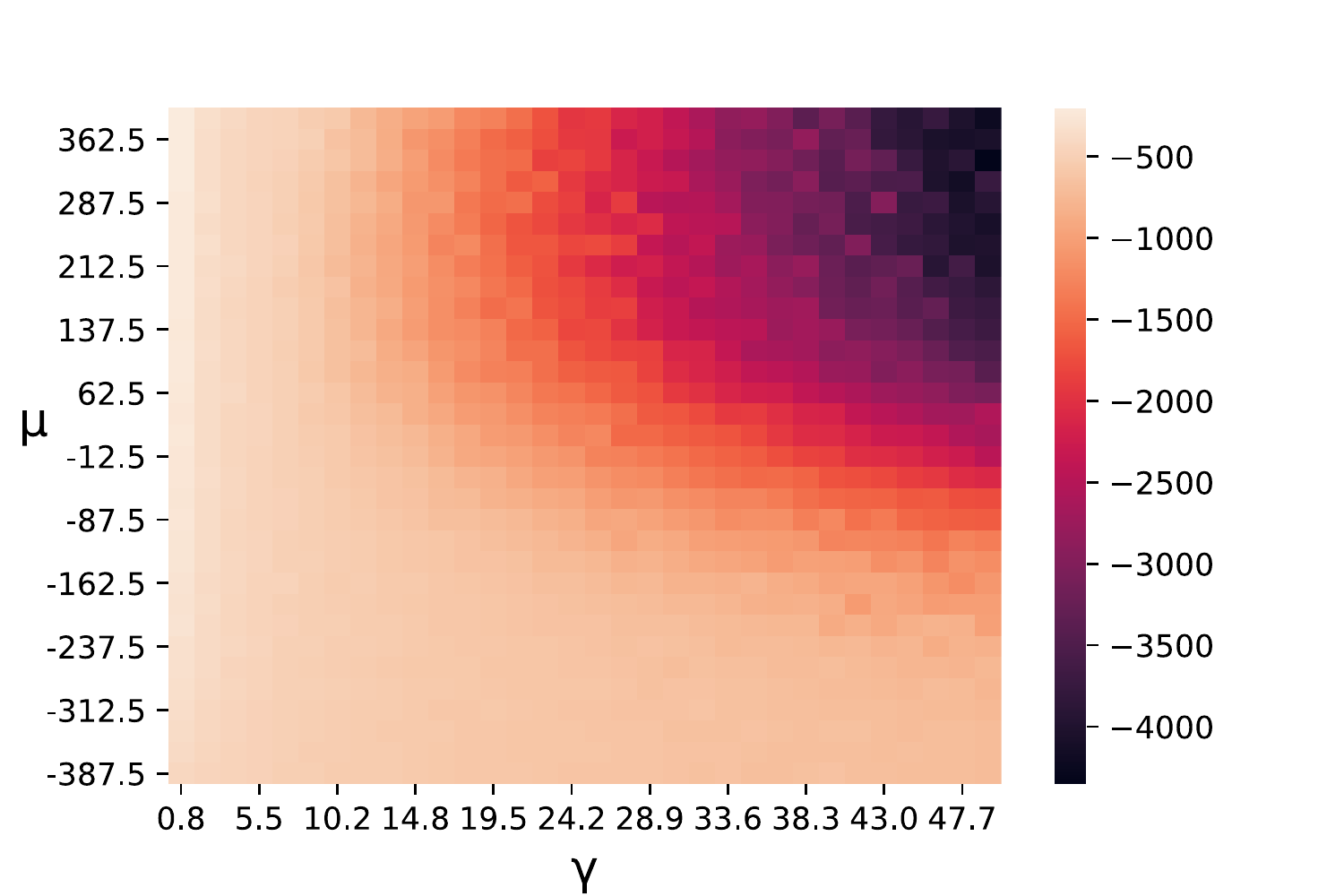} }
		\hfill
    \subfloat[$X_{\min}$ for $\beta=0.95$]{\includegraphics[width=0.48\columnwidth,viewport=5 0 390 255,clip]{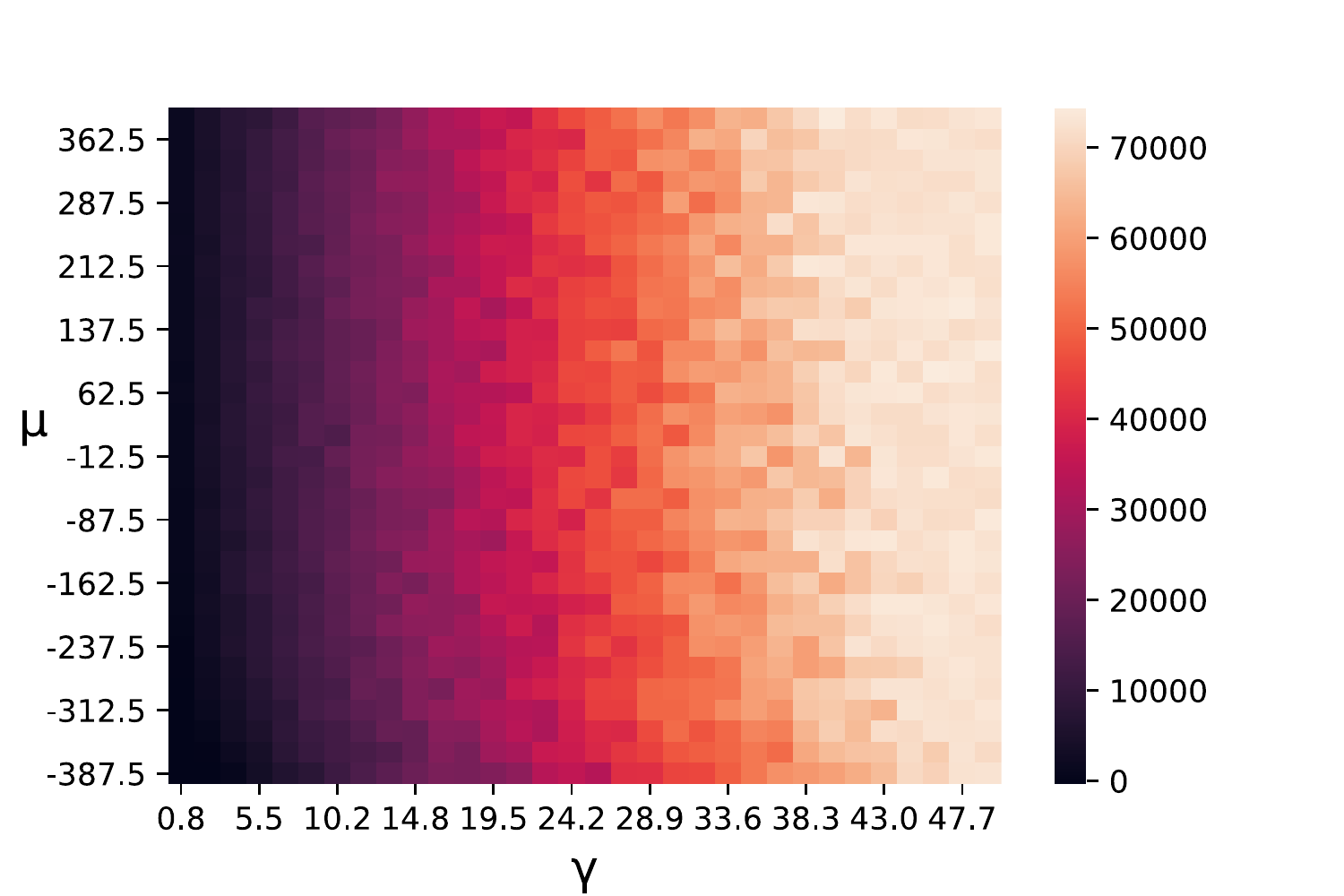} }
    \caption{Effect of $\mu$ and $\gamma$ on extreme opinions (maximal and minimal,  $X_{\max}$, $X_{\min}$) recorded at the end of sufficiently long simulations. Two scenarios are examined: $\beta = \{0.05, 0.95\}$.}
		\label{fig:min-max-opinion}
\end{figure}

\inlinetitle{I.~Steering mechanism drives the model's dynamics}~%
In the following, we call as maximal (resp. minimal) opinion the maximum (resp. minimum) agents' opinion at the end of a trial, and denote it by $X_{\max}$ (resp. $X_{\min}$). More formally:
\begin{align}
X_{\max}&=\max_{i\in V} X_{i,\infty},\\
X_{\min}&=\min_{i\in V}X_{i,\infty}.
\end{align}
In practice, we estimate these quantities empirically using sufficiently long simulations. The heatmaps in \Fig{fig:min-max-opinion} show how the $X_{\max}$ and $X_{\min}$ opinions among agents vary as a function of the strength of the GSM (controlled by $\gamma$ and $\mu$) in two distinct cases. We observe that: for $\beta=0.05$ both the minimum and the maximum opinions decrease, while for $\beta=0.95$ both increase. Previously, we identified such a behavior for the \textit{average} opinion, however the generalization to the nodes' opinions was not obvious: increasing the steering mechanism, in particular for $\beta>\frac{1}{2}$, could have implied an increasing maximum and a decreasing minimum (since the negative reaction of negative reacting nodes is fueled by the behavior of positively reacting nodes). This suggests the interaction between the GSM and the OPM: the OPM spreads those dynamics impulsed by the GSM to the majority, towards the rest of the population. This highlights how the GSM together with the OPM drive the dynamics of our model.

\newcommand{\figratio}{1}
\newcommand{\figpanelsmall}{0.12}
\newcommand{\figpanel}{0.3}

\begin{figure*}[h]
    \vspace{-1em}
\begin{minipage}{1\linewidth}
\scriptsize %
\hspace{13.5em} \textbf{Barab\'asi-Albert} \hspace{16em}   \textbf{Watts-Strogatz} \hspace{14.5em} \textbf{Stochastic Block Model}
\end{minipage}
\phantom{\hspace{4.14em}} {\rule[1ex]{5.325cm}{0.5pt}}%
\phantom{\hspace{0.99em}} {\rule[1ex]{5.325cm}{0.5pt}}%
\phantom{\hspace{0.99em}} {\rule[1ex]{5.325cm}{0.5pt}}%
\phantom{\hspace{-0.99em}}

\vspace{-1em}

\hspace{-2.25em}
\begin{minipage}{\figpanelsmall\linewidth}\centering 
{\scriptsize $N=100$}
\end{minipage}%
\begin{minipage}{\figpanel\linewidth}\centering 
		\subfloat{%
		\!\!\!\!\!\includegraphics[width=\fppeval{0.364*\figratio}\columnwidth, viewport=0 0 266 260, clip]{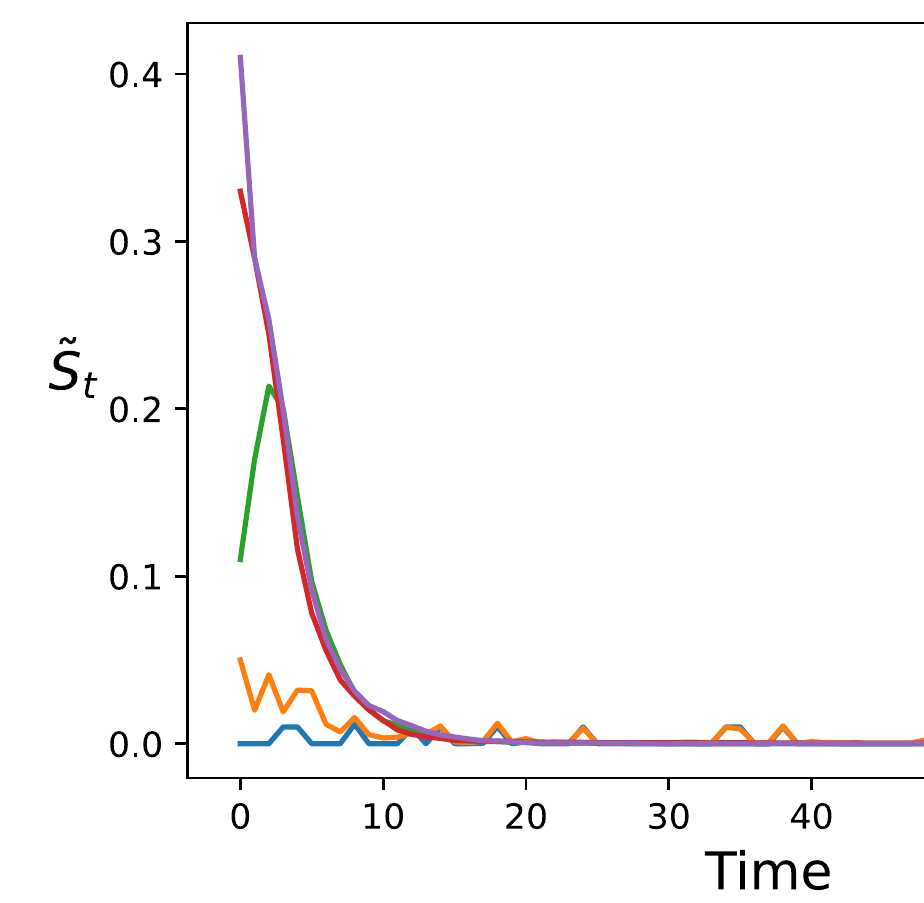}}
    \ 
    \subfloat{\includegraphics[width=\fppeval{0.32*\figratio}\columnwidth, viewport=32 0 266 260, clip]{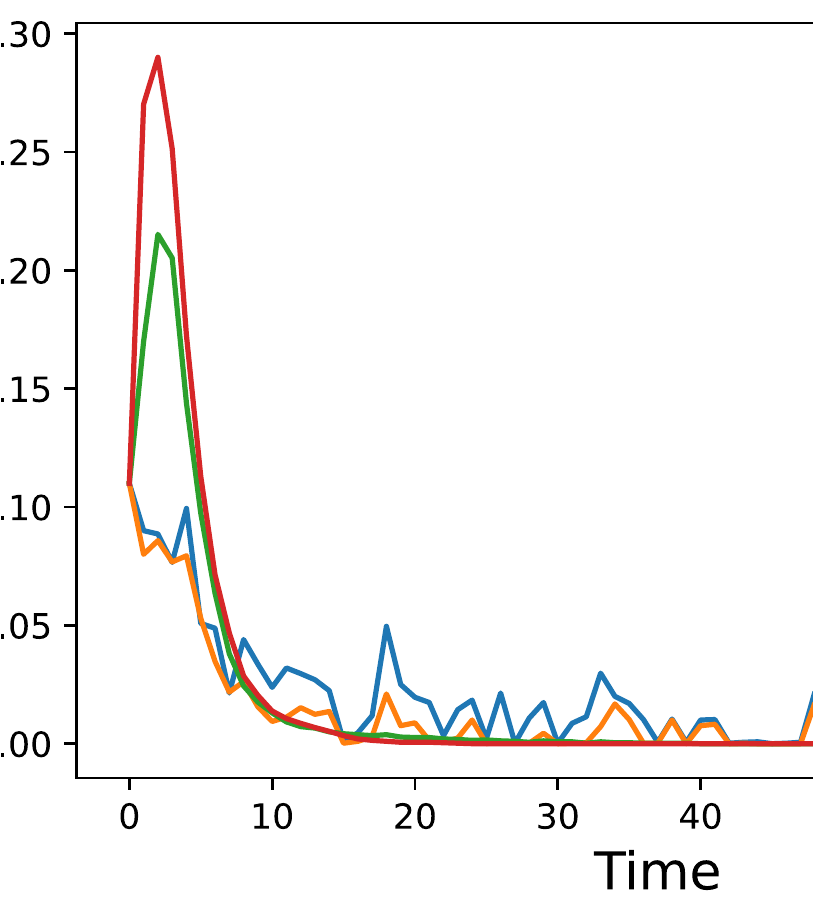}}
    \ 
    \subfloat{\includegraphics[width=\fppeval{0.32*\figratio}\columnwidth, viewport=32 0 266 260, clip]{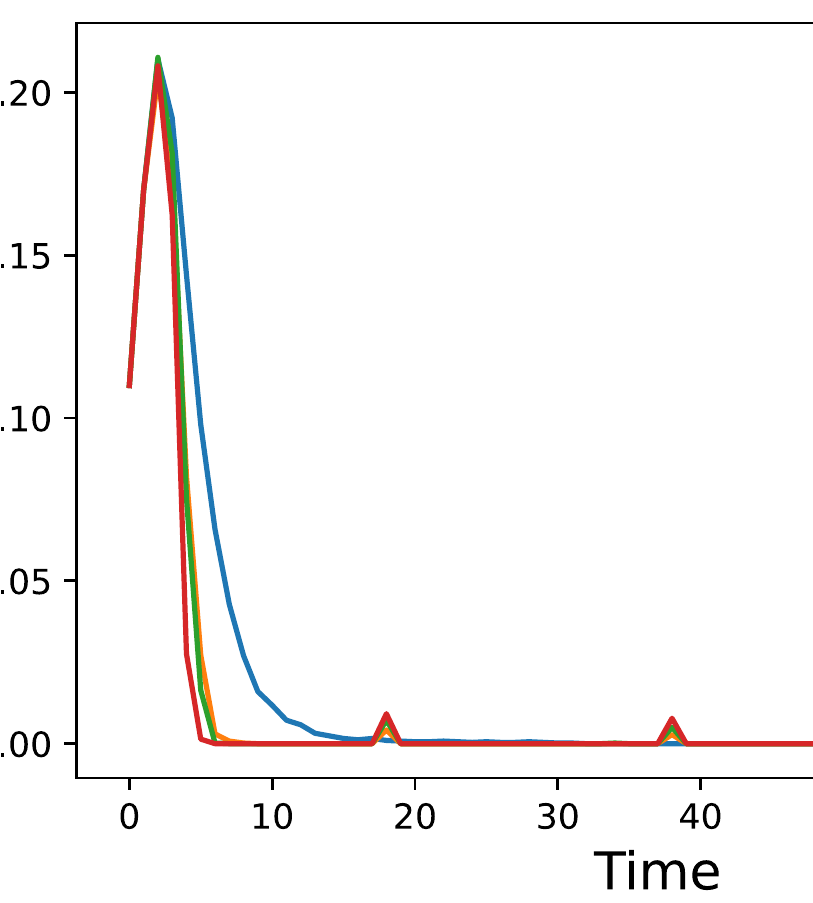}}
\end{minipage}\quad%
\begin{minipage}{\figpanel\linewidth}\centering 
		\subfloat{%
\!\!\!\!\!\includegraphics[width=\fppeval{0.364*\figratio}\columnwidth, viewport=0 0 266 260, clip]{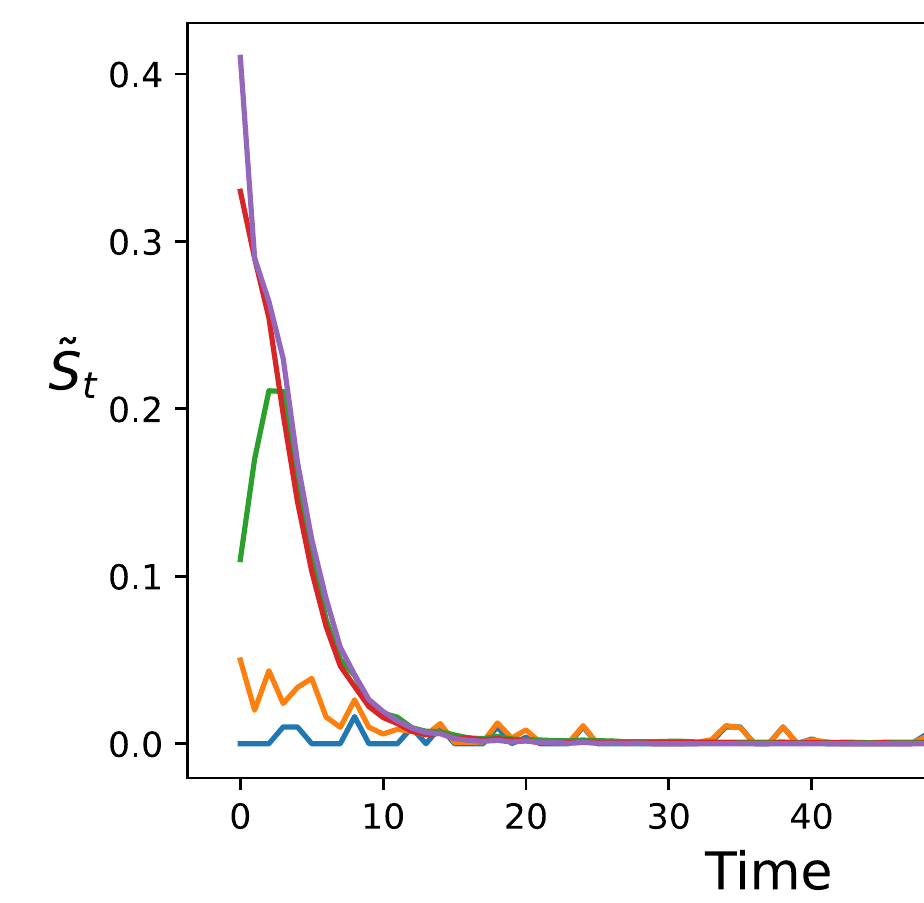}}
		\ 
    \subfloat{\includegraphics[width=\fppeval{0.32*\figratio}\columnwidth, viewport=32 0 266 260, clip]{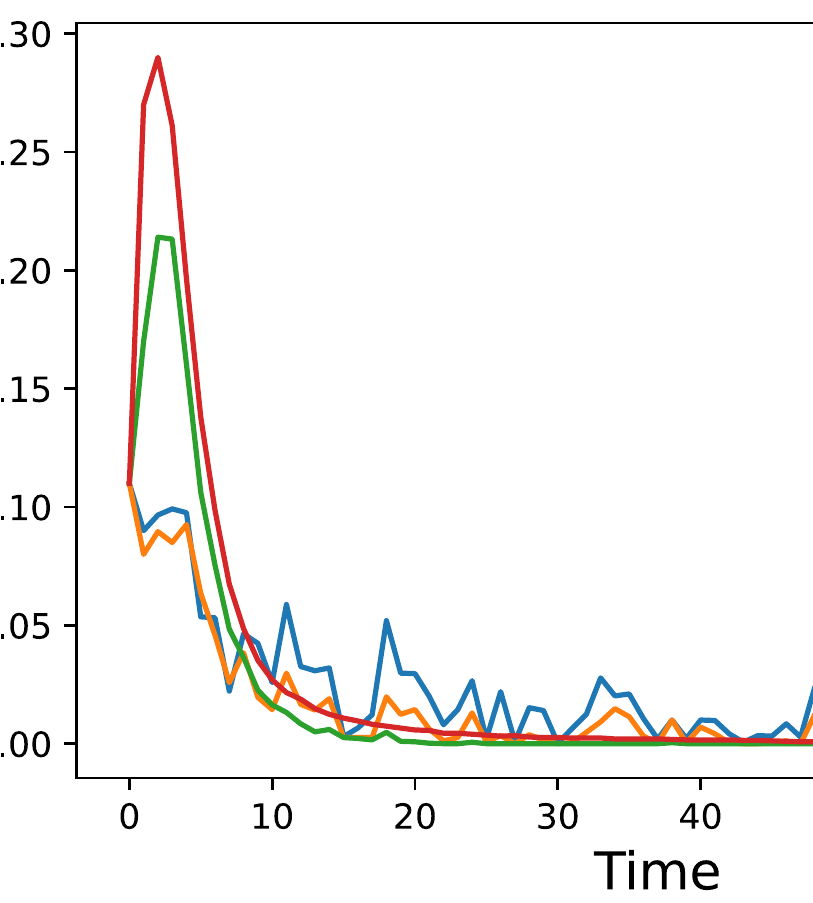}}
		\ 
    \subfloat{\includegraphics[width=\fppeval{0.32*\figratio}\columnwidth, viewport=32 0 266 260, clip]{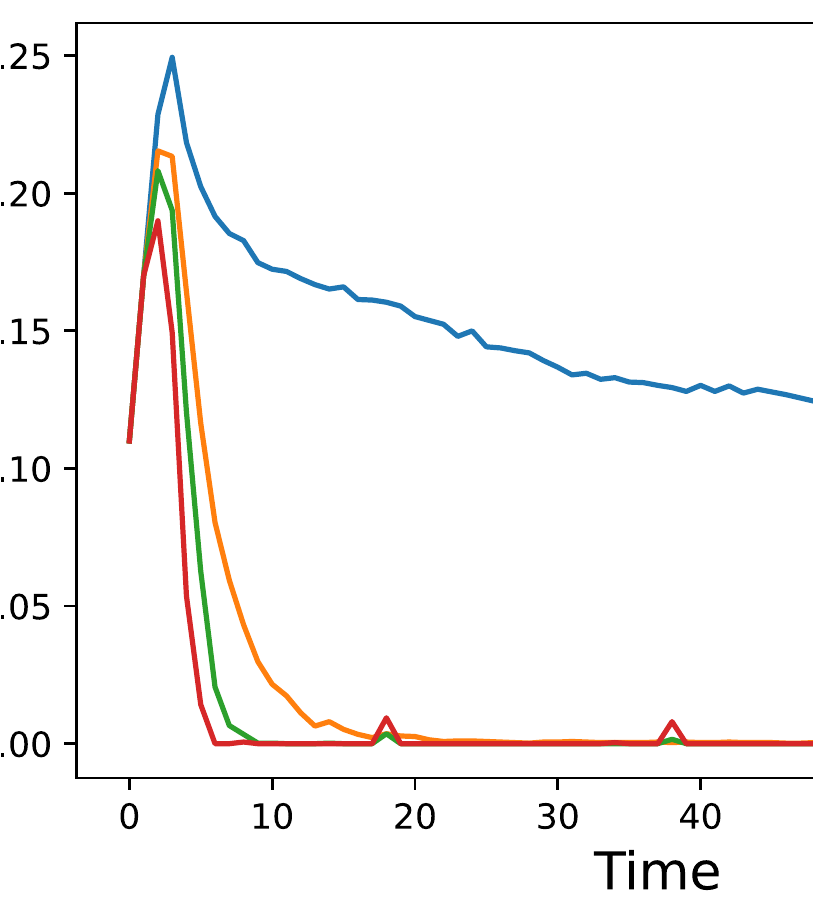}}
\end{minipage}\quad%
\begin{minipage}{\figpanel\linewidth}\centering 
		\subfloat{%
\!\!\!\!\!\includegraphics[width=\fppeval{0.364*\figratio}\columnwidth, viewport=0 0 266 260, clip]{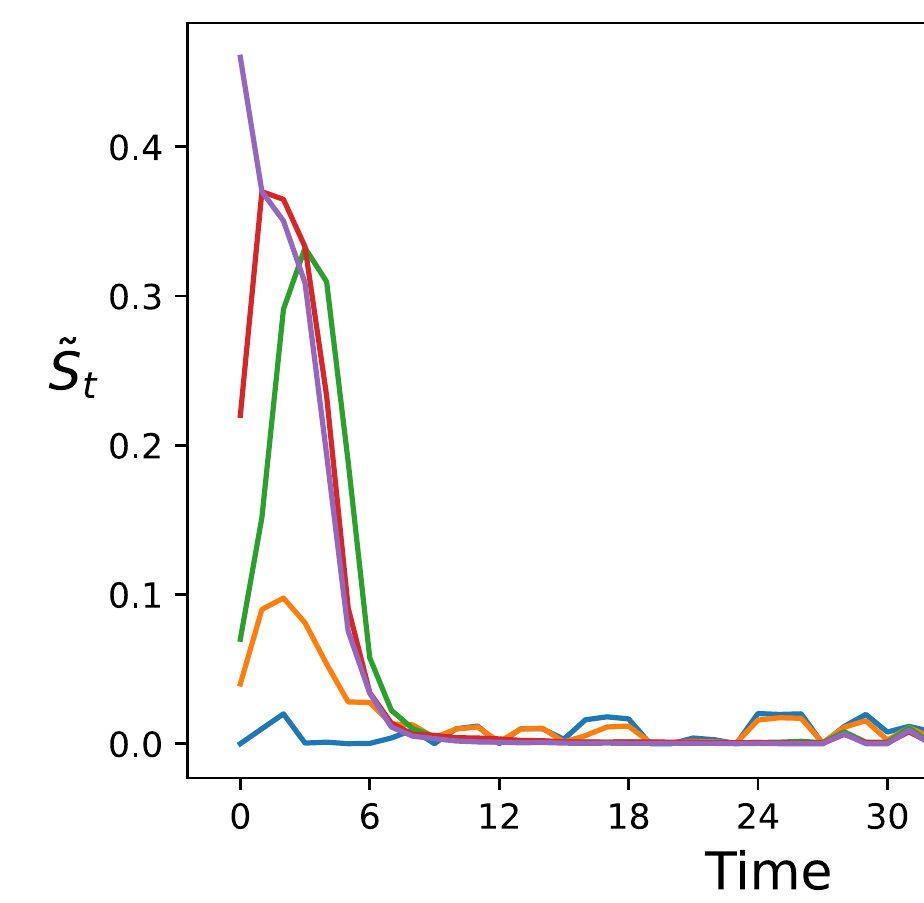}}
		\ 
    \subfloat{\includegraphics[width=\fppeval{0.32*\figratio}\columnwidth, viewport=32 0 266 260, clip]{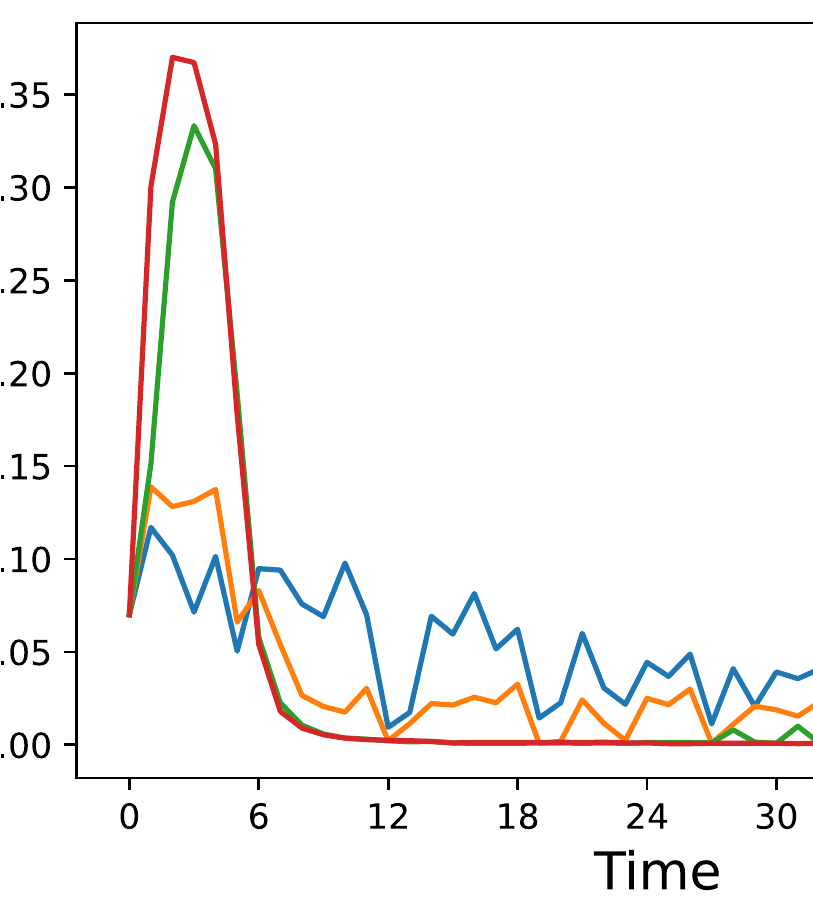}}
		\ 
    \subfloat{\includegraphics[width=\fppeval{0.32*\figratio}\columnwidth, viewport=32 0 266 260, clip]{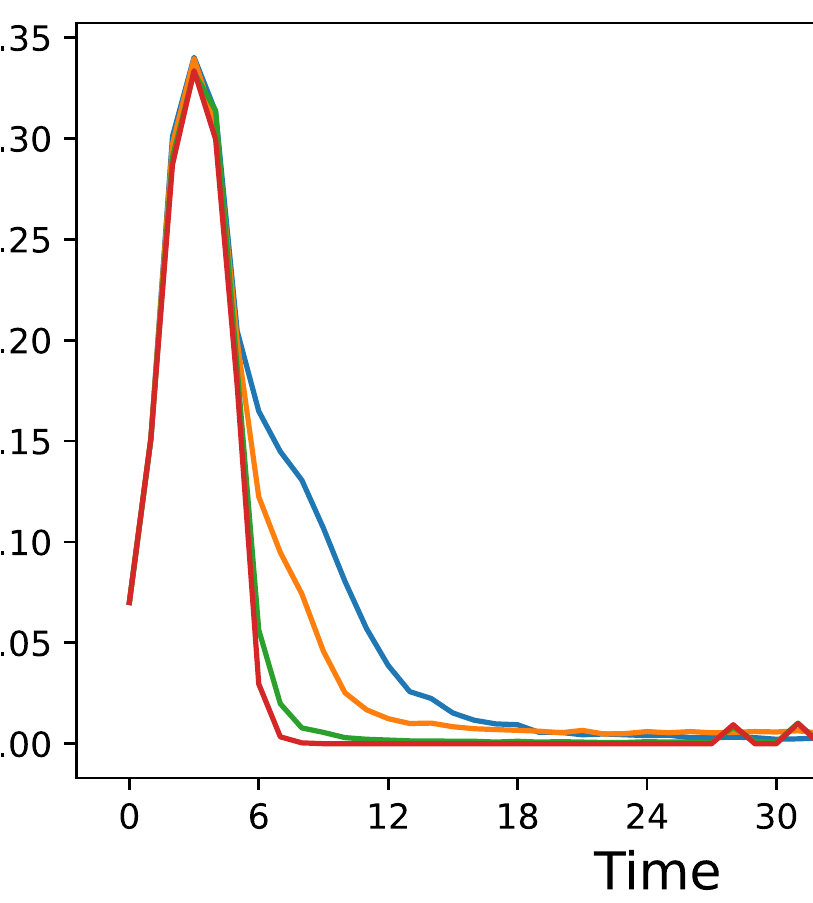}}
\end{minipage}

\vspace{-1.5em}

\hspace{-2.25em}
\begin{minipage}{\figpanelsmall\linewidth}\centering
{\scriptsize $N=1000$}
\end{minipage}%
\begin{minipage}{\figpanel\linewidth}\centering 
		\subfloat{%
		\!\!\!\!\!\includegraphics[width=\fppeval{0.364*\figratio}\columnwidth, viewport=0 0 266 260, clip]{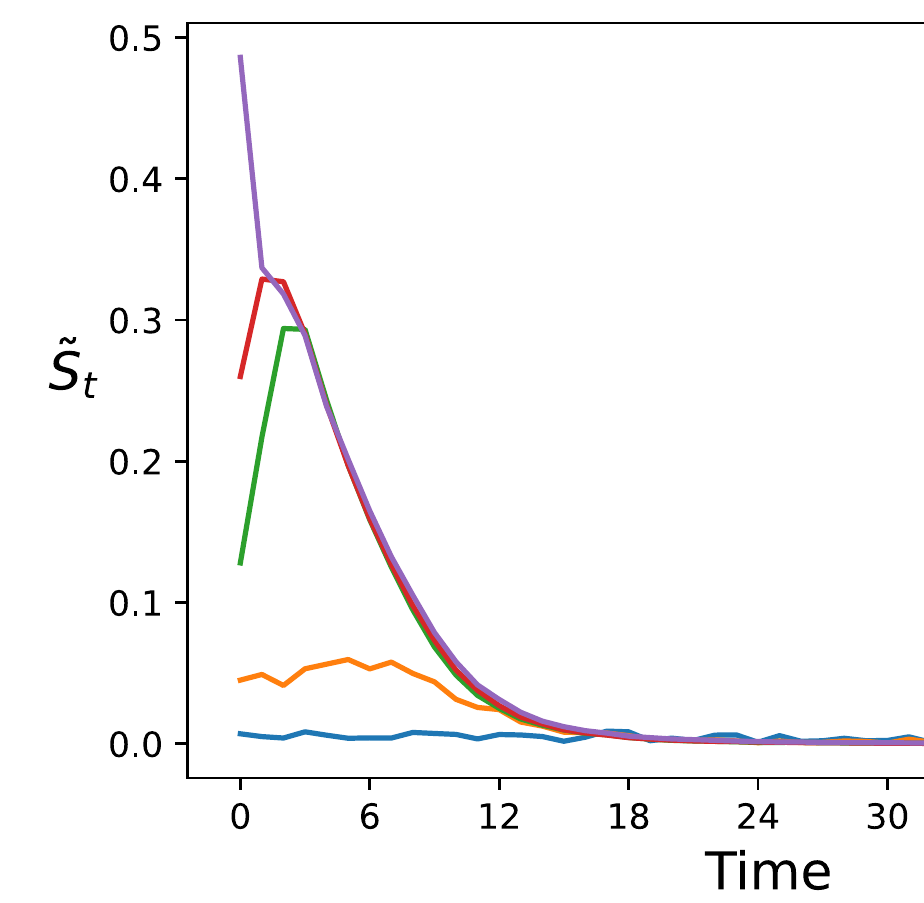}}
		\ 
    \subfloat{\includegraphics[width=\fppeval{0.32*\figratio}\columnwidth, viewport=32 0 266 260, clip]{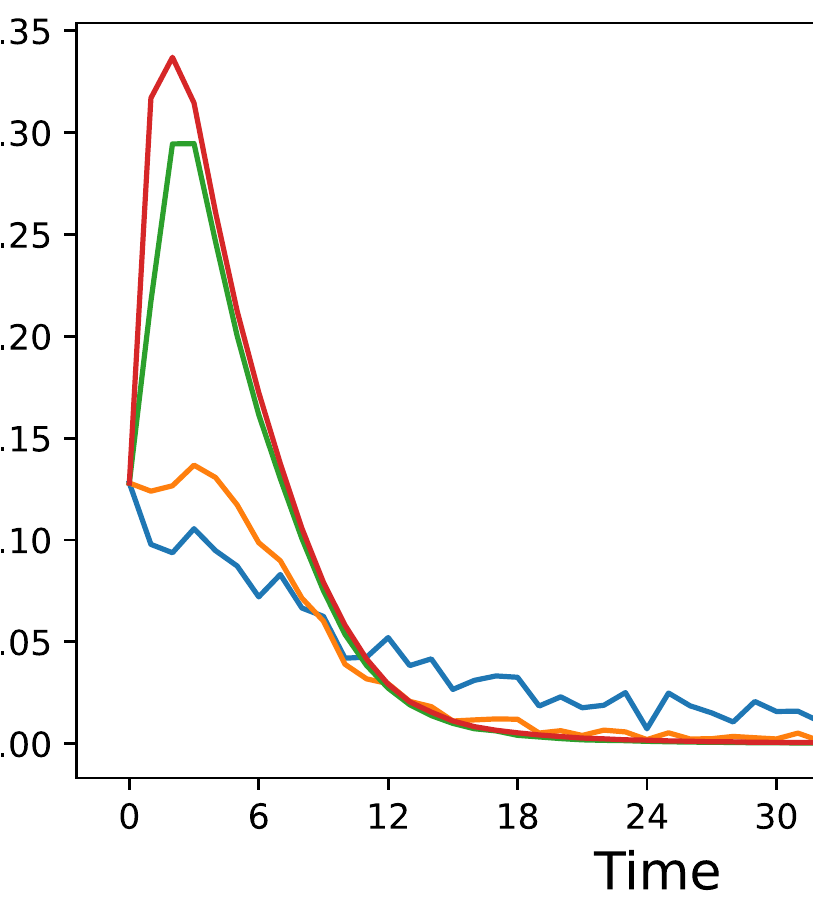}}
		\ 
    \subfloat{\includegraphics[width=\fppeval{0.32*\figratio}\columnwidth, viewport=32 0 266 260, clip]{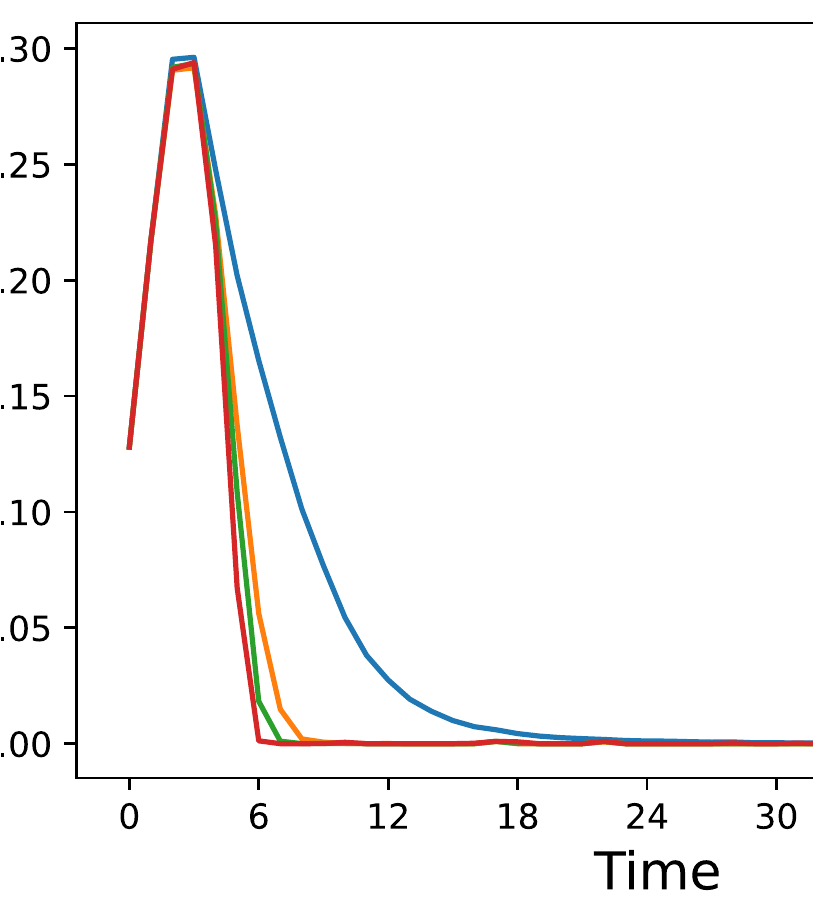}}
\end{minipage}\quad%
\begin{minipage}{\figpanel\linewidth}\centering 
		\subfloat{%
\!\!\!\!\!\includegraphics[width=\fppeval{0.364*\figratio}\columnwidth, viewport=0 0 266 260, clip]{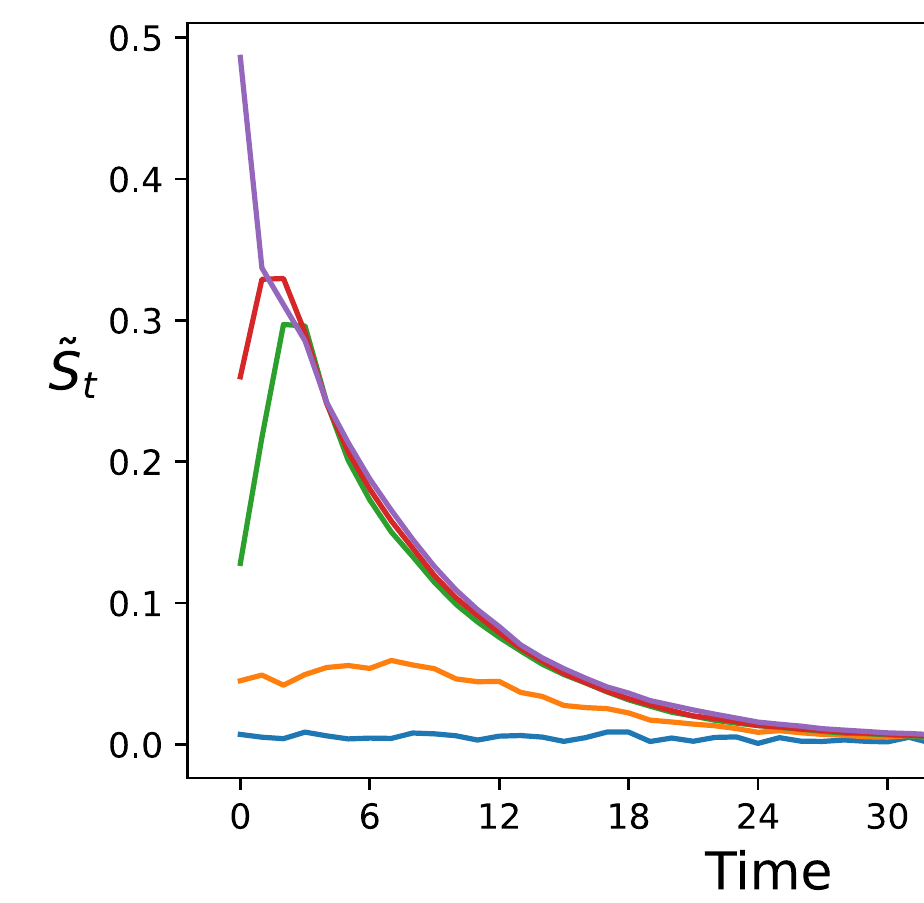}}
		\ 
    \subfloat{\includegraphics[width=\fppeval{0.32*\figratio}\columnwidth, viewport=32 0 266 260, clip]{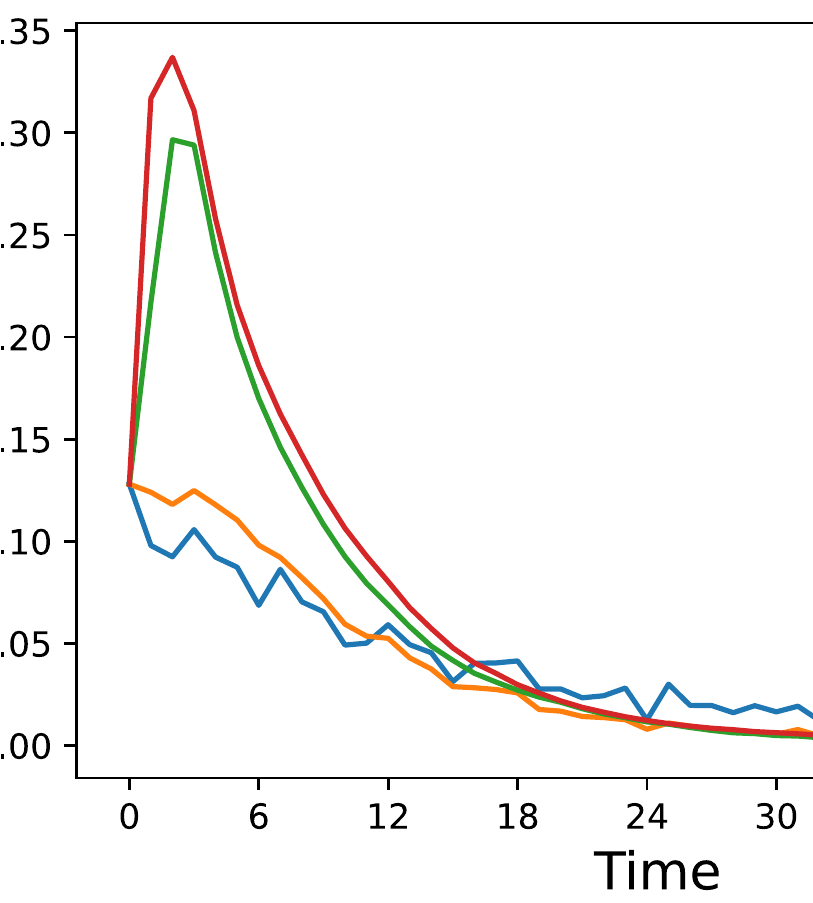}}
		\ 
    \subfloat{\includegraphics[width=\fppeval{0.32*\figratio}\columnwidth, viewport=35 0 266 260, clip]{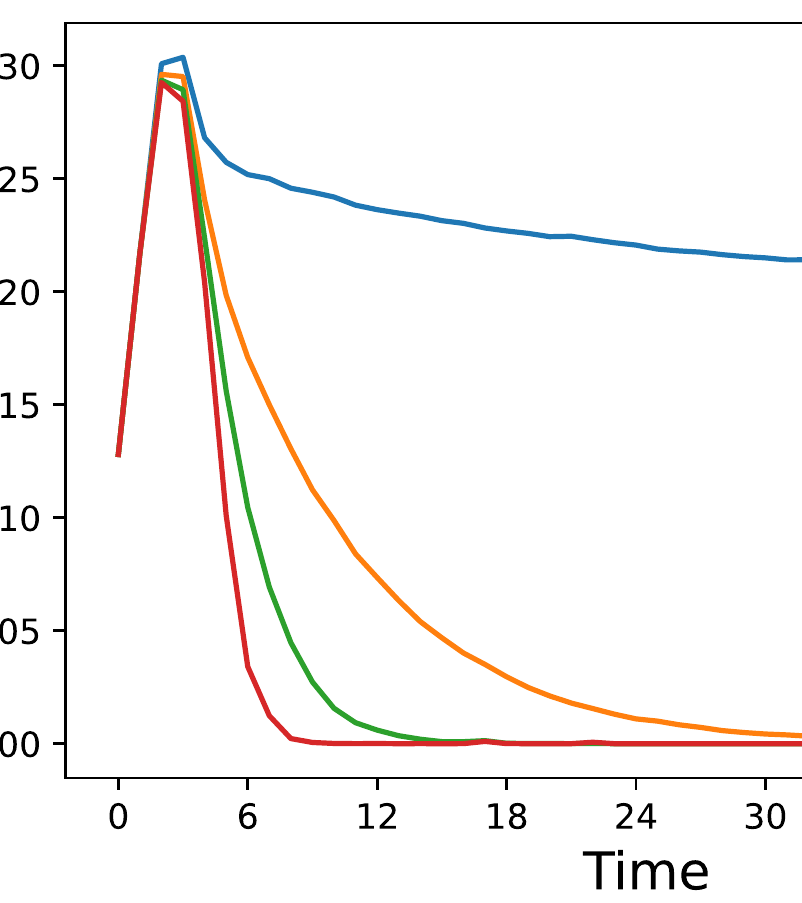}}
\end{minipage}\quad%
\begin{minipage}{\figpanel\linewidth}\centering 
		\subfloat{%
\!\!\!\!\!\includegraphics[width=\fppeval{0.364*\figratio}\columnwidth, viewport=0 0 266 260, clip]{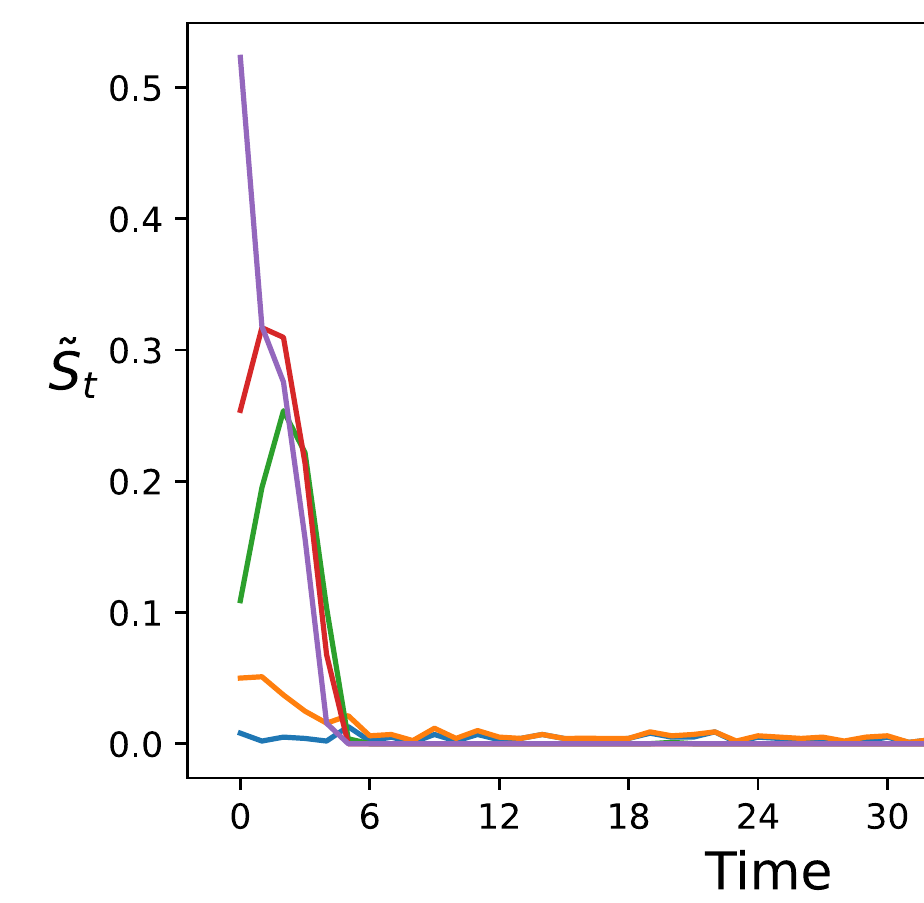}}
		\ 
    \subfloat{\includegraphics[width=\fppeval{0.32*\figratio}\columnwidth, viewport=32 0 266 260, clip]{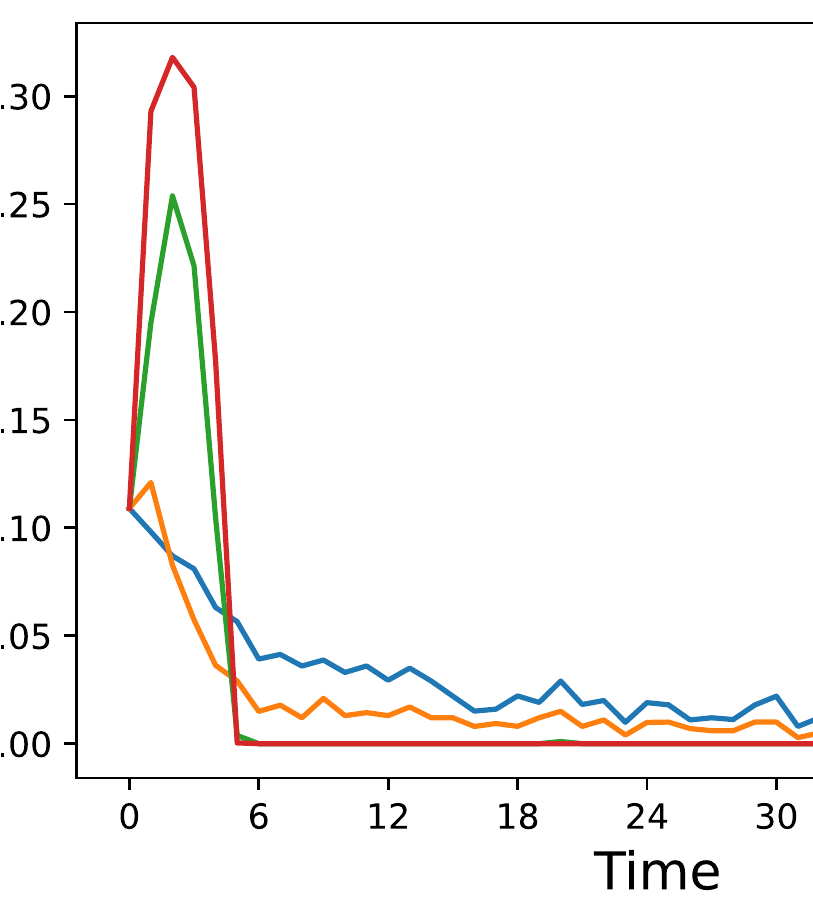}}
		\ 
    \subfloat{\includegraphics[width=\fppeval{0.32*\figratio}\columnwidth, viewport=32 0 266 260, clip]{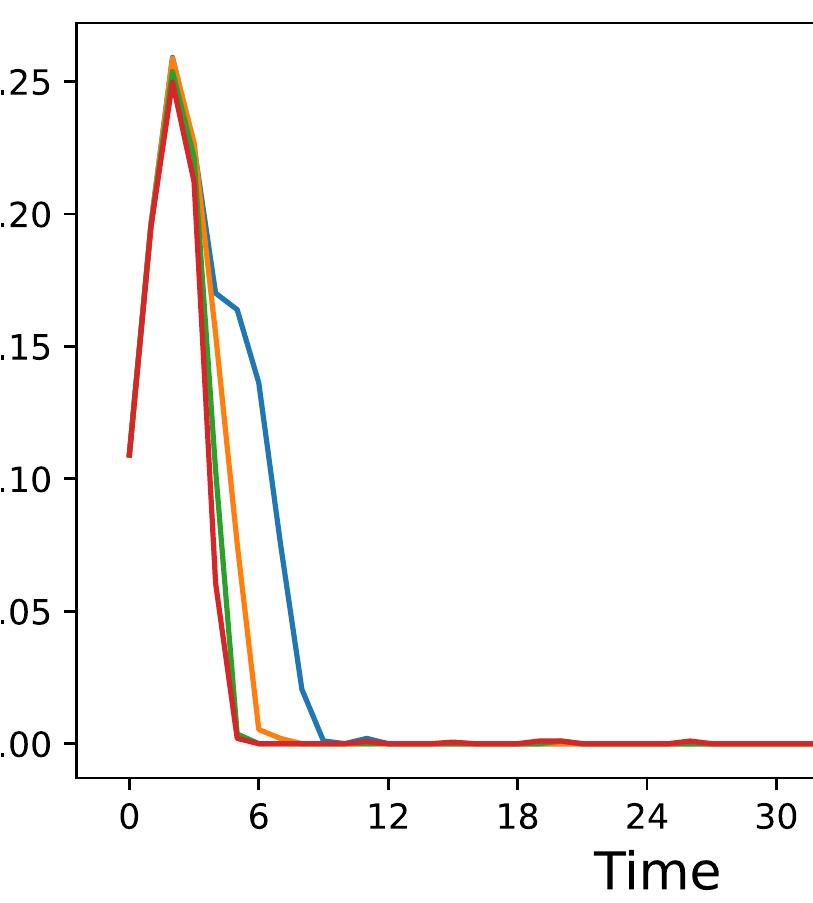}}
\end{minipage}

\vspace{-1.5em}

\hspace{-2.48em}
\setcounter{subfigure}{0}
\begin{minipage}{\figpanelsmall\linewidth}\centering
{\scriptsize $N=10000$}
\end{minipage}%
\begin{minipage}{\figpanel\linewidth}\centering 
		\subfloat[Effect of $\mu$]{%
		\!\!\!\!\!\includegraphics[width=\fppeval{0.364*\figratio}\columnwidth, viewport=0 0 266 260, clip]{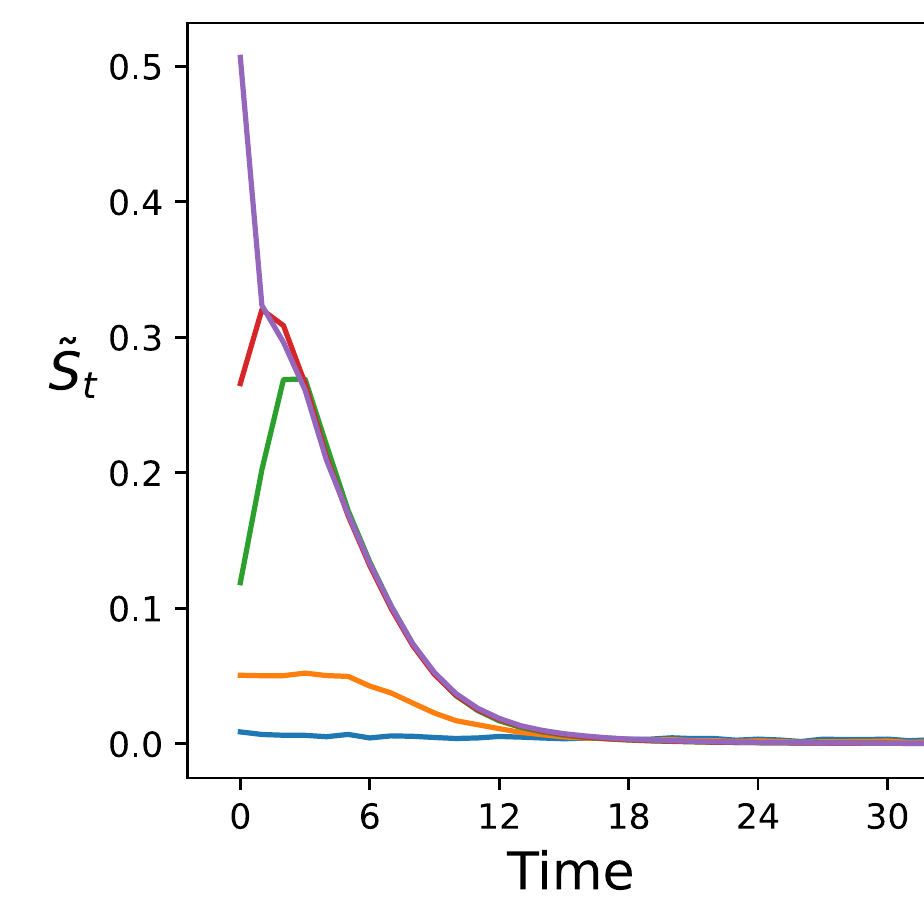}}
		\ 
    \subfloat[Effect of $\gamma$]{\includegraphics[width=\fppeval{0.32*\figratio}\columnwidth, viewport=32 0 266 260, clip]{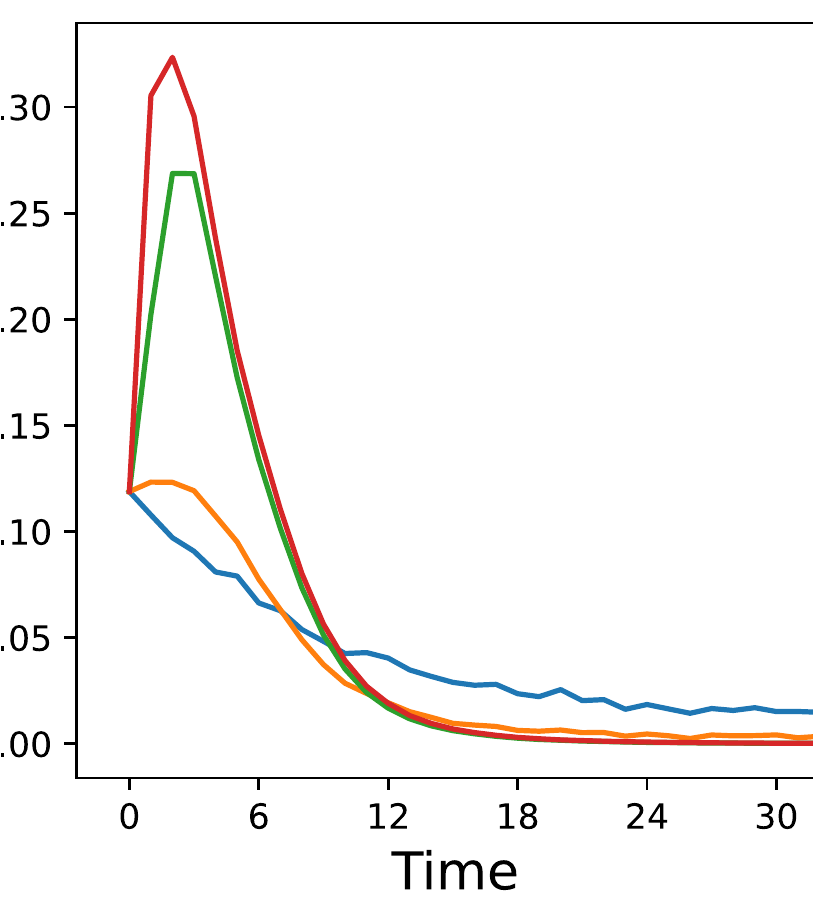}}
		\ 
    \subfloat[Effect of $m$]{\includegraphics[width=\fppeval{0.32*\figratio}\columnwidth, viewport=32 0 266 260, clip]{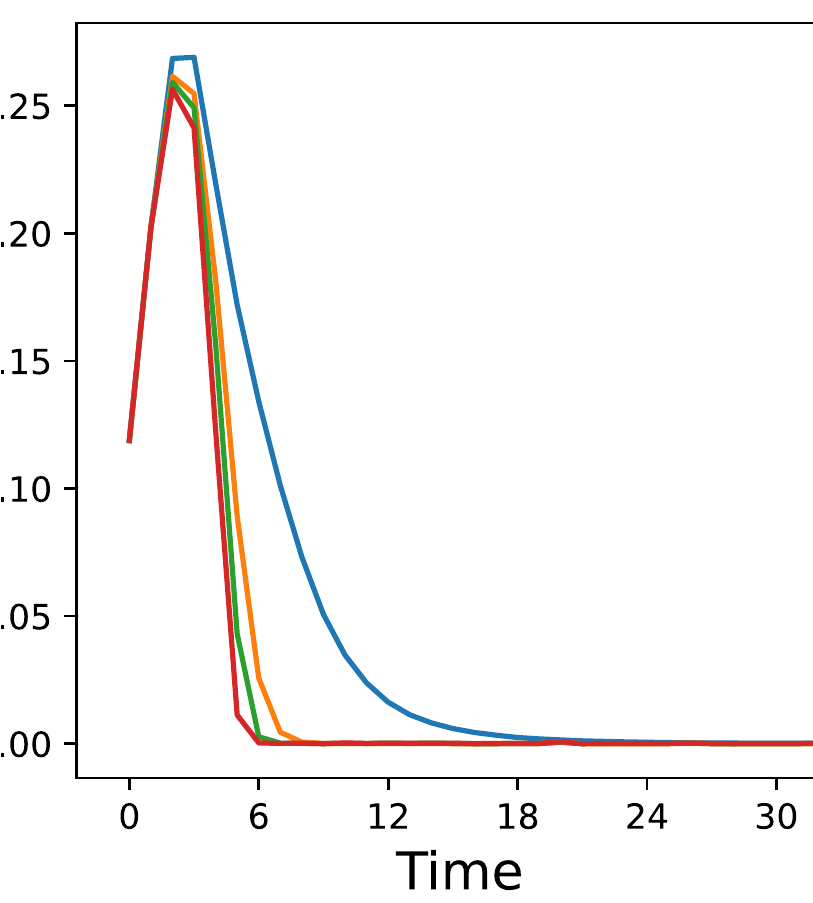}}
\end{minipage}\quad%
\begin{minipage}{\figpanel\linewidth}\centering 
		\subfloat[Effect of $\mu$]{%
\!\!\!\!\!\includegraphics[width=\fppeval{0.364*\figratio}\columnwidth, viewport=0 0 266 260, clip]{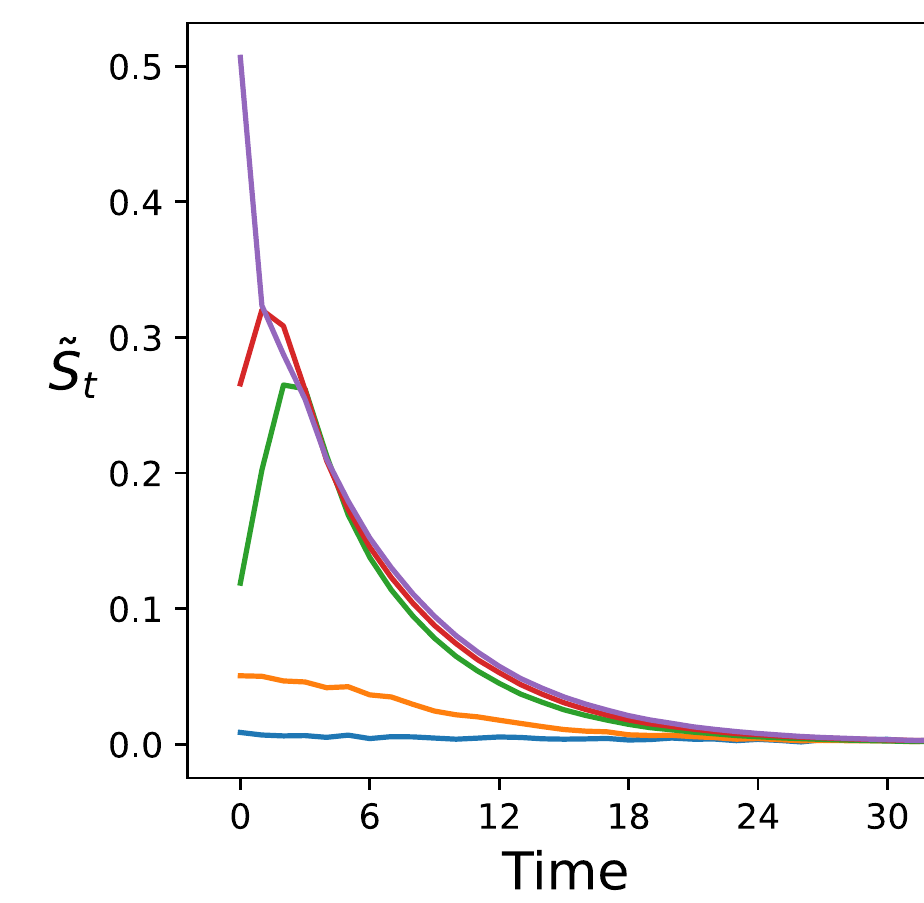}}
		\ 
    \subfloat[Effect of $\gamma$]{\includegraphics[width=\fppeval{0.32*\figratio}\columnwidth, viewport=32 0 266 260, clip]{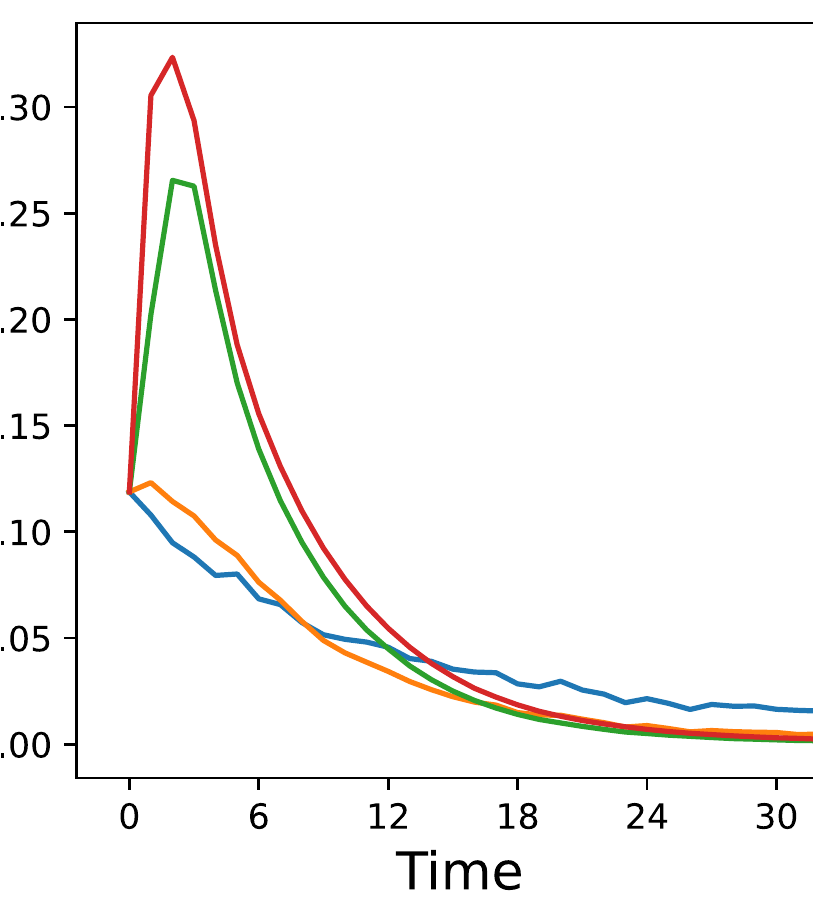}}
		\ 
    \subfloat[Effect of $k$]{\includegraphics[width=\fppeval{0.32*\figratio}\columnwidth, viewport=32 0 266 260, clip]{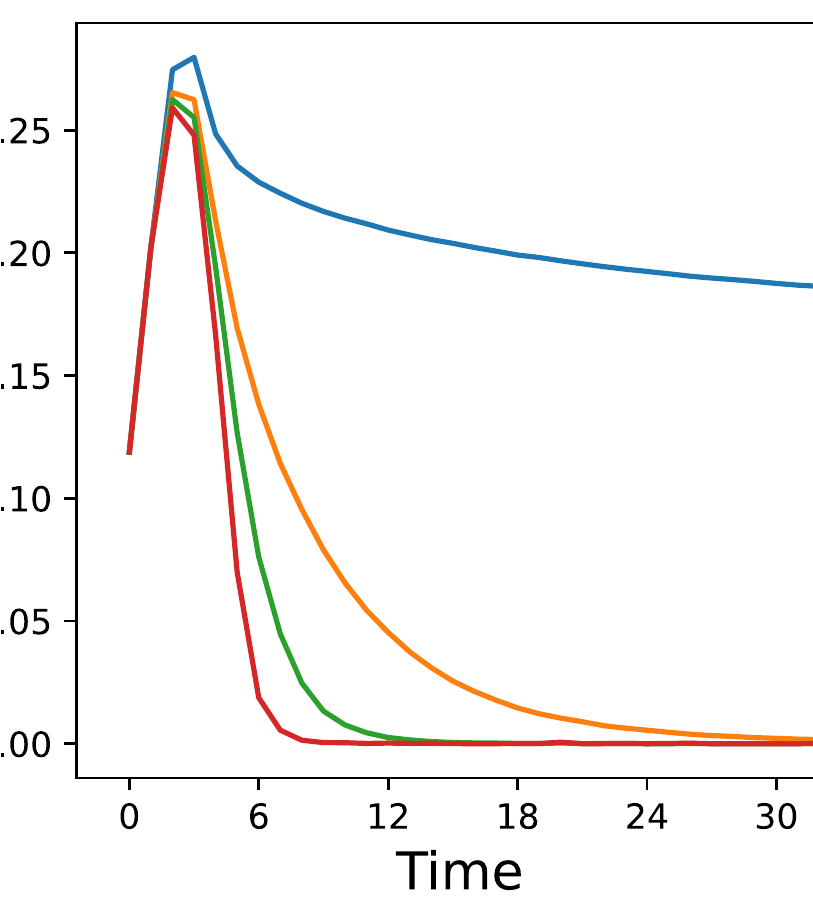}}
\end{minipage}\quad%
\begin{minipage}{\figpanel\linewidth}\centering 
		\subfloat[Effect of $\mu$]{%
\!\!\!\!\!\includegraphics[width=\fppeval{0.364*\figratio}\columnwidth, viewport=0 0 266 260, clip]{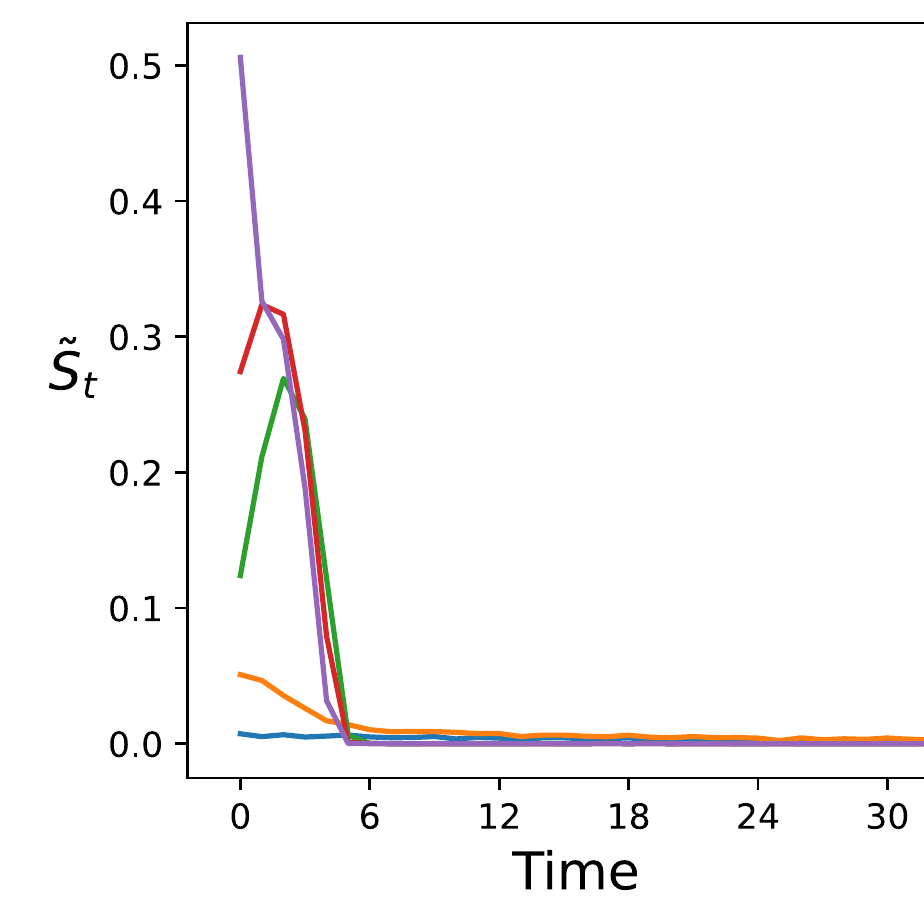}}
		\ 
    \subfloat[Effect of $\gamma$]{\includegraphics[width=\fppeval{0.32*\figratio}\columnwidth, viewport=32 0 266 260, clip]{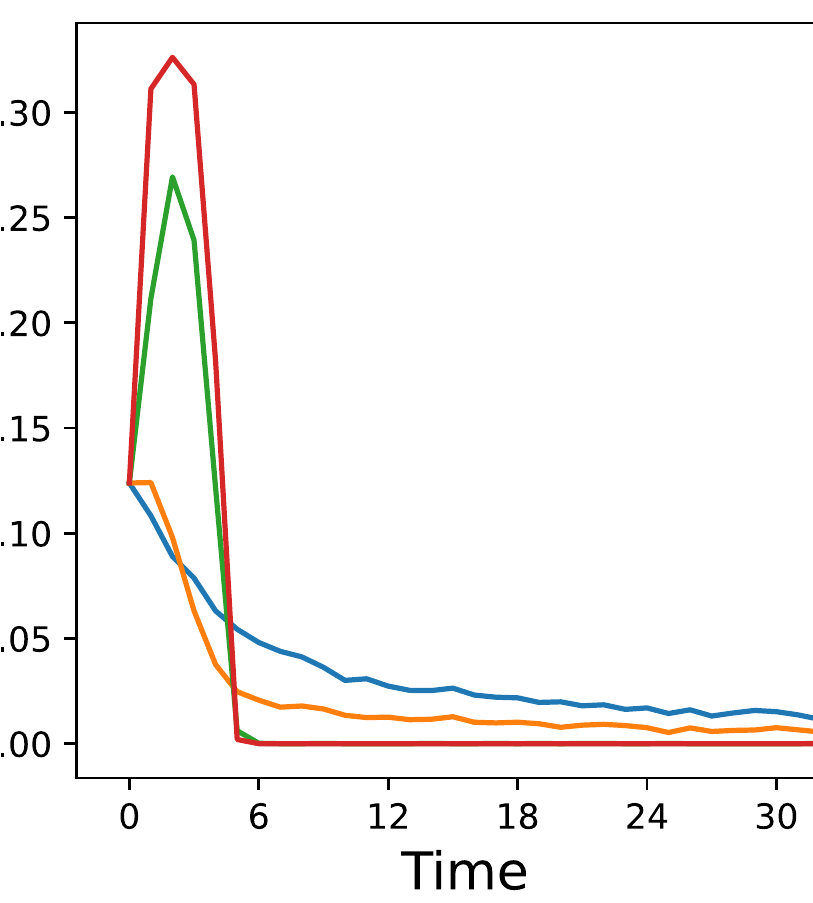}}
		\ 
    \subfloat[Effect of $r$]{\includegraphics[width=\fppeval{0.32*\figratio}\columnwidth, viewport=32 0 266 260, clip]{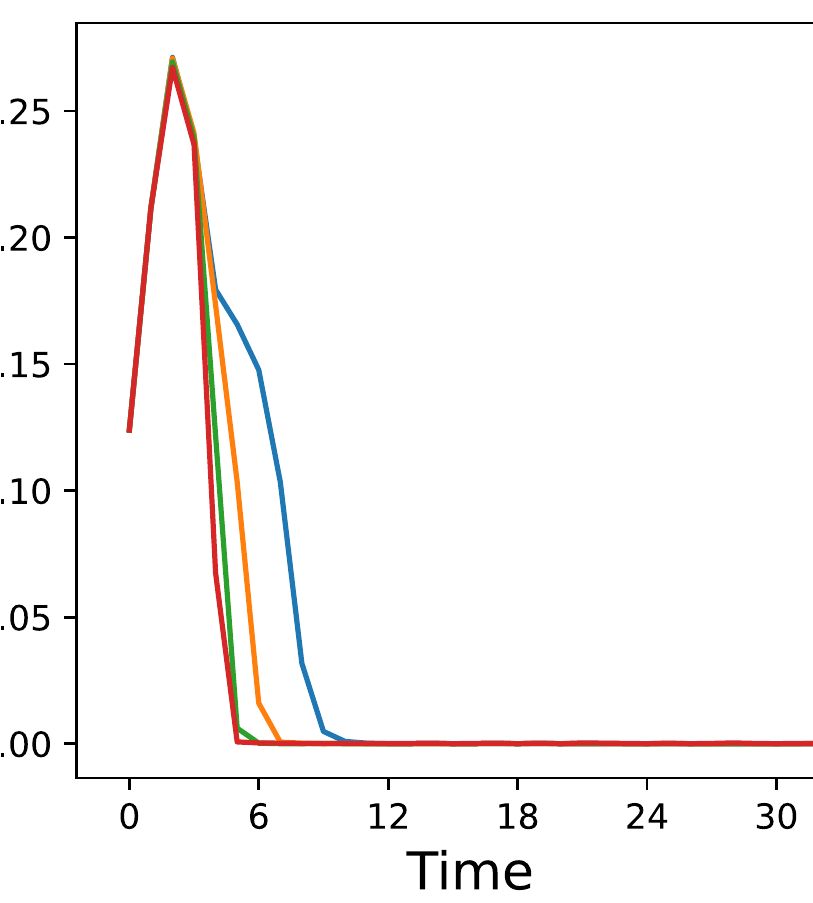}}
\end{minipage}
\caption{Simulations on networks of different node sizes ($N$) generated by the Barab\'asi-Albert, Watts-Strogatz, and SBM models. The same parameter value (same line color) has been used across different network scales and for different graph models. Comparing the plots column-wise for each parameter in effect, we can see that the network scale has little effect to the \modelname's behavior.}
     \label{fig:network-scale}
\end{figure*}

\begin{figure*}[t]%
\vspace{-1em}
\centering%
	\hspace{-8mm}%
	\subfloat[$X_{\min}$]{ \includegraphics[width=0.25\textwidth, viewport=0 0 400 255, clip]{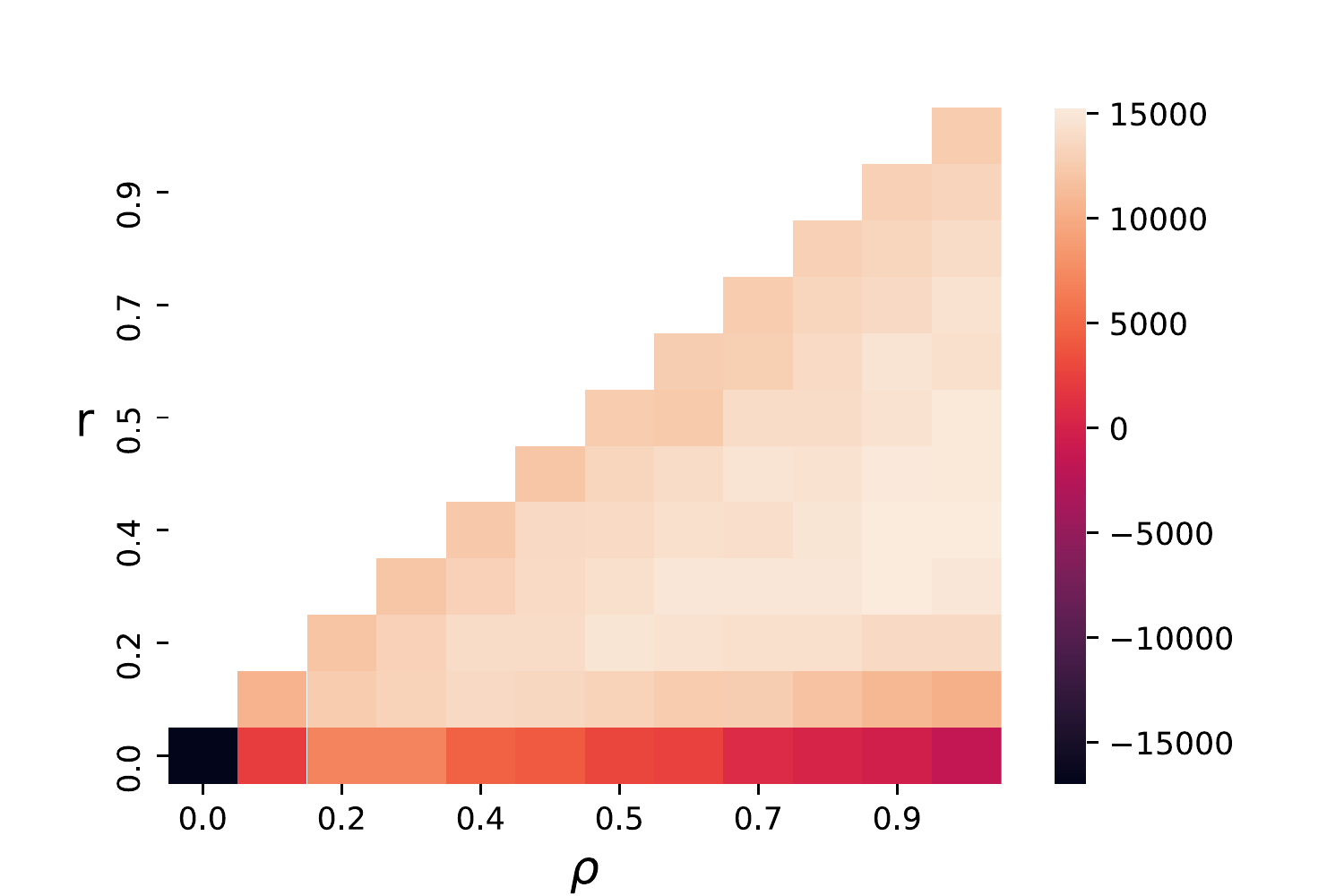} }%
		\hspace{-3.5mm}%
    \subfloat[$X_{\max}$]{\includegraphics[width=0.25\textwidth, viewport=0 0 400 255, clip]{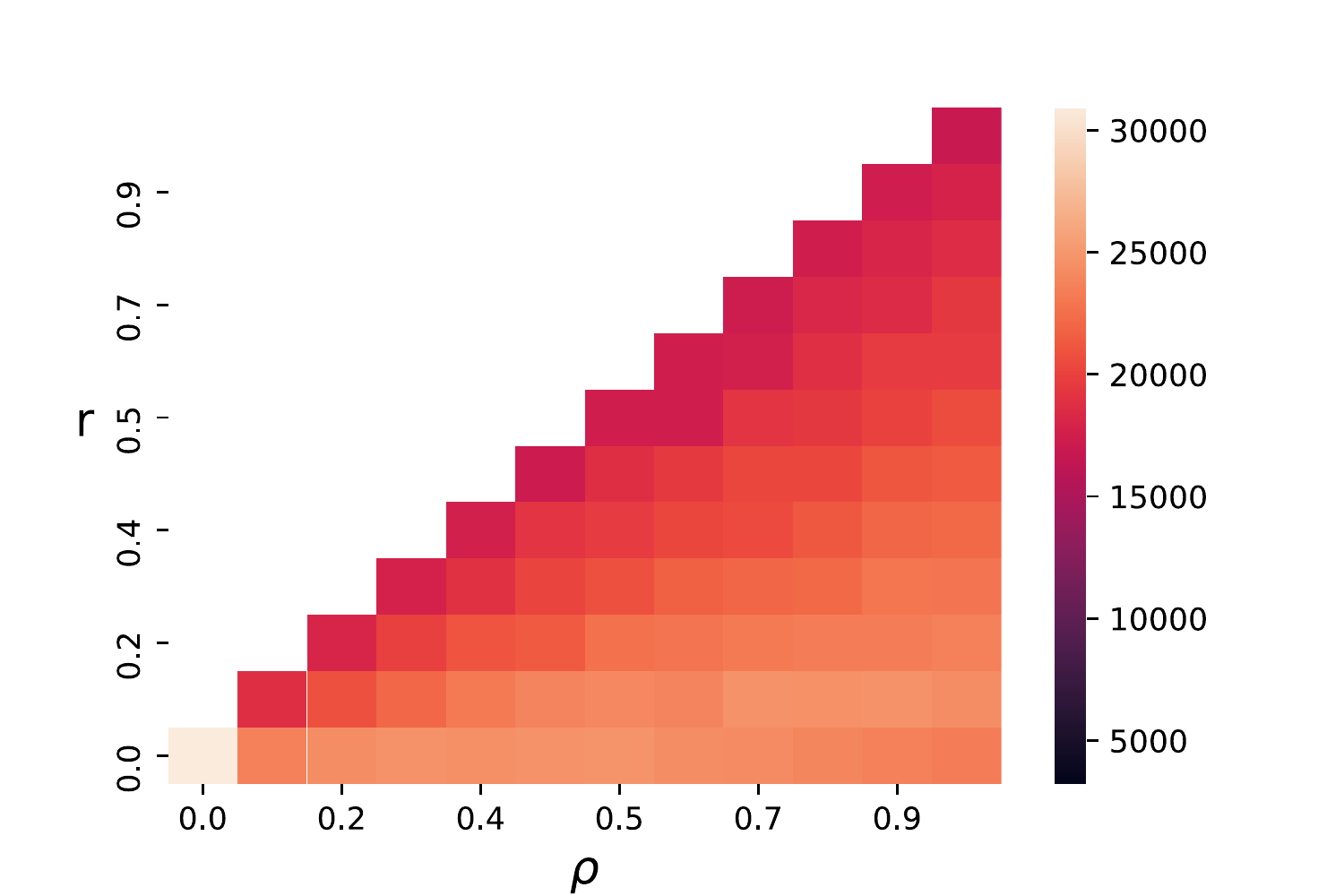} }%
		\hspace{-3.5mm}%
	\subfloat[$D_{\max}$]{ \includegraphics[width=0.25\textwidth, viewport=0 0 400 255, clip]{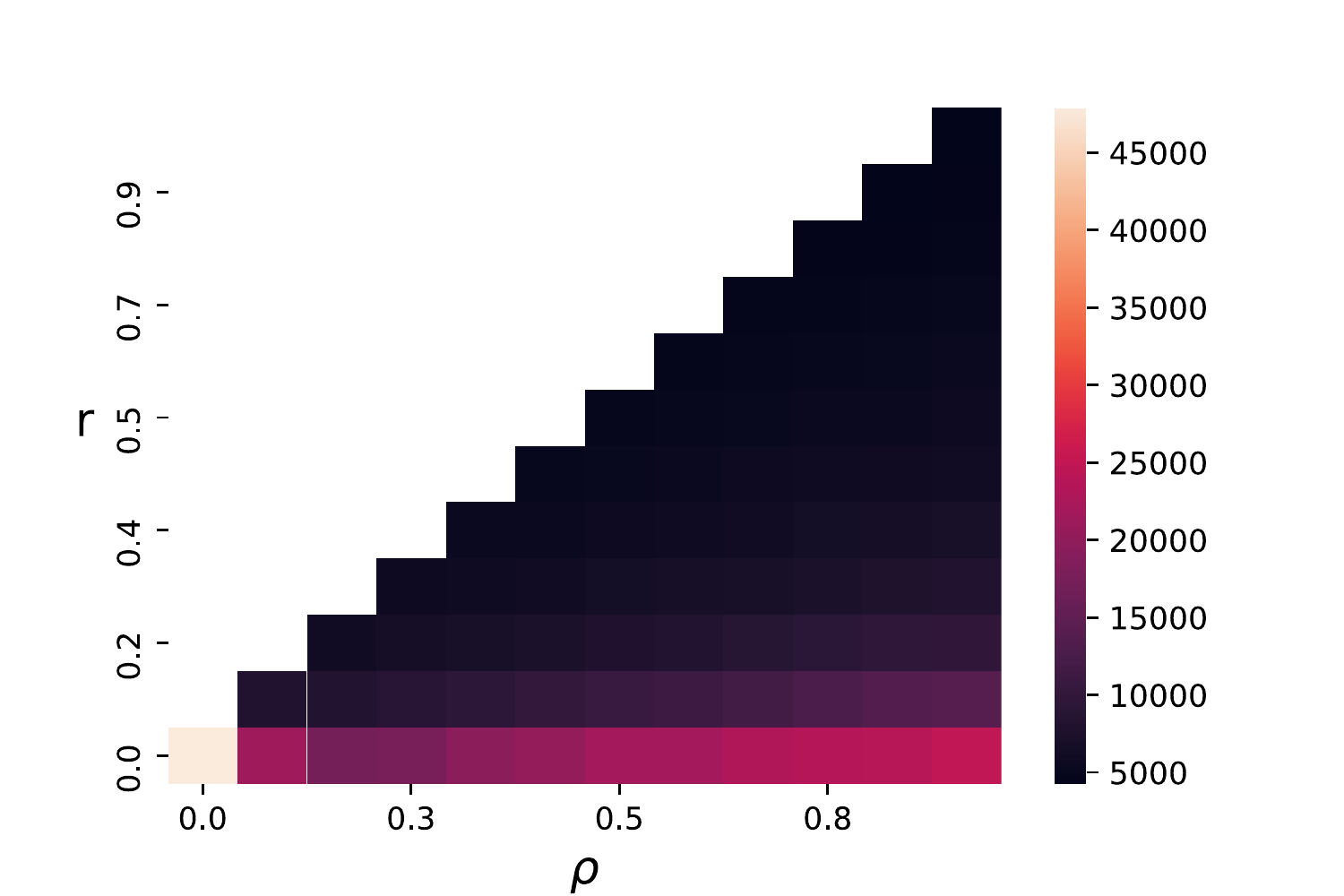} }%
	\hspace{-5mm}%
	
		\vspace{3mm}
		\ \hspace{0.2mm}\subfloat[$X_{\min}$]{\hspace{-8.95mm} \includegraphics[width=0.25\textwidth, viewport=0 0 400 255, clip]{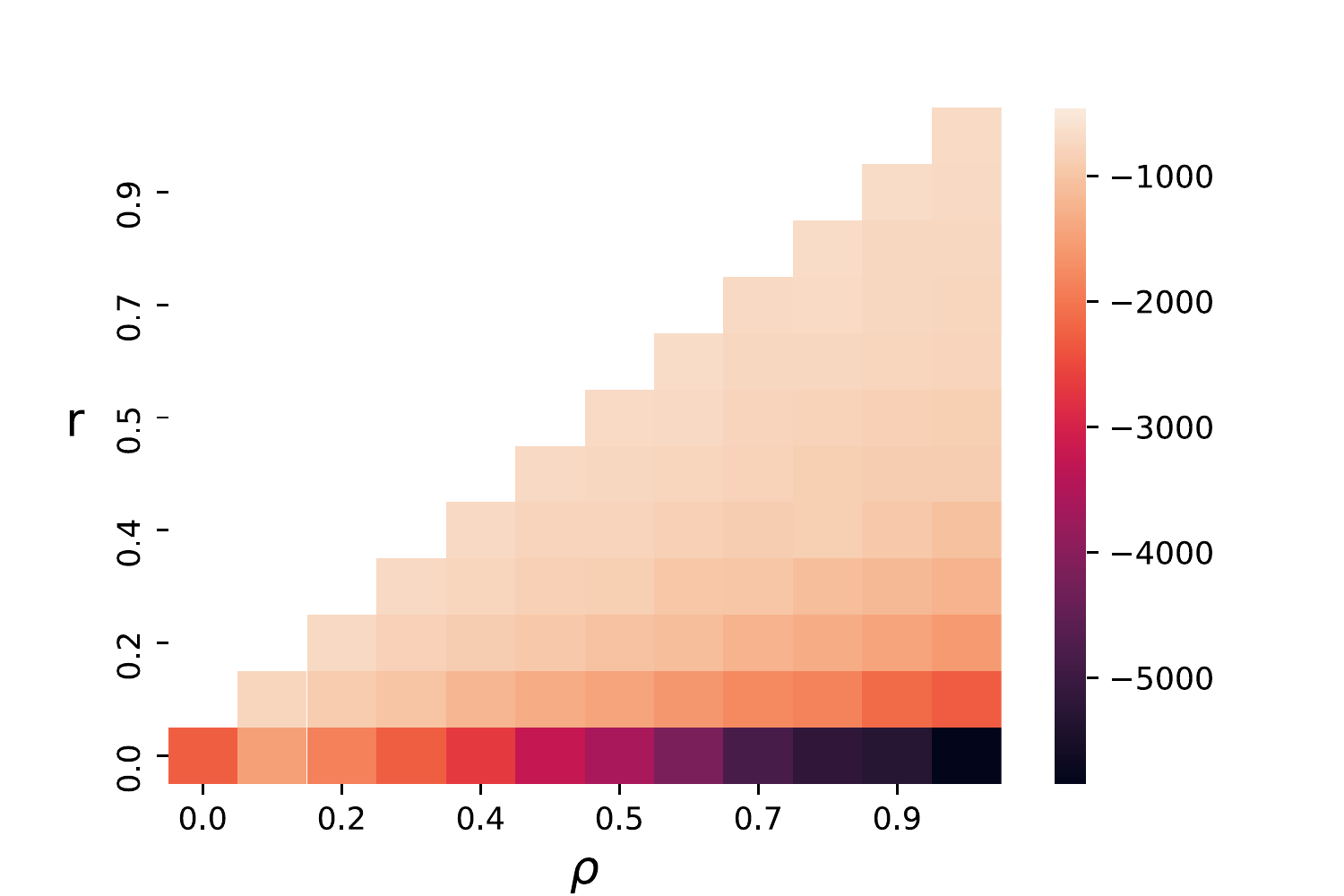} }%
			\hspace{-3.5mm}%
    \subfloat[$X_{\max}$]{\includegraphics[width=0.25\textwidth, viewport=0 0 400 255, clip]{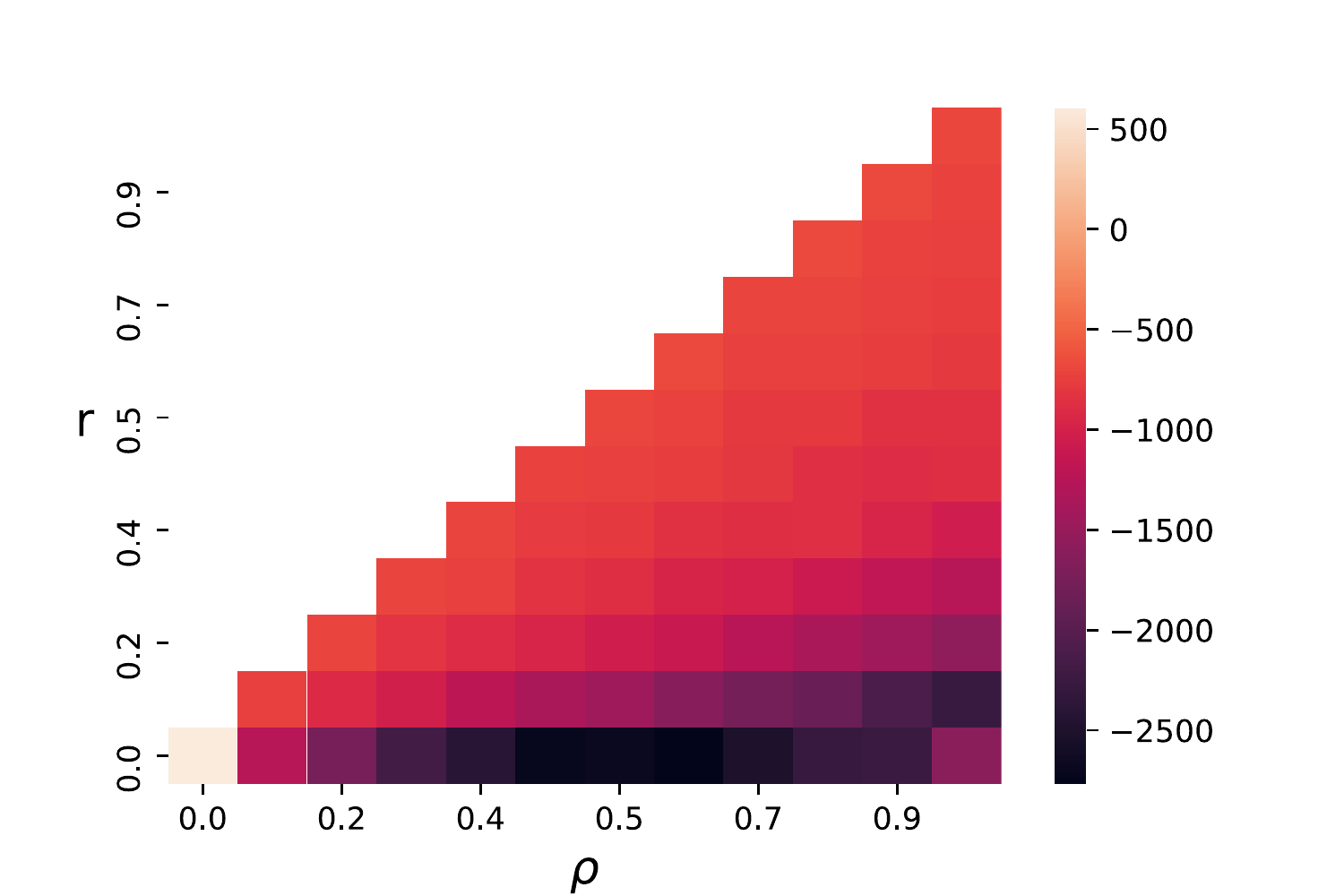}}%
			\hspace{-1.6mm}%
    \subfloat[$D_{\max}$]{\includegraphics[width=0.25\textwidth, viewport=0 0 400 255, clip]{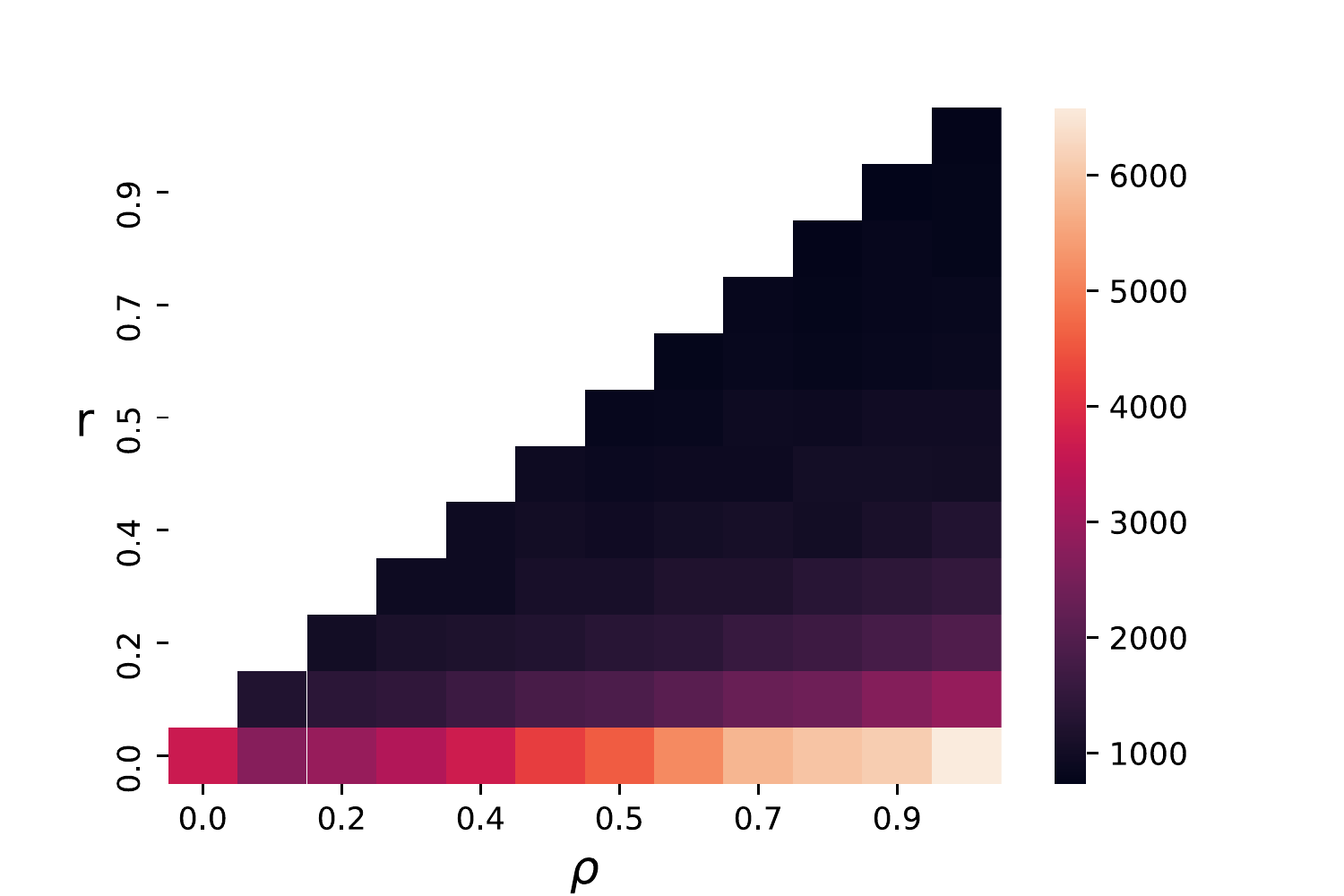} }%
			\hspace{-5mm}%
\caption{Heatmaps for the minimal and maximal opinions ($X_{\min}$, $X_{\max}$), as well as the maximal polarization ($D_{\max}$ of \Sec{sec:interplay}) observed when using graphs generated by two SBM configurations. \textbf{Top row (a-c):} first case, SBM with $\beta^{(1)}=0.958$, $\beta^{(2)}=0.041$. \textbf{Bottom row (d-f):} second case, SBM with $\beta^{(1)}=0.041$, $\beta^{(2)}=0.958$.}%
\label{fig:min-max-pol_heatmaps}
\end{figure*}
\begin{figure}[!t]
    \centering
		\vspace{-1em}%
    \subfloat{\!\!\!\!\!\includegraphics[width=0.495\columnwidth,viewport=5 0 390 270,clip]{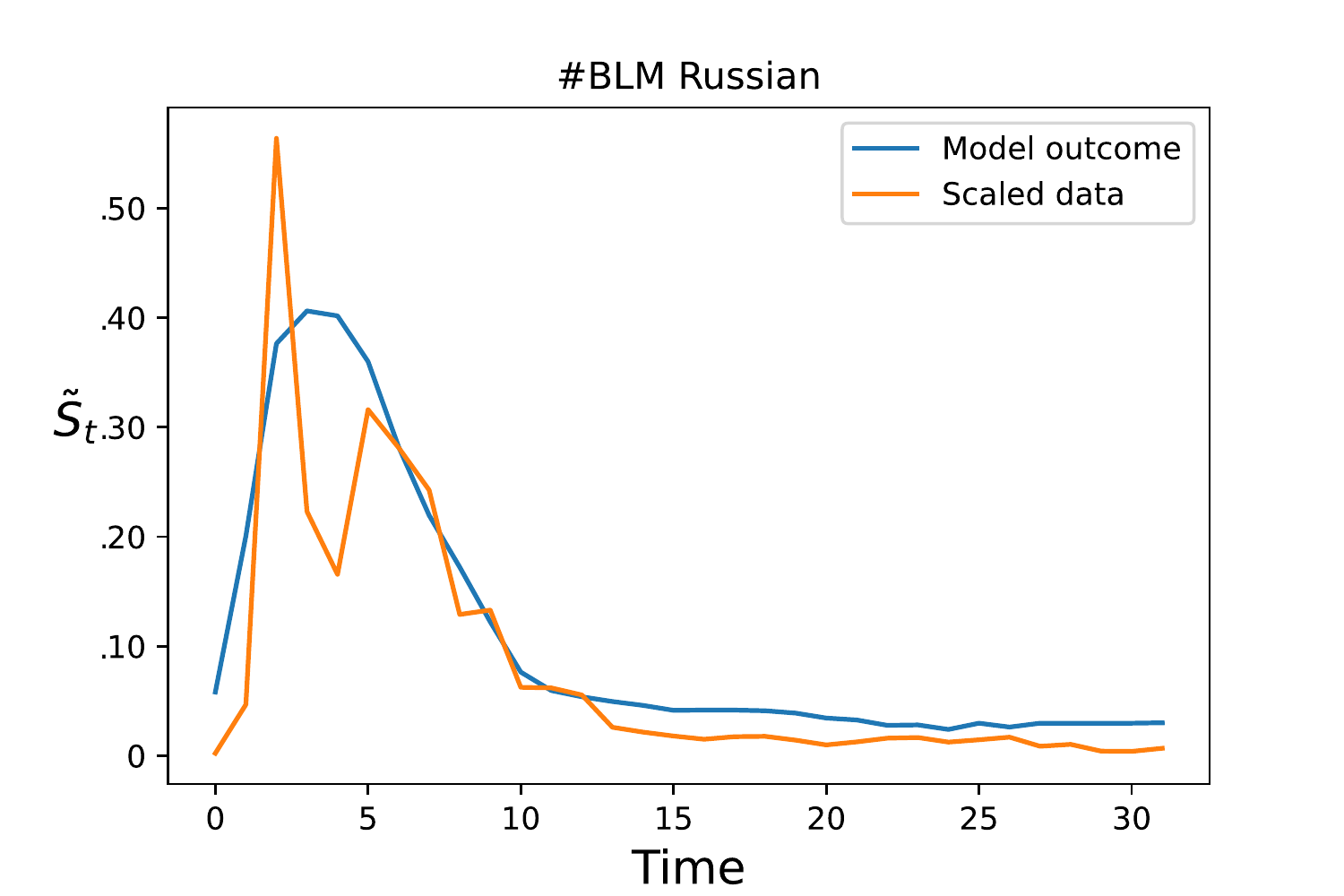}}%
		\ 
    \subfloat{\includegraphics[width=0.495\columnwidth,viewport=5 0 390 270,clip]{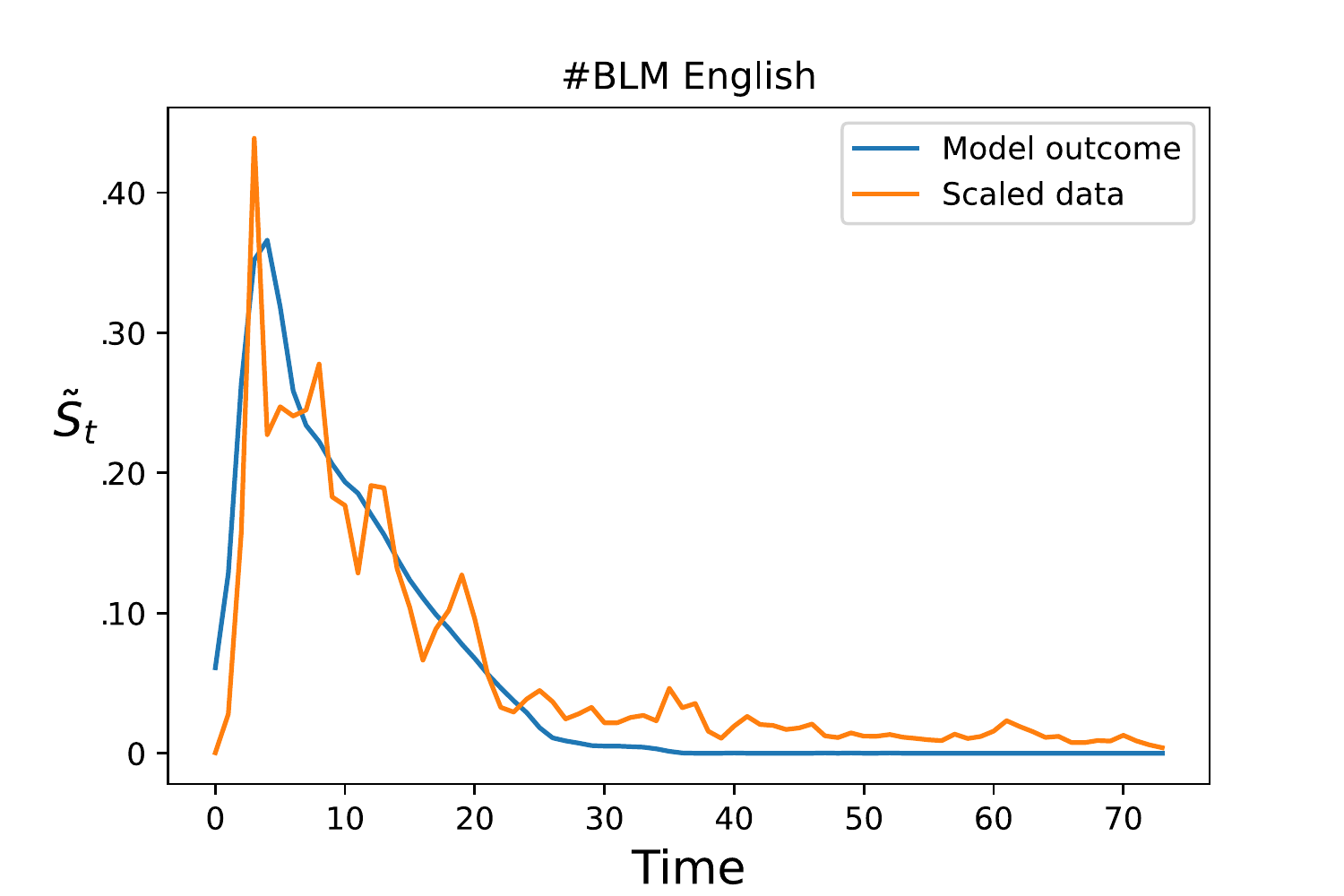}}
    \caption{Two examples of fitted models on scaled data, for %
		the Black Lives Matter (\#BLM) hashtag in Russian and English tweets.}
    \label{fig:fit-two-example}
\end{figure}

\inlinetitle{II.~Effect of network size and graph model parameters}~%
Earlier, for mere computational reasons, we have used relatively small networks of $N=100$ nodes for demonstrating the properties of the \modelname model, as well as for fitting it to real data. Here, we justify empirically this choice by showing that the model behavior does not present any significant change when the network size increases from $100$ nodes and up to two orders of magnitude larger, except from making the line plots a little smoother locally. We specifically run additional experiments to compare the model behavior on networks of size $N= \{100, 1000, 10000\}$ nodes. %
We use various topologies produced by standard random graph models: Barab\'asi-Albert, Watts-Strogatz, and Stochastic Block Model \cite{newman2002}. The edge weights were generated as described in \Sec{sec:properties}. \Fig{fig:network-scale} presents the results. We observed similar results for Erdös-Rényi and Grid graphs (omitted results). Therefore, the essence of our empirical analysis and the conclusions we make in this work are valid also for $N\gg100$.

\inlinetitle{III.~The ambiguous role of inter-cluster propagation}~%
Here we take a deeper look into the implications of the inter-cluster propagation mechanism in the SBM setting, which is a special sort of propagation mechanism. The SBM graphs are generated randomly and characterized by two parameters: $\rho$ is the probability with which two nodes of the same cluster are connected, $r$ is the probability with which two nodes of different clusters are connected. Thus, when $r$ gets closer to $\rho$, the model gets closer to a structure-less model with $\bar{\beta}\approx c^{(1)}\beta^{(1)}+c^{(2)}\beta^{(2)}$ ($c^{(1)}$ and $c^{(2)}$ are defined in \Sec{sec:properties}). We refer to \textit{increasing isolation} when $r$ gets closer to $0$ and $\rho$ closer to $1$, and we will be looking at the maximal and minimal opinions in two different cases: $\beta^{(1)}=0.958$, $\beta^{(2)}=0.041$ (i.e.~self-exciting regime since $\bar{\beta}=0.683>\frac{1}{2}$), $\beta^{(1)}=0.041$, $\beta^{(2)}=0.958$ (i.e.~self-cooling regime since $\bar{\beta}=0.316<\frac{1}{2}$).

Increasing isolation has the effect of radicalizing dynamics inside each cluster, and a weaker interaction with the steering mechanism: this creates different dynamics depending on which case we find ourselves in. In the first case (overall self-exciting, displayed in the top row of \Fig{fig:min-max-pol_heatmaps}), increasing isolation causes the dynamics in the majority cluster to be stronger, which in return feeds the dynamic of the minority one. As a consequence, the maximal opinion increases whereas the minimal opinion decreases. In the second case (overall self-cooling, displayed in the bottom row of \Fig{fig:min-max-pol_heatmaps}), increasing isolation causes the dynamics in the negatively-reacting majority cluster to get stronger, which decreases the minimal opinion but also decreases the strength of the steering mechanism. In return, this decreases the maximal opinion by calming the dynamics in the positively reacting minority cluster. 

This is a good example of how complex the dynamics created by the interaction of steering mechanism and propagation mechanism can get (in particular the inter-cluster propagation mechanism in the SBM setting), while they are quite simple when considered separately.

\inlinetitle{IV.~Fitting method}~%
Let $S=\{S_{t}\}_{t\le T}$ be the time-series of the data, and $m=\{m_{t}(\mu, \gamma, r)\}_{t\le T}
$ be the time-series issued by the model using a specific triplet of parameter values ($\mu, \gamma, r$). $T$ is simply the length of the time-series, we will use model outcomes with the same observation length. Finding the best way to reproduce the data using the model is expressed as:
\begin{equation}
(\mu^{*}, \gamma^{*}, r^{*})=\arg \min_{(\mu, \gamma, r)}d(S,m(\mu,\gamma,r)),
\end{equation}where $d$ is a scale-invariant distance (similarly to \cite{yang2011patterns}), so that 
\begin{equation}
d(S,S')=\frac{1}{\norm{S}_2}\min_{\lambda}\sqrt{{\textstyle\sum_{t=1}^{T}}(S_{t}-\lambda S'_{t})^2}. %
\end{equation}
The optimal scaling coefficient is $\lambda^{*}=\frac{\langle S ,\, S' \rangle }{\norm{S'}_2}$. Given the stochastic nature of our model and the absence of analytic form for $\{m_{t}(\mu, \gamma, r)\}_{t\le T}$, there is no way for us to use gradient-based optimization method, so that we will primarily rely on simulated annealing \cite{kirkpatrick1983optimization} for the fitting. As our model is stochastic, and we use a scale-free distance, it is likely that in noisy regions of the parameter space the model generates similar time-series as the data without being informative. To cope with this issue, we add a variance indicator to the distance in order to avoid too noisy regions.

Our optimization process takes place in the parameter space $\Omega = [\underline{\mu},\bar{\mu}]\times [\underline{\gamma}, \bar{\gamma}] \times [\underline{r}, \bar{r}]$, with bounds $\underline{\mu}=-500$, $\bar{\mu}=500$, $\underline{\gamma}=0$, $\bar{\gamma}=50$, $\underline{r}=0$, $\bar{r}=\frac{1}{2}$. The strategy we follow uses a first exploration phase, during which we get the model's fitting error to the data as well as an estimation of the variance at the center of each cell of the parameter grid, and then run $K$ instances of simulated annealing using the $K$ best points (i.e.~solutions: triplets of parameter values) found earlier as initialization points. The simulated annealing requires us to define a neighborhood for any point of the parameter space. For a point $x$, we define its neighborhood $\mathcal{N}(x)$ as a cuboid centered at $x$, and of volume $vol<1$ times the total volume of the space $\Omega$. We also opt for a continuously decreasing temperature such that $\forall t, \text{temp}_{t}=\eta\text{temp}_{t-1}$. We set the hyperparameters at $T_{0}=10$, $\eta=0.95$, $vol=0.001$. A visual example of model fittings is shown in \Fig{fig:fit-two-example}.

\begin{figure}[t]
    \centering
		\includegraphics[width=\columnwidth, viewport=53 5 390 275,clip]{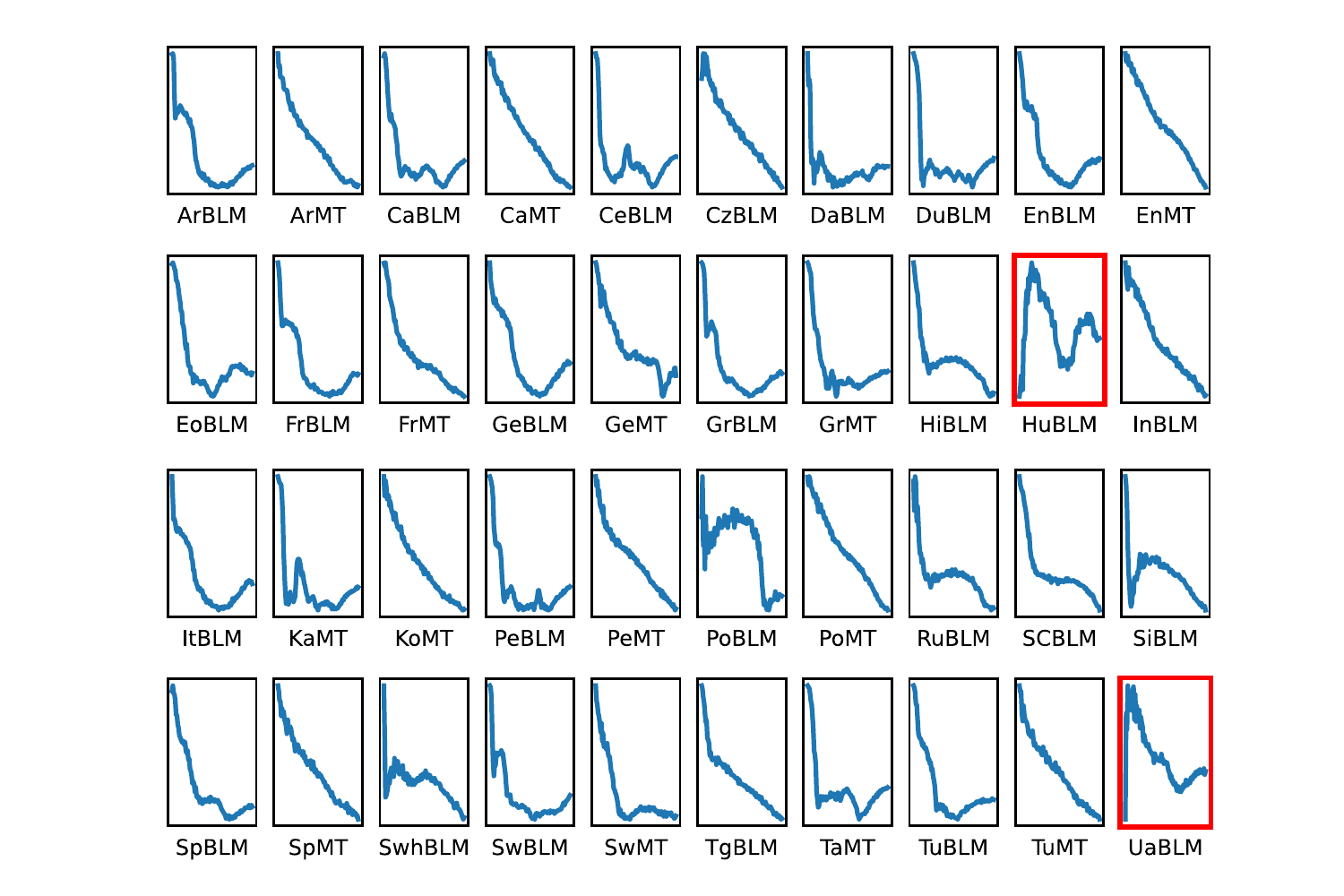}
		\vspace{-4mm}
    \caption{The bootstrap metric $\chi(q)$ (y-axis) as a function of $q$ (x-axis) for each language. The metric decreases rapidly (with two exceptions indicated by red frames: HuBLM and UaBLM) as $q$ increases and a wider $q$-quantiles of model parametrizations are included in the top performing set $P^*(q)$. This provides evidence for the extent of our model's identifiability in these scenarios.}
    \label{fig:bootstrap}
\end{figure}

\inlinetitle{V.~Model identifiability}~%
When it comes to the outcome of the optimization process, i.e.~the learned parameter values, an important point is the \emph{model identifiability}. More specifically, for each use-case, we would like to empirically verify that the combination of parameter values (i.e.~specific ($\mu, \gamma, r$) triplets) that lead to good fitting of our model to the data lie in a relatively compact region of the parameter space (as opposed to being arbitrarily scattered therein, which would mean that multiple largely different parametrizations lead to equally good local optima). It is convenient to denote by $P^*(q)$ the top-$q$ parametrizations, as to be the top $q$-quantile of triplets leading to the lowest data fitting error, where $q\in[0,1]$ is a chosen threshold. Similarly, we denote by $P(q)$ a bootstrap subset of the same size, containing randomly drawn triplets. 

The main idea we employ aiming to investigate thoroughly model identifiability, is the comparison of these two sets, $P^*(q)$ and $P(q)$, in terms of a notion of variance, which we express as $\text{Variance}(P(q)) = \frac{1}{|P(q)|}\sum_{p\in P(q)}\lVert p - \bar{p}(q) \rVert _{2}$, where $\bar{p}(q)$ is the barycenter of $P(q)$  (similarly we compute for $P^*(q)$). Then, the procedure we follow is detailed below:
\begin{itemize}
\item[1.] We first explore the parameter space by computing the fitting error of our model in each cell of a sufficiently fine-grained grid over $\Omega$. Let the number of cells be %
$|\Omega|_\text{grid}$.

\item[2.] We let $q$ vary in $[10^{-4}, 10^{-2}]$, and for each %
$q$ value: 
\begin{itemize}
\item[2a.] we compile $P^*(q)$ with the $\lfloor q|\Omega|_\text{grid}\rfloor$ best triplets leading to the best model fitting;
\item[2b.] we compile multiple $P_i(q)$, $i=1,...,B$, by drawing each time the same number of triplets, but selecting them uniformly at random. We set the number of bootstrap samples at $B=10$. 
\end{itemize} 

\item[3.] We use the following metric:%
\vspace{-0.2em}
\begin{equation}
\!\!\!\!\!\chi(q) = \frac{1}{B}\sum_{i=1}^B \!\Big(\!\text{Variance}(P_i(q)) - \text{Variance}(P^*(q)) \!\Big).\!\!\!\!
\end{equation}
This measures the statistical significance of the concentration (i.e.~low variance) of the best triplets in the parameter space, compared to bootstrap subsets of triplets. $\chi(q)$ is expected to give a high positive value when the top performing triplets are concentrated, and a smaller positive or even negative value when there is no statistical evidence for concentration.
\end{itemize}

The results of \Fig{fig:bootstrap} show that $\chi(q)$ (y-axis) is always decreasing in $q$ (x-axis) for the vast majority of cases (only $2$ exceptions out of $41$ languages). This general decreasing tendency of $\chi(q)$ indicates that the good performing triplets of parameter values lie at a relatively compact and delimited region of the parameter space, which indeed provides evidence for model identifiability.

\bibliographystyle{IEEEtran}
%\bibliography{bibliography}
% Generated by IEEEtran.bst, version: 1.14 (2015/08/26)

%\section{Biography Section}
%...

%

\end{document}